\documentclass[a4paper]{article}
\usepackage{amsmath,latexsym,amsfonts,pstricks}
\newcommand{\scr}{\scriptscriptstyle}
\newcommand{\eqa}{\begin{eqnarray}}
\newcommand{\neqa}{\end{eqnarray}}
\newcommand{\equ}{\begin{equation}}
\newcommand{\nequ}{\end{equation}}

\newcommand{\di}{\textrm{dim}}

 \newcommand{\Ref}[1]{(\ref{#1})}
 
\begin{document}

\title{\bf{\LARGE The complete LQG propagator \\
I. Difficulties with the Barrett-Crane vertex}} 
\author{\large
Emanuele Alesci${}^{ab}$, Carlo Rovelli${}^b$
 \\[3mm]
\em
\normalsize
{${}^a$ Dipartimento di Fisica Universit\`a di Roma Tre, I-00146 Roma EU}
\\ \em\normalsize{${}^b$Centre de Physique Th\'eorique de Luminy%
\footnote{Unit\'e mixte de recherche (UMR 6207) du CNRS et des
Universit\'es de Provence (Aix-Marseille I), de la Mediterran\'ee
(Aix-Marseille II) et du Sud (Toulon-Var); laboratoire affili\'e \`a
la FRUMAM (FR 2291).}, Universit\'e de la M\'editerran\'ee, F-13288
Marseille EU}}

\date{\small\today}
\maketitle \vspace{-.6cm}

\begin{abstract}

\noindent Some components of the graviton two-point function have been 
recently computed in the context of loop quantum gravity, using the spinfoam 
Barrett-Crane vertex.  We complete the calculation of the remaining components. 
We find that, under our assumptions, the Barrett-Crane vertex 
does \emph{not} yield the correct long distance limit. 
We argue that the problem is general 
and can be traced to the intertwiner-independence of the Barrett-Crane vertex, 
and therefore to the well-known mismatch between the Barrett-Crane formalism
and the standard canonical spin networks.  In a companion paper we illustrate
the asymptotic behavior of a vertex amplitude that can correct this difficulty. 
\end{abstract}

\section{Introduction}

A key problem in loop quantum gravity (LQG) \cite{lqg1,book,lqg2} is to derive 
low--energy quantities from the full background independent theory. 
A strategy for addressing this problem was presented in 
\cite{scattering1} and some components of the graviton propagator of linearized
quantum general relativity
\begin{equation}
G^{abcd}(x,y)=\left\langle 0| h^{ab}(x)h^{cd}(y) |0\right\rangle 
\label{partenza-1}
\end{equation}
($h^{ab}(x), a,b=1,...4,$ is the linearized 
gravitational field) were computed in \cite{scattering2} (at first
order) and  \cite{scattering3}  (to higher order)
starting from the
background-independent theory and using a suitable
expansion. More precisely, 
the ``diagonal" components  $G^{aacc}(x,y)$ have been computed 
in the large-distance limit.  This result has
been extended to the three-dimensional theory in \cite{3d}; an improved form
of the boundary states used in the calculation has been considered in 
\cite{boundarystate}; and the exploration of 
some Planck-length corrections to the propagator of the linear theory has begun in \cite{corrections}. See also \cite{Dittrich:2007wm}.

Here we complete the calculation of the propagator. We 
compute the nondiagonal terms of $G^{abcd}(x,y)$, those where 
$a\ne b$ or $c\ne d$, and therefore derive the full tensorial structure
of the propagator.  The nondiagonal terms are important because 
they involve the 
\emph{intertwiners} of the spin networks.  Avoiding the complications
given by the intertwiners' algebra was indeed 
the rational behind the relative simplicity of the 
diagonal terms. 

The dependence of the 
vertex from the intertwiners is a crucial aspect of the definition of the
quantum dynamics.  The particular version of the dynamics used 
in \cite{scattering2} and \cite{scattering3}, indeed, is defined by the 
Barrett-Crane (BC) vertex \cite{BC}, where the dependence on the intertwiners 
is trivial.  This is an aspect of the BC dynamics that has long been seen as 
suspicious (see for instance \cite{baez}); and it is directly tested here.  

We find that under our assumptions the BC 
vertex \emph{fails} to give the correct tensorial structure 
of the propagator in the large-distance limit.  We argue that this
result is general, and cannot be easily corrected, say by a different
boundary state. 
This result is of interest for a number of reasons.  First, it indicates that the 
propagator calculations are nontrivial; in particular they are not governed just by
dimensional analysis, as one might have worried, and they do test the
dynamics of the theory.   Second, it reinforces the  
expectation that the BC model fails to yield classical GR in the long-distance limit. 
Finally, and more importantly, it opens the possibility of studying the conditions that an 
alternative vertex must satisfy, in order to yield the correct long-distance 
behavior.   This analysis is presented in the companion paper \cite{II}.

The BC model exists in a number of variants \cite{book,variants}; the results
presented here are valid for all of them. 
Alternative models have been considered, see for instance \cite{Reisenberger:1997sk}.   
Recently, a vertex amplitude that modifies the 
BC amplitude, and which addresses precisely the problems that
we find here, has 
been proposed \cite{vertice nuovo1,Livine:2007vk}, see also \cite{sergei}.  
It would be of great interest to repeat the calculation presented here for 
the new vertex proposed in those papers. 

This paper is organized as follows.  In section 2 we formulate the 
problem and we compute the action of the field
operators on the intertwiner spaces. This calculation is a technical result
with an interest in itself. Here we will use only part of this result, the rest will be
relevant for the companion paper.  In section 3 we discuss the
form of the boundary state needed to describe a semiclassical geometry
to the desired approximation. Section 4 contains the main calculation.
In Section 5 we discuss the interpretation of our result. 

This paper is not self-contained: for full background, see \cite{scattering3}.
For an introduction to the general ideas and the formalism, see the book \cite{book}.  
However, we include here a detailed Appendix, with all basic equations of the recoupling 
theory needed for the calculations. The Appendix corrects some imprecisions in
previous formularies and can be useful as a tool for further developments. We work
entirely in the euclidean theory. 

\section{The propagator in LQG}

We refer to \cite{scattering3} for the notation and the basic definitions. We want to compute
\begin{equation}
{\mathbf G} _{\mathbf q}^{abcd}(x, y) =  \langle W | h^{ab}( x)\  h^{cd}( y) | 
\Psi_{\mathbf q} \rangle
\label{partenza0}
\end{equation}
to first order in $\lambda$.  Here $\Psi_{\mathbf q}$ is a state peaked on $\mathbf q$,
which is  the (intrinsic and extrinsic)  3d geometry of the boundary of a spherical 
4-ball of radius $L$ in $R^4$, $x$ and $y$ are two points in this geometry, 
$\scriptstyle  ab$ are tangent indices at $x$ and $\scriptstyle  cd$ tangent indices at $y$ in this 
geometry.  That is, ${\mathbf G} _{\mathbf q}^{abcd}(x, y)$ is a quantity that
transforms covariantly under  3d diffeomorphisms acting \emph{conjointly} 
on $x, y$, on the indices $\scriptstyle  abcd$, \emph{and} on $\mathbf q$. 
$h^{ab}(x)$ is the fluctuation of the gravitational field over the euclidean metric.
$W$ is the boundary functional, that defines the dynamics; it is assumed to be
given here by a Barrett-Crane GFT \cite{variants,GFT} with coupling constant $\lambda$.
The expansion parameter $\lambda$ is a cut-off in the degrees of freedom of 
wavelength smaller than $L$.  Degrees of freedom of wavelength larger than 
$L$ do not enter the problem.   We normalize 
here $\Psi_{\mathbf q} $  by $\langle W | \Psi_{\mathbf q} \rangle=1$.  

Consider the $s$-knot (abstract spin network) basis $|s\rangle=|\Gamma,  {\mathbf j}, {\mathbf i}\rangle$, where $\Gamma$ is an abstract 
graph, $n,m...$ label the nodes of $\Gamma$,  $ {\mathbf j}=\{j_{mn}\}$ are the spins and ${\mathbf i}=\{i_n\}$ the intertwiners of a spin network with graph $\Gamma$.  Insert a resolution of the identity in (\ref{partenza0})
\begin{equation}
{\mathbf G} _{\mathbf q}^{abcd}(x, y) = \sum_s \langle W |s\rangle \langle s |  h^{ab}( x)\   h^{cd}( y) | \Psi_{\mathbf q} \rangle.
\label{partenza00nonmipiace}
\end{equation}
To first order in $\lambda$, $\langle W |s\rangle =W[s]=W[\Gamma,  {\mathbf j}, {\mathbf i}]$ is different from zero only if $\Gamma$ is the pentagonal graph, that is, for the s-knot  
\begin{equation}s = \hspace{-4em}\setlength{\unitlength}{0.0005in} \begin{picture}(5198,1100)(5000,-4330)\thicklines\put(8101,-5161){\circle*{68}}\put(8401,-3961){\circle*{68}}\put(6601,-3961){\circle*{68}}\put(7501,-3361){\circle*{68}}\put(6901,-5161){\circle*{68}}\put(8101,-5161){\line(1,4){300}}\put(7501,-3361){\line( 3,-2){900}}\put(6601,-3961){\line( 3, 2){900}}\put(8101,-5161){\line(-1, 0){1200}}\put(6901,-5161){\line(-1, 4){300}}\put(7876,-4479){\line( 1,-3){225.800}}\put(7089,-4351){\line(-6, 5){491.312}}\put(7801,-3961){\line( 1, 0){600}}\put(6901,-5161){\line( 1, 3){383.100}}\put(7321,-3871){\line( 1, 3){173.100}}\put(7501,-3354){\line( 1,-3){325.500}}\put(7456,-4726){\line(-4,-3){581.920}}\put(8394,-3961){\line(-5,-4){828.902}}\put(7569,-4629){\line( 0,-1){  7}}\put(6601,-3969){\line( 1, 0){1020}}\put(8109,-5161){\line(-5, 4){867.073}}\put(7456,-3264){\makebox(0,0)[lb]{\smash{${}_{i_1}$}}}\put(8499,-3961){\makebox(0,0)[lb]{\smash{${}_{i_2}$}}}\put(8229,-5326){\makebox(0,0)[lb]{\smash{${}_{i_3}$}}}\put(6300,-3969){\makebox(0,0)[lb]{\smash{${}_{i_5}$}}}\put(6714,-5349){\makebox(0,0)[lb]{\smash{${}_{i_4}$}}}\put(7951,-3551){\makebox(0,0)[lb]{\smash{${}_{j_{12}}$}}}\put(8349,-4644){\makebox(0,0)[lb]{\smash{${}_{j_{23}}$}}}\put(7464,-5356){\makebox(0,0)[lb]{\smash{${}_{j_{34}}$}}}\put(6504,-4696){\makebox(0,0)[lb]{\smash{${}_{j_{45}}$}}}\put(6929,-3521){\makebox(0,0)[lb]{\smash{${}_{j_{51}}$}}}\put(7569,-4261){\makebox(0,0)[lb]{\smash{${}_{j_{13}}$}}}\put(7339,-4462){\makebox(0,0)[lb]{\smash{${}_{j_{35}}$}}}\put(7266,-4284){\makebox(0,0)[lb]{\smash{${}_{j_{14}}$}}}\put(7576,-4423){\makebox(0,0)[lb]{\smash{${}_{j_{24}}$}}}\put(7397,-4110){\makebox(0,0)[lb]{\smash{${}_{j_{52}}$}}}\end{picture}\!\!\!\!\!\!\!\!\!\!\!\!\!\!\!\!\!\!.\label{s}\vspace{1.5cm}\end{equation}
In this case, and from now on, we have five intertwiners  ${\mathbf i}=\{i_n\}$, labeled by $n,m,... =1,...,5$ and ten spins   $ {\mathbf j}=\{j_{mn}\}$. We use equally the indices $i, j,k,... =1,...,5$ to indicate the nodes.   Since the operators $h^{ab}(x)$ do not change the graph (they are operators acting on the spin and intertwiners variables ${\mathbf j}, {\mathbf i}$)
\begin{equation}
{\mathbf G} _{\mathbf q}^{abcd}(x, y) = \sum_{ {\mathbf j}, {\mathbf i}}\  W( {\mathbf j}, {\mathbf i})\  h^{ab}( x)\, h^{cd}(y)\,  \Psi( {\mathbf j}, {\mathbf i}) 
\label{partenza01}
\end{equation}
where $W({\mathbf j}, {\mathbf i}) =W[\Gamma_5, {\mathbf j}, {\mathbf i}]$ and $\Psi({\mathbf j}, {\mathbf i}) =\Psi_{\mathbf q}[\Gamma_5, {\mathbf j}, {\mathbf i}]=\langle\Gamma_5, {\mathbf j}, {\mathbf i}|\Psi_{\mathbf q}\rangle$, and the sum is over the fifteen variables $({\mathbf j}, {\mathbf i})=(j_{nm},i_n)$.  (We use the physicists notation $h^{cd}(y)\, \Psi( {\mathbf j}, {\mathbf i}) $ for $[h^{cd}(y)\,  \Psi]( {\mathbf j}, {\mathbf i})$.)

Following \cite{scattering3}, we choose the form of  $\Psi({\mathbf j}, {\mathbf i})$ by identifying $\Gamma_5$ with the dual skeleton of a regular triangulation of the three-sphere.  Each node $n=1,...,5$ corresponds to a tetrahedron $t_1....t_5$ and we choose the points $x$ and $y$ to be the centers  $x_n$ and $x_m$ of the two tetrahedra $t_n$ and $t_m$.   We consider
\begin{equation}
{\mathbf G} _{{\mathbf q}\, n,m}^{\scriptscriptstyle ij,kl}:= {\mathbf G} _{\mathbf q}^{abcd}(x_n,x_m)\ n^{\scriptscriptstyle(ni)}_a n^{\scriptscriptstyle(nj)}_b\  n^{\scriptscriptstyle(mk)}_c n^{\scriptscriptstyle(ml)}_d,\qquad  \label{proiezione}
\end{equation}
where  $n_a^{(ij)}$ is the one--form normal to the triangle  that bounds the tetrahedra $t_i$ and $t_j$.  From now on, we assume $n\ne m$. Since $h^{ab}=g^{ab}-\delta^{ab}=E^{ai}E^{b}_i-\delta^{ab}$, this is given by 
\begin{eqnarray}
{\mathbf G} _{{\mathbf q}\, n,m}^{\scriptscriptstyle ij,kl} &=& \langle W | \big(E^{\scriptscriptstyle(ni)}_n \cdot E^{\scriptscriptstyle(nj)}_n-n^{\scriptscriptstyle(ni)}\cdot n^{\scriptscriptstyle(nj)}\big)
\big(E^{\scriptscriptstyle(mk)}_m  \cdot E^{\scriptscriptstyle(ml)}_m-n^{\scriptscriptstyle(mk)}\cdot n^{\scriptscriptstyle(ml)}\big) |\Psi_{\mathbf q} \rangle  \nonumber \\  
 &=& \sum_{{\mathbf j}, {\mathbf i}} W({\mathbf j}, {\mathbf i})   \big(E^{\scriptscriptstyle(ni)}_n \cdot E^{\scriptscriptstyle(nj)}_n-n^{\scriptscriptstyle(ni)}\!\!\cdot n^{\scriptscriptstyle(nj)}\big)\big(E^{\scriptscriptstyle(mk)}_m\cdot  E^{\scriptscriptstyle(ml)}_m-n^{\scriptscriptstyle(mk)}\!\!\cdot n^{\scriptscriptstyle(ml)}\big) \Psi({\mathbf j}, {\mathbf i}).
\label{partenzaA}
\end{eqnarray}
where,  $E^{\scriptscriptstyle(ml)}_n =E^{a}(\vec{x})n^{\scriptscriptstyle(ml)}_a$ is valued in the $su(2)$ algebra and, with abuse of notation, the scalar product between the triad fields indicates the product in the $su(2)$  algebra (in the internal space); while the scalar product among the one forms $n^{\scriptscriptstyle(ij)}$ is the one defined by  the background metric $\delta^{ab}$. In the rest of the paper,  we compute the right hand side of (\ref{partenzaA}).

\subsection{Linearity conditions}

Before proceedings to the actual computation of \Ref{partenzaA}, let us pause to consider the following question.   The four normal one-forms of a tetrahedron sum up to zero.  Thus
\begin{equation}
\sum_{i\ne n}  \ n_a^{\scriptscriptstyle (ni)}=0.
\label{dipendenzalin}
\end{equation}
This determines a set of linear conditions that must be satisfied by ${\mathbf G} _{{\mathbf q}\, n,m}^{\scriptscriptstyle ij,kl}$. In fact, from the last equation it follows immediately that 
\begin{equation}
\sum_{i\ne n}  \ {\mathbf G} _{{\mathbf q}\, n,m}^{\scriptscriptstyle ij,kl}=0.
\label{linear}
\end{equation}
(The existence of conditions of this kind, of course, is necessary, since the four one forms $n_a^{\scriptscriptstyle (ni)}$ (for fixed $n$) span a three-dimensional space, namely the space tangent to the boundary surface at $x_n$, and therefore the quantities
${\mathbf G}_{{\mathbf q}\, n,m}^{\scriptscriptstyle ij,kl}$ are determined by the restriction of the bi-tensor ${\mathbf G}_{\mathbf q}^{abcd}$ to these tangent spaces.)  How is it possible that the linear conditions \Ref{linear} are satisfied by the expression \Ref{partenzaA}?

The answer is interesting.  The operator $E^{\scriptscriptstyle(ni)}_n \cdot E^{\scriptscriptstyle(nj)}_n$ acts on the space of the intertwiners of the node $n$. This is the $SU(2)$ invariant part of the tensor product of the four $SU(2)$ irreducible representations determined by the four spins $j_{ni}$.  In particular, $E^{\scriptscriptstyle(ni)}_n$ is the generator of $SU(2)$ rotations in the representation $j_{ni}$. Therefore 
\begin{equation}
J = \sum_{i\ne n}   \ E^{\scriptscriptstyle(ni)}_n
\end{equation}
is the generator of  $SU(2)$ rotations in the tensor product of these representations.  But the intertwiners space is precisely the  $SU(2)$ invariant part of the tensor product. Therefore $J=0$  on the intertwiner space. Inserting this in \Ref{partenzaA}, equation \Ref{linear} follows immediately.  Therefore the linearity conditions between the projections of the propagator in the space tangent to the boundary surface are implemented by the $SU(2)$ invariance at the nodes. 

\subsection{Operators}

We begin by computing the action of the field operator $E^{\scriptscriptstyle(ni)}_n \cdot E^{\scriptscriptstyle(nj)}_n$ on the state. This operator acts on the intertwiner space at the node $n$. 
It acts as a ``double grasping"  \cite{lqg2} operator  that inserts a virtual link (in the spin-one representation) at the node, connecting the links labelled $ni$ and $nj$. 
The state of each node $n$ ($n=1,..,5$) is determined by five quantum numbers: the four spins $j_{nj}$ ($n\neq j$, $j={1,..,5})$ that label the links adjacent to the node and a quantum number $i_n$  of the virtual link that specifies the value of the intertwiner.   In this section we study the action of this operator on a single node $n$; hence we drop for clarity the index $n$ and write the intertwiner quantum number as $i$, the adjacent spins as $j_i,j_j,j_p,j_q$, and the operator as $E^{\scriptscriptstyle(i)} \cdot E^{\scriptscriptstyle(j)}$.
We use the graphic notation of $SU(2)$ recoupling theory to compute the action of the operators on the spin network states  (see \cite{book}).  The basics of this notation are given in  Appendix \ref{recoupling} and the details of the derivation of the action of the operator are given in  Appendix \ref{operators}. 
Choose a given pairing at the node, say $(i,j)(p,q)$ (and fix the orientation, say clockwise, of each of the two trivalent vertices). We represent the node in the form 
\begin{eqnarray}
i &=& 
\begin{array}{c}\setlength{\unitlength}{1 pt}
\begin{picture}(50,40)          \put(-6,0){$j_{j}$}\put(-6,30){$j_{i}$}          \put(45,0){$j_{p}$}\put(45,30){$j_{q}$}          \put(10,10){\line(1,1){10}}\put(10,30){\line(1,-1){10}}          \put(30,20){\line(1,1){10}}\put(30,20){\line(1,-1){10}}          \put(20,20){\line(1,0){10}}\put(22,25){$i$}          \put(20,20){\circle*{3}}\put(30,20){\circle*{3}}\end{picture} , 
\end{array} 
\label{Apairings}
\neqa 
where we use the same notation $i$ for the intertwiner and the spin of the virtual link that determines it.  This  basis diagonalizes the operator
$E^{\scriptscriptstyle(i)} \cdot E^{\scriptscriptstyle(j)}$, but not the operators 
$E^{\scriptscriptstyle(i)} \cdot E^{\scriptscriptstyle(q)}$ and $E^{\scriptscriptstyle(i)} \cdot E^{\scriptscriptstyle(p)}$. 
We consider the action of these three  ``doublegrasping" operator on this basis. The simplest is the action of $E^{\scr(i)} \cdot E^{\scr(i)}$. Using the formulas in Appendix \ref{operators} we have easily 
\begin{equation}
	E^{\scr(i)} \cdot E^{\scr(i)}\;
	\left|  
	\begin{array}{c}\setlength{\unitlength}{1 pt}
\begin{picture}(50,40)          \put(-6,0){$j_{j}$}\put(-6,30){$j_{i}$}          \put(45,0){$j_{p}$}\put(45,30){$j_{q}$}          \put(10,10){\line(1,1){10}}\put(10,30){\line(1,-1){10}}          \put(30,20){\line(1,1){10}}\put(30,20){\line(1,-1){10}}          \put(20,20){\line(1,0){10}}\put(22,25){$i$}        
 \put(20,20){\circle*{3}}\put(30,20){\circle*{3}}\end{picture} 
\end{array} \;\right\rangle 
 = -(N^i)^2 
 \left|\begin{array}{c}\setlength{\unitlength}{1 pt}
\begin{picture}(50,40)         \put(-6,0){$j_{j}$}\put(-6,30){$j_{i}$}          \put(45,0){$j_{p}$}\put(45,30){$j_{q}$}          \put(5,35){\circle*{3}}\put(5,35){\qbezier(0,0)(10,0)(10,-10)}\put(15,25){\circle*{3}
}\put(0,0){\line(1,1){20}}\put(0,40){\line(1,-1){20}}          \put(30,20){\line(1,1){20}}\put(30,20){\line(1,-1){20}}          \put(20,20){\line(1,0){10}}\put(22,25){$i$}\put(13,35){$1$}           \put(20,20){\circle*{3}}\put(30,20){\circle*{3}}\end{picture} 
 \end{array}\;\right\rangle
 =C^{\scriptscriptstyle ii}\left|  
	\begin{array}{c}\setlength{\unitlength}{1 pt}
\begin{picture}(50,40)          \put(-6,0){$j_{j}$}\put(-6,30){$j_{i}$}          \put(45,0){$j_{p}$}\put(45,30){$j_{q}$}          \put(10,10){\line(1,1){10}}\put(10,30){\line(1,-1){10}}          \put(30,20){\line(1,1){10}}\put(30,20){\line(1,-1){10}}          \put(20,20){\line(1,0){10}}\put(22,25){$i$}          \put(20,20){\circle*{3}}\put(30,20){\circle*{3}}\end{picture} 
\end{array} \;\right\rangle , 
\label{ii}
\end{equation}
 where 
 \begin{equation}
	C^{\scriptscriptstyle ii}=C^2(j_{i}).
\end{equation}
with $C^2(a)=a(a+1)$ is the Casimir of the representation $a$.  Just slightly more complicated is the  action of $E^{\scr(i)} \cdot E^{\scr(j)}$
  \begin{equation}
	E^{\scr(i)} \cdot E^{\scr(j)}\;
	\left|  
	\begin{array}{c}\setlength{\unitlength}{1 pt}
\begin{picture}(50,40)          \put(-6,0){$j_{j}$}\put(-6,30){$j_{{i}}$}          \put(45,0){$j_{p}$}\put(45,30){$j_{q}$}          \put(10,10){\line(1,1){10}}\put(10,30){\line(1,-1){10}}          \put(30,20){\line(1,1){10}}\put(30,20){\line(1,-1){10}}          \put(20,20){\line(1,0){10}}\put(22,25){$i$}          \put(20,20){\circle*{3}}\put(30,20){\circle*{3}}\end{picture} 
\end{array} \;\right\rangle 
 =-N^iN^j \left|\begin{array}{c}\setlength{\unitlength}{1 pt}
\begin{picture}(50,40)          \put(-6,0){$j_{j}$}\put(-6,30){$j_{i}$}          \put(45,0){$j_{p}$}\put(45,30){$j_{q}$}          \put(10,30){\circle*{3}}\put(10,10){\circle*{3}}\put(10,10){\line(0,1){20}}\put(0,0){\line(1,1){20}}\put(0,40){\line(1,-1){20}}          \put(30,20){\line(1,1){20}}\put(30,20){\line(1,-1){20}}          \put(20,20){\line(1,0){10}}\put(22,25){$i$} \put(4,18){$1$}         \put(20,20){\circle*{3}}\put(30,20){\circle*{3}}\end{picture} 
\end{array}\;\right\rangle
=D^{\scriptscriptstyle ij}\left|  
	\begin{array}{c}\setlength{\unitlength}{1 pt}
\begin{picture}(50,40)          \put(-6,0){$j_{j}$}\put(-6,30){$j_{i}$}          \put(45,0){$j_{p}$}\put(45,30){$j_{q}$}          \put(10,10){\line(1,1){10}}\put(10,30){\line(1,-1){10}}          \put(30,20){\line(1,1){10}}\put(30,20){\line(1,-1){10}}          \put(20,20){\line(1,0){10}}\put(22,25){$i$}          \put(20,20){\circle*{3}}\put(30,20){\circle*{3}}\end{picture} 
\end{array} \;\right\rangle, 
\label{ij}
\end{equation}
where 
\begin{equation}
	D^{\scriptscriptstyle ij}=\frac{C^2(i)-C^2(j_{i})-C^2(j_{j})}{2}.
	\label{D} 
\end{equation}

In these two cases the action of the operator is diagonal. If, instead, the grasped links are \emph{not} paired together, the action of the operator is not diagonal in this basis.  In this case, the recoupling theory in the Appendix gives 
\begin{equation}
\begin{split}
		&E^{\scr(i)} \cdot E^{\scr(q)}\;
		\left|  
		\begin{array}{c}\setlength{\unitlength}{1 pt}
	\begin{picture}(50,40)          \put(-6,0){$j_{j}$}\put(-6,30){$j_{i}$}          \put(45,0){$j_{p}$}\put(45,30){$j_{q}$}          \put(10,10){\line(1,1){10}}\put(10,30){\line(1,-1){10}}          \put(30,20){\line(1,1){10}}\put(30,20){\line(1,-1){10}}          \put(20,20){\line(1,0){10}}\put(22,25){$i$}           \put(20,20){\circle*{3}}\put(30,20){\circle*{3}}\end{picture} 
	\end{array} \;\right\rangle 
	 =-N^iN^q \left|\begin{array}{c}\setlength{\unitlength}{1 pt}
	\begin{picture}(50,40)          \put(-6,0){$j_{j}$}\put(-6,30){$j_{i}$}          \put(45,0){$j_{p}$}\put(45,30){$j_{q}$}          \put(10,30){\circle*{3}}\put(10,30){\qbezier(0,0)(15,15)(30,0)}\put(40,30){\circle*{3}}\put(0,0){\line(1,1){20}}\put(0,40){\line(1,-1){20}}          \put(30,20){\line(1,1){20}}\put(30,20){\line(1,-1){20}}          \put(20,20){\line(1,0){10}}\put(22,25){$i$}\put(23,40){$1$}          \put(20,20){\circle*{3}}\put(30,20){\circle*{3}}\end{picture} 
	\end{array}\;\right\rangle= \\
	&=X^{\scriptscriptstyle iq} 
	\left|  
		\begin{array}{c}\setlength{\unitlength}{1 pt}
	\begin{picture}(50,40)          \put(-6,0){$j_{j}$}\put(-6,30){$j_{i}$}          \put(45,0){$j_{p}$}\put(45,30){$j_{q}$}          \put(10,10){\line(1,1){10}}\put(10,30){\line(1,-1){10}}          \put(30,20){\line(1,1){10}}\put(30,20){\line(1,-1){10}}          \put(20,20){\line(1,0){10}}\put(22,25){$i$}          \put(20,20){\circle*{3}}\put(30,20){\circle*{3}}\end{picture} 
	\end{array} \;\right\rangle
	-Y^{\scriptscriptstyle iq}
	\left|  
		\begin{array}{c}\setlength{\unitlength}{1 pt}
	\begin{picture}(65,40)          \put(-6,0){$j_{j}$}\put(-6,30){$j_{i}$}          \put(60,0){$j_{p}$}\put(60,30){$j_{q}$}          \put(10,10){\line(1,1){10}}\put(10,30){\line(1,-1){10}}          \put(45,20){\line(1,1){10}}\put(45,20){\line(1,-1){10}}          \put(20,20){\line(1,0){25}}\put(22,25){$i-1$}          \put(20,20){\circle*{3}}\put(45,20){\circle*{3}}\end{picture} 
	\end{array} \;\right\rangle
	-Z^{\scriptscriptstyle iq}
	\left|  
		\begin{array}{c}\setlength{\unitlength}{1 pt}
	\begin{picture}(65,40)          \put(-6,0){$j_{j}$}\put(-6,30){$j_{i}$}          \put(60,0){$j_{p}$}\put(60,30){$j_{q}$}          \put(10,10){\line(1,1){10}}\put(10,30){\line(1,-1){10}}          \put(45,20){\line(1,1){10}}\put(45,20){\line(1,-1){10}}          \put(20,20){\line(1,0){25}}\put(22,25){$i+1$}          \put(20,20){\circle*{3}}\put(45,20){\circle*{3}}\end{picture} 
	\end{array} \;\right\rangle, 
	\end{split}
	\label{iq}
\end{equation}
where
\begin{equation}
	X^{\scriptscriptstyle iq}=
	-\frac{\left(C^2(i)+C^2(j_{\scr{i}})-C^2(j_{\scr{j}})\right)
	\left(C^2(i)+C^2(j_{\scr{q}})-C^2(j_{\scr{p}})\right)}{4\, C^2(i)},
\end{equation}
\begin{equation}
\begin{split}
		Y^{\scriptscriptstyle iq}=-
&\frac{1}{4 i \dim(i)}\sqrt{(j^{\scr{i}}+j^{\scr{j}}+i+1)
(j^{\scr{i}}-j^{\scr{j}}+i)
(-j^{\scr{i}}+j^{\scr{j}}+i)
(j^{\scr{i}}+j^{\scr{j}} -i+1)} \;\cdot\\
&\qquad\qquad\quad\cdot\sqrt{(j^{\scr{p}}+j^{\scr{q}}+i+1)(j^{\scr{p}}-j^{\scr{q}}+i)(-j^{\scr{p}}+j^{\scr{q}}+i)(j^{\scr{p}}+j^{\scr{q}}-i+1)}, 
\end{split}
\end{equation}
\begin{equation}
\begin{split}
		Z^{\scriptscriptstyle iq}=-
&\frac{1}{4 (i+1) \dim(i)}\sqrt{(j^{\scr{i}}+j^{\scr{j}}+i+2)
(j^{\scr{i}}-j^{\scr{j}}+i+1)
(-j^{\scr{i}}+j^{\scr{j}}+i+1)
(j^{\scr{i}}+j^{\scr{j}} -i)} \;\cdot\\
&\qquad\qquad\qquad\quad	\cdot\sqrt{(j^{\scr{p}}+j^{\scr{q}}+i+2)(j^{\scr{p}}-j^{\scr{q}}+i+1)(-j^{\scr{p}}+j^{\scr{q}}+i+1)(j^{\scr{p}}+j^{\scr{q}}-i)}. 
\end{split}
\end{equation}
The last possibility is
\begin{equation}
\begin{split}
		&E^{\scr(i)} \cdot E^{\scr(p)}\;
		\left|  
		\begin{array}{c}\setlength{\unitlength}{1 pt}
	\begin{picture}(50,40)          \put(-6,0){$j_{j}$}\put(-6,30){$j_{i}$}          \put(45,0){$j_{p}$}\put(45,30){$j_{q}$}          \put(10,10){\line(1,1){10}}\put(10,30){\line(1,-1){10}}          \put(30,20){\line(1,1){10}}\put(30,20){\line(1,-1){10}}          \put(20,20){\line(1,0){10}}\put(22,25){$i$}          \put(20,20){\circle*{3}}\put(30,20){\circle*{3}}\end{picture} 
	\end{array} \;\right\rangle 
	 =-N^i N^p\left|\begin{array}{c}\setlength{\unitlength}{1 pt}
\begin{picture}(50,40)          \put(-6,0){$j_{j}$}\put(-6,30){$j_{i}$}          \put(45,0){$j_{p}$}\put(45,30){$j_{q}$}          \put(10,30){\circle*{3}}\put(10,30){\qbezier(0,0)(20,20)(30,-20)}\put(40,10){\circle*{3}}\put(0,0){\line(1,1){20}}\put(0,40){\line(1,-1){20}}          \put(30,20){\line(1,1){20}}\put(30,20){\line(1,-1){20}}          \put(20,20){\line(1,0){10}}\put(22,25){$i$}          \put(20,20){\circle*{3}}\put(30,20){\circle*{3}}\end{picture} 
\end{array}\;\right\rangle= \\
&=X^{\scriptscriptstyle ip} 
	\left|  
		\begin{array}{c}\setlength{\unitlength}{1 pt}
	\begin{picture}(50,40)          \put(-6,0){$j_{j}$}\put(-6,30){$j_{i}$}          \put(45,0){$j_{p}$}\put(45,30){$j_{q}$}          \put(10,10){\line(1,1){10}}\put(10,30){\line(1,-1){10}}          \put(30,20){\line(1,1){10}}\put(30,20){\line(1,-1){10}}          \put(20,20){\line(1,0){10}}\put(22,25){$i$}          \put(20,20){\circle*{3}}\put(30,20){\circle*{3}}\end{picture} 
	\end{array} \;\right\rangle
	+Y^{\scriptscriptstyle ip}
	\left|  
		\begin{array}{c}\setlength{\unitlength}{1 pt}
	\begin{picture}(65,40)          \put(-6,0){$j_{j}$}\put(-6,30){$j_{i}$}          \put(60,0){$j_{p}$}\put(60,30){$j_{q}$}          \put(10,10){\line(1,1){10}}\put(10,30){\line(1,-1){10}}          \put(45,20){\line(1,1){10}}\put(45,20){\line(1,-1){10}}          \put(20,20){\line(1,0){25}}\put(22,25){$i-1$}          \put(20,20){\circle*{3}}\put(45,20){\circle*{3}}\end{picture} 
	\end{array} \;\right\rangle
	+Z^{\scriptscriptstyle ip}
	\left|  
		\begin{array}{c}\setlength{\unitlength}{1 pt}
	\begin{picture}(65,40)          \put(-6,0){$j_{j}$}\put(-6,30){$j_{i}$}          \put(60,0){$j_{p}$}\put(60,30){$j_{q}$}          \put(10,10){\line(1,1){10}}\put(10,30){\line(1,-1){10}}          \put(45,20){\line(1,1){10}}\put(45,20){\line(1,-1){10}}          \put(20,20){\line(1,0){25}}\put(22,25){$i+1$}          \put(20,20){\circle*{3}}\put(45,20){\circle*{3}}\end{picture} 
	\end{array} \;\right\rangle.
\end{split}
\end{equation}
Note that $X^{\scriptscriptstyle ip}$ is exactly $X^{\scriptscriptstyle iq}$ with $p$ and $q$ switched and $Y^{\scriptscriptstyle ip}=Y^{\scriptscriptstyle iq}$, $Z^{\scriptscriptstyle ip}=Z^{\scriptscriptstyle iq}$. 

Finally, we have to take care of the orientation.  As shown in the Appendix, the sign of the non diagonal terms is influenced by the orientations: in the planar representation that we are using, there is a $+$ sign if the added link intersect the virtual one and a $-1$ otherwise.

Summarizing, in a different notation and reinserting explicitly the index ${}_n$ of the node, we have the following action of the $EE$ operators. If the grasped links are paired together we have the diagonal action
\begin{equation}
	E^{\scr(ni)} \cdot E^{\scr(nj)}|\Gamma_5, {\mathbf j}, i_1,..,i_n,..,i_5 \rangle=S_n^{\scriptscriptstyle ij}\ |\Gamma_5, {\mathbf j}, i_1,..,i_n,..,i_5 \rangle,  \label{operatori}
\end{equation}
where 
\begin{equation}\label{diag}
	S_n^{\scriptscriptstyle ij}=\begin{cases}
C^{\scr{ii}}=C^2(j_{ni})& \text{if $i=j$},\\
D^{\scr{ij}}=\frac{C^2(i_n)-C^2(j_{ni})-C^2(j_{nj})}{2}& \text{if $i\neq j$}.
	\end{cases}
\end{equation}\\[1mm]
If the grasped links are not paired together, we have the non-diagonal action
\begin{eqnarray}
	&&\hspace{-3em}E^{\scr(ni)} \cdot E^{\scr(nq)}|\Gamma_5, {\mathbf j}, i_1,..,i_n,..,i_5 \rangle=\nonumber\\
	&&=\begin{cases}
	X_n^{\scriptscriptstyle iq}\ |\Gamma_5, {\mathbf j}, i_1,..,i_n,..,i_5 \rangle+Y_n^{\scriptscriptstyle iq}\ |\Gamma_5, {\mathbf j}, i_1,..,i_n\!\!-\!\!1,..,i_5 \rangle\\ 
	\hspace{3em}+Z_n^{\scriptscriptstyle iq}\ |\Gamma_5, {\mathbf j}, i_1,..,i_n\!+\!1,..,i_5 \rangle &\text{if $i$ opposite to $q$},\\
	X_n^{\scriptscriptstyle iq}\ |\Gamma_5, {\mathbf j}, i_1,..,i,..,i_5 \rangle-Y_n^{\scriptscriptstyle iq}\ |\Gamma_5, {\mathbf j}, i_1,..,i_n\!\!-\!\!1,..,i_5 \rangle\\ 
	\hspace{3em}-Z_n^{\scriptscriptstyle iq}\ |\Gamma_5, {\mathbf j}, i_1,..,i_n\!+\!1,..,i_5 \rangle &\text{otherwise}.
\end{cases}
\end{eqnarray}\\[1mm]
This completes the calculation of the action of the gravitational field operators. 

\section{The boundary state}

The boundary state utilized in  \cite{scattering3} was assumed to have a gaussian dependence on the spins, and to be peaked on a particular intertwiner. This intertwiner was assumed 
to project trivially onto the BC intertwiner of the BC vertex. This was a simplifying assumption permitting to avoid dealing with the intertwiners, motivated by the fact that intertwiners play no role for the diagonal terms. However, it was also pointed out in  \cite{scattering3} that this procedure is not well defined, because of the mismatch between $SO(4)$ linearity and $SU(2)$ linearity (see the discussion in the Appendix of \cite{scattering3}).  Here we face the problem squarely, and consider the intertwiner dependence of the boundary state explicitly. 

A  natural  generalization of the gaussian state used in  \cite{scattering3}, whith a well-defined and  non-trivial intertwiner dependence, is the state
\begin{equation}
\begin{split}
\Phi({\mathbf j},{\mathbf i}) =\ &C\exp\left\{-\frac{1}{2j_0} \sum_{\scr{(ij)(mr)}}\alpha_{\scr{(ij)(mr)}}\
	  (j_{ij} - j_0) (j_{mr}- j_0)+ i \Phi \sum_{\scr{(ij)}}  j_{ij}\right\}\cdot\\
	  &\cdot\exp\left\{-\sum_n\left(\frac{(i_n - i_0)^2}{4\sigma}+ \sum_{p\neq n} \phi(j_{np} - j_0)(i_n - i_0)+i \chi (i_n - i_0) \right)\right\}
\end{split}
  \label{cbis}
\end{equation}
The first line of this equation is precisely the spin dependence of the state used in \cite{scattering3}. The second line contains a gaussian dependence on the intertwiner variables.
More precisely, it includes a diagonal gaussian term, a nondiagonal gaussian spin-intertwiner term, and a phase factor.  We do not include non-diagonal intertwiner-intertwiner terms here.
These will be considered in the companion paper.  

Let us fix some of the constants appearing in (\ref{cbis}), by requiring the state to be peaked on the expected geometry. The constant $j_0$ determines the background area $A_0$ of the faces, 
via $C(j_{nm})=A_{nm}$. As in \cite{scattering3}, we leave $j_0$ free to determine the overall scale. The constant $\Phi$ determines the background values of the angles between the normals to the tetrahedra. As in  \cite{scattering3}, we fix them to those of a regular four-simplex, namely  $\cos\Phi=-1/4$.   

The constant $i_0$ is the background value of the intertwiner variable. As shown in  \cite{scattering3}, the spin of the virtual link $i_n$ is the quantum number of the \emph{angle} between the normals of two triangles.  More precisely, the Casimir $C(i_n)$ of the representation $i_n$ is the operator corresponding to the classical quantity 
\begin{equation}
C^2(i_n)= A_{ni} +A_{nj}+2\ \vec n^{\scriptscriptstyle(ni)}\cdot \vec n^{\scriptscriptstyle(nj)}, 
\label{constr2}
\end{equation}
where $i$ and $j$  are the paired links at the node $n$ and $A_{ni}$ is the area of the triangle dual to the link $(ni)$. 
The scalar product of the normals to the triangles can therefore be related to the Casimirs of spins and intertwiners: 
\begin{equation}
	n^{\scriptscriptstyle(ni)}\cdot n^{\scriptscriptstyle(nj)}=
	\frac{C(i_n)-C(j_{ni})-C(j_{nj})}{2}. 
	\label{normali di background}
\end{equation}
For each node, the state must therefore be peaked on a value $i_{0}$ such that 
\begin{equation}
i_0(i_0+1)= A_0 +A_0+2 A_0 A_0 \cos{\theta_{ij}},
\label{constr20}
\end{equation}
where $\cos{\theta_{ij}}$ is the 3d dihedral angle between the faces of the tetrahedron.
For the regular 4-symplex, in the large distance limit we have $A_{ij}=j_0$, $\cos{\theta_{ij}}=-\frac{1}{3}$, which gives 
\begin{equation}
	i_0=\frac{2}{\sqrt{3}}\ j_0. 
\end{equation}
This fixes $i_0$. 
Notice that in \cite{scattering3} equation \eqref{constr2} refers to the Casimir of an $SO(4)$ simple representation and follows from the quantization of the Plebanski 2 form $B^{IJ}=e^I\wedge e^J$ associated with the discretized geometry. Exactly the same result follows from  equation \eqref{ij} directly from $LQG$.   

Fixing $i_0$ in this manner determines only the mean value of the angle $\theta_{ij}$ between the two triangles that are paired together in the chosen pairing. What about the mean value of the angles between faces that are not paired together, such as $\theta_{iq}$?  It is shown in \cite{sr} that a state of the form $e^{(i-i_0)^2/\sigma}$ is peaked on  $\theta_{iq}=0$, which is not what we want; but the mean value of  $\theta_{iq}$, can be modified by adding a phase to the state. This is the analog of the fact that a phase changes the mean value of the momentum of the wave packet of a non relativistic particle, without affecting the mean value of the position.  In particular, it was shown in  \cite{sr} that by choosing the phase and the width of the Gaussian to be
\begin{equation}
	 \chi=\frac{\pi}{2},   \quad  \quad \quad \sigma=\frac{ j_0}{3}, 
	\label{tunings1}
\end{equation}
we obtain a state whose mean value and variance for all angles is the same.

Let us therefore adopt here these values.  Still, the present situation is more complicated than the case considered in  \cite{sr}, because the tetrahedron considered there had fixed {\em and equal} values of the external spins; while here the spins can take arbitrary values around the peak symmetric configuration $j_{nm}=j_0$.  As a consequence, when repeating the calculation in  \cite{sr}, one finds additional spin-intertwiner gaussian terms.   These, however can be corrected by fixing the spin-intertwiner gaussian terms in (\ref{cbis}).  A detailed calculation (see below), shows indeed 
that in the large $j_0$ limit, the state (\ref{cbis}) transforms under change of pairing into a state with the same intertwiner mean value and the same variance $\sigma$, provided we also choose
\begin{equation}
	\phi=-i\frac{3}{4j_0},
	\label{tunings2}
\end{equation}
which we assume from now on.  With these values and 
introducing the difference variables  $\delta i_n=i_n-i_0$ and $\delta j_{mr}=j_{mr}-j_0$ the wave functional, given in (\ref{cbis}),  reads
\begin{equation}
	\Phi({\mathbf j},{\mathbf i})\ 
		= C\ 
		e^{-\frac{1}{2j_0}\sum \alpha_{\scr{(ij)(mr)}}\delta j_{ij}\delta j_{mr}+i\Phi\sum_{ij} \delta j_{ij}}
		e^{-\sum_{n}\left(\frac{3(\delta i_n)^2}{4j_0}-i\left(\sum_a\frac{3}{4 j_0} \delta j_{an}-\frac{\pi}{2}\right) \delta i_n\right)}.
	\label{Cpartenza}
\end{equation}
This state, however, presents a problem, which we discuss in the next section. 

\subsection{Pairing independence}

It is natural to require that the state respects the symmetries of the problem. A moment of reflection shows that the state (\ref{Cpartenza}) does not.  The reason is that the variables $i_n$ are the spin of the virtual links {\em in one specific pairing}, and this breaks the symmetry of the four-simplex.  The phases and variances chosen assure that the mean values are the desired ones, hence symmetric; but an explicit calculation confirms that the relative fluctuations of the angle variables determined by the state (\ref{Cpartenza}) depend on the pair  chosen.  

To correct the problem, recall that there are three natural bases in each intertwiner space,  determined by the three possible pairings of these links. Denote them as follows. 
\begin{eqnarray}
i^x &=& 
\begin{array}{c}\setlength{\unitlength}{1 pt}
\begin{picture}(50,40)          \put(-6,0){$j_{j}$}\put(-6,30){$j_{i}$}          \put(45,0){$j_{p}$}\put(45,30){$j_{q}$}          \put(10,10){\line(1,1){10}}\put(10,30){\line(1,-1){10}}          \put(30,20){\line(1,1){10}}\put(30,20){\line(1,-1){10}}          \put(20,20){\line(1,0){10}}\put(22,25){$i^x$}          \put(20,20){\circle*{3}}\put(30,20){\circle*{3}}\end{picture} 
\end{array} \qquad
 i^y = 
\begin{array}{c}\setlength{\unitlength}{1 pt}
\begin{picture}(50,40)          \put(-6,0){$j_{j}$}\put(-6,30){$j_{i}$}          \put(45,0){$j_{p}$}\put(45,30){$j_{q}$}          \put(10,0){\line(1,1){10}}\put(10,30){\line(1,-1){10}}          \put(20,20){\line(1,1){10}}\put(20,10){\line(1,-1){10}}          \put(20,10){\line(0,1){10}}\put(23,13){$i^y$}          \put(20,20){\circle*{3}}\put(20,10){\circle*{3}}\end{picture} 
\end{array}\qquad
i^z = 
\begin{array}{c}\setlength{\unitlength}{1 pt}
\begin{picture}(50,40)          \put(-6,0){$j_{j}$}\put(-6,30){$j_{i}$}          \put(45,0){$j_{p}$}\put(45,30){$j_{q}$}          \put(10,0){\line(1,1){30}}\put(10,30){\line(1,-1){30}}                  \put(20,20){\line(1,0){10}}\put(22,25){$i^z$}          \put(20,20){\circle*{3}}\put(30,20){\circle*{3}}\end{picture} 
\end{array},
\label{3pairings}
\neqa 
where we conventionally denote $i^x\equiv i$ the basis in the pairing chosen as reference. 
These bases diagonalize the three non commuting operators
$E^{\scriptscriptstyle(i)} \cdot E^{\scriptscriptstyle(j)}$, $E^{\scriptscriptstyle(i)} \cdot E^{\scriptscriptstyle(q)}$ and $E^{\scriptscriptstyle(i)} \cdot E^{\scriptscriptstyle(p)}$, respectively. 
Furthermore a spin-network state is specified by the orientation of the three-valent nodes \cite{spinnetworks}; we fix this orientation by giving an ordering to the links.  We write for instance 
\begin{equation}
	i^{x\scr{(+,-)}} = 
\begin{array}{c}\setlength{\unitlength}{1 pt}
\begin{picture}(50,40)          \put(-6,0){$j_{j}$}\put(-6,30){$j_{i}$}          \put(45,0){$j_{p}$}\put(45,30){$j_{q}$}          \put(10,10){\line(1,1){10}}\put(10,30){\line(1,-1){10}}          \put(30,20){\line(1,1){10}}\put(30,20){\line(1,-1){10}}          \put(20,20){\line(1,0){10}}\put(22,25){$i^x$}          \put(20,20){\circle*{3}}    \put(30,20){\circle*{3}}  \put(15,10){$\scr{+}$}\put(30,10){$\scr{-}$}\end{picture} 
\end{array}
\end{equation}
where the plus sign + (-) means anticlockwise (clockwise) ordering of the links in the two nodes. 
A complete basis in the space of the spin networks on $\Gamma_5$ is specified giving the pairing and the orientation at each node.  
In order to label the different bases, introduce at each node $n$ a variable $m_n$ that takes the values $m_n=x,y,z$, namely that ranges over the three possible pairings at the node. Similarly, 
introduce a variables $o_n=\{{\scriptstyle(++),(+-),(-+),(--)}\}$ that labels the possible orientations.  
To correct the pairing dependence of the state (\ref{cbis}), let us first rewrite it in the notation 
\begin{equation}
\begin{split}
|\Phi_{\mathbf q}\rangle_{x++}=\  & \sum_{{\mathbf j},{\mathbf i^{x++}}} \  \Phi[{\mathbf j},{\mathbf i^{x++}}] \  |{\mathbf j},{\mathbf i^{x++}}\rangle. 
\end{split}
  \label{c1}
\end{equation}
where the suffix ${}_{x++}$ to the ket emphasizes the fact that the state has been defined with the
chosen pairing and orientation at each node. 
We can now consider a new state obtained by summing (\ref{c1}) over all choices of 
pairings and orientations.   That is, we change the definition of the boundary state to 
\begin{equation}
\begin{split}
|\Psi_{\mathbf q}\rangle=\  & \sum_{m_n, o_n}\  
|\Phi_{\mathbf q}\rangle_{m_no_n} . 
\end{split}
  \label{ctris}
\end{equation}
where $\sum_{m_n, o_n}= \sum_{m_1... m_5}\  \sum_{o_1... o_5}$ and
\begin{equation}
\begin{split}
|\Phi_{\mathbf q}\rangle_{m_no_n}=\  & \sum_{{\mathbf j},{\mathbf i^{m_no_n}}} \  \Phi[{\mathbf j},{\mathbf i^{m_no_n}}] \  |{\mathbf j},{\mathbf i^{m_no_n}}\rangle. 
\end{split}
  \label{c2}
\end{equation}
namely $|\Phi_{\mathbf q}\rangle_{m_no_n}$ is the same as the state $|\Phi_{\mathbf q}\rangle_{x++}$, but defined with a different choice of pairing at each node. 

Since (by assumption) (\ref{cbis}) does not depend on the orientation, the sum over the orientation of the node (say) 1, in (\ref{ctris}) reduces to a term proportional to 
	\begin{equation}
	\begin{split}
	&\sum_{o}	\left| \mathbf j,i^{o}_1,i_2,i_3,i_4,i_5\right\rangle \sim  \\ & \ \ \ \ \ 
	\left| 
	\begin{array}{c}\setlength{\unitlength}{1 pt}
	\begin{picture}(50,40)     \put(-6,0){$j_{12}$}\put(-6,30){$j_{13}$}          \put(45,0){$j_{15}$}\put(45,30){$j_{14}$}          \put(10,10){\line(1,1){10}}\put(10,30){\line(1,-1){10}}          \put(30,20){\line(1,1){10}}\put(30,20){\line(1,-1){10}}          \put(20,20){\line(1,0){10}}\put(22,25){$i_1$}          \put(15,10){$\scr{+}$}\put(30,10){$\scr{+}$}\put(20,20){\circle*{3}}\put(30,20){\circle*{3}}\end{picture} 
	\end{array}\, \right\rangle
	+
	\left| 
	\begin{array}{c}\setlength{\unitlength}{1 pt}
	\begin{picture}(50,40)          \put(-6,0){$j_{12}$}\put(-6,30){$j_{13}$}          \put(45,0){$j_{15}$}\put(45,30){$j_{14}$}          \put(10,10){\line(1,1){10}}\put(10,30){\line(1,-1){10}}          \put(30,20){\line(1,1){10}}\put(30,20){\line(1,-1){10}}          \put(20,20){\line(1,0){10}}\put(22,25){$i_1$}          \put(15,10){$\scr{+}$}\put(30,10){$\scr{-}$}\put(20,20){\circle*{3}}\put(30,20){\circle*{3}}\end{picture} 
	\end{array}\, \right\rangle
	+
	\left| 
	\begin{array}{c}\setlength{\unitlength}{1 pt}
	\begin{picture}(50,40)          \put(-6,0){$j_{\scr{12}}$}\put(-6,30){$j_{13}$}          \put(45,0){$j_{15}$}\put(45,30){$j_{14}$}          \put(10,10){\line(1,1){10}}\put(10,30){\line(1,-1){10}}          \put(30,20){\line(1,1){10}}\put(30,20){\line(1,-1){10}}          \put(20,20){\line(1,0){10}}\put(22,25){$i_1$}          \put(15,10){$\scr{-}$}\put(30,10){$\scr{+}$}\put(20,20){\circle*{3}}\put(30,20){\circle*{3}}\end{picture} 
	\end{array}\, \right\rangle
	+
	\left| 
	\begin{array}{c}\setlength{\unitlength}{1 pt}
	\begin{picture}(50,40)          \put(-6,0){$j_{12}$}\put(-6,30){$j_{13}$}          \put(45,0){$j_{15}$}\put(45,30){$j_{14}$}          \put(10,10){\line(1,1){10}}\put(10,30){\line(1,-1){10}}          \put(30,20){\line(1,1){10}}\put(30,20){\line(1,-1){10}}          \put(20,20){\line(1,0){10}}\put(22,25){$i_1$}          \put(15,10){$\scr{-}$}\put(30,10){$\scr{-}$}\put(20,20){\circle*{3}}\put(30,20){\circle*{3}}\end{picture} 
	\end{array}\, \right\rangle.
	\end{split}
\end{equation}
As shown in the Appendix, the change in orientation of a vertex produces the sign $(-1)^{a+b+c}$, where $a,b,c$ are the three addiacent spins.  Hence
	\begin{equation}
	\begin{split}
	&\sum_{o}	\left| \mathbf j,i^{o}_1,i_2,i_3,i_4,i_5\right\rangle \sim  \\ & \ \ \ \ \ 
=\left(1+(-1)^{j_{\scr{14}}+j_{\scr{15}}+i_1}+(-1)^{j_{\scr{12}}+j_{\scr{13}}+i_1}+(-1)^{j_{\scr{12}}+j_{\scr{13}}+j_{\scr{14}}+j_{\scr{15}}+2\, i_1}\right)	\left| \mathbf j,i^{++}_1,i_2,i_3,i_4,i_5\right\rangle \\ &  \ \ \ \ \ 
	=\begin{cases}4\left| \mathbf j,i^{++}_1,i_2,i_3,i_4,i_5,\right\rangle
  \ \ \ \textrm{if} \ \ \big(  
  \   j_{12}+j_{13}+i^{m_1}_1=2n_1\quad{\rm and}\quad 
	j_{14}+j_{15}+i^{m_1}_1=2n_2\  \big), 	\\
	0 \hspace{10em}\textrm{otherwise}.
	\end{cases}
	\label{esistenza dei j}
	\end{split}
\end{equation}
We can therefore trade the sum over orientations in (\ref{ctris}) with a condition on the spins 
summed over:  at all trivalent vertices, the sum of the two external spins and 
the virtual spin, must be an even integer.  (The factor 4 is absorbed in the 
normalization factor $C$.) With this understanding, we drop the sum over 
orientations in (\ref{ctris}), which now reads 
\begin{equation}
\begin{split}
|\Psi_{\mathbf q}\rangle=\  & \sum_{m_n}\  
|\Phi_{\mathbf q}\rangle_{m_n} ,
\end{split}
  \label{c}
\end{equation}
where all orientations are fixed.  
This state can of course also be expressed in terms of a single basis 
\begin{equation}
\begin{split}
|\Psi_{\mathbf q}\rangle=\  
\sum_{{\mathbf j},{\mathbf i}} 
 \Psi_{\mathbf q}({\mathbf j},{\mathbf i}) \  
 	  |{\mathbf j},{\mathbf i}\rangle, 
\end{split}
  \label{c22}
\end{equation}
where we have returned to the notation $i_n=i_n^{x,++}$. Its components are 
\begin{equation}
\begin{split}
\Psi({\mathbf j},{\mathbf i}) =\  
\langle {\mathbf j},{\mathbf i}|\Psi_{\mathbf q} \rangle
=  \sum_{m_n}\ \Phi({\mathbf j},{\mathbf i^{m_n}}) 
\ \langle {\mathbf j},{\mathbf i}|{\mathbf j},{\mathbf i^{m_n}}\rangle.
\end{split}
  \label{c23}
\end{equation}
The matrices of the change of basis $\langle {\mathbf j},{\mathbf i}|{\mathbf j},{\mathbf i^{m_n}}\rangle$ are (products of five) 6-$j$ Wigner-symbols, as given  by standard recoupling theory. 

The state (\ref{c}) is the boundary state we shall use.  The complication of the sum over
pairings is less serious than what could seem at first sight, due to a key technicality that we prove in the next section: the components of (\ref{c}) become effectively orthogonal in the large distance limit. 

\subsubsection{Orthogonality of the terms in different bases in the large $j_0$ limit}

Suppose we want to compute the norm of the boundary state, in the 
limit of large $j_0$.  From (\ref{c}), this is given by 
\begin{equation}
\begin{split}
|\Psi|^2=\  
 \sum_{m_n}\  
 \sum_{m'_n}\  
 {}_{m_n}\!\langle \Phi_{\mathbf q} |
 \Phi_{\mathbf q} \rangle_{m'_n}
\end{split}
  \label{d}
\end{equation}
We now show that in the large $j_0$ limit the non-diagonal terms of this sum 
(those with $m_n\ne m'_n$) vanish.  Consider one of these terms, say
\begin{equation}
I =  {}_{m_n}\!\langle \Phi_{\mathbf q} | \Phi_{\mathbf q} \rangle_{m'_n}
= 
\sum_{{\mathbf j}{\mathbf i^{m_n}}}\  
\sum_{{\mathbf j'}{\mathbf i^{m'_n}}}\  
\overline{\Phi({\mathbf j},{\mathbf i^{m_n}})}\ 
\Phi[{\mathbf j'},{\mathbf i^{m'_n}}]\ 
\langle {\mathbf j}{\mathbf i^{m_n}} | {\mathbf j'}{\mathbf i^{m'_n}} \rangle
\end{equation}
where, say,  $m_n=(x,x,x,x,x)$ and $m'_n=(y,x,x,x,x)$. 
The scalar product is diagonal in the spins $\mathbf j$ and is given by 6-$j$ symbol
in the intertwiners quantum numbers. Hence
\begin{equation}
 I = 
\sum_{{\mathbf j}}\  
\sum_{{\mathbf i^{m_n}}}\  
\sum_{{\mathbf i^{m'_n}}}\  
\overline{\Phi({\mathbf j},{\mathbf i^{m_n}})}\ 
\Phi[{\mathbf j},{\mathbf i^{m'_n}}]\ 
\langle {i^x_1} | i^{y}_1 \rangle, 
\end{equation}
where (see Appendix \ref{snormalization}), 
\begin{equation}
\langle {i^x_1} | i^y_1 \rangle =  (-1)^{j_{\scr{13}}+j_{\scr{14}}+i^{\scr{x}}_1+i^{\scr{y}}_1} 
\    \ \sqrt{d_{ i^{\scr{x}}_1}d_{ i^{\scr{y}}_1}}\ \ 
  \left\{\begin{array}{ccc}                      j_{\scr{12}}  & j_{\scr{13}} & i^{\scr{x}}_1 \\
		                                                j_{\scr{15}} & j_{\scr{14}} & i^{\scr{y}}_1        
		 \end{array}\right\} . 
\end{equation}
In the large $j_0$ limit, this sum can be approximated by an integral, as in  \cite{scattering3}.
Both the spin and the interwiner sums become gaussian integrals, peaked respectively on 
$j_0$ and $i_0$. The range of the sum over intertwiners is finite for finite $j_0$, because
of the Clebsh Gordan conditions at the two trivalent node;  but this range is
much larger than the width of the Gaussian in the limit, and therefore the integral over
the intertwiner variables too can be taken over the entire real line.  In the limit, 
the 6-$j$ symbol has the asymptotic value \cite{Freidel:2002mj}
\begin{equation}
	\left\{\begin{array}{ccc}                      j_{\scr{12}}  & j_{\scr{13}} & i^{\scr{x}}_1 \\
	                                                j_{\scr{15}} & j_{\scr{14}} & i^{\scr{y}}_1        
\end{array}\right\} \approx \ \  \frac{e^{i (S_R+\frac{\pi}{4}) }+e^{-i (S_R+\frac{\pi}{4}) }}{\sqrt{12\pi V}}, 
\label{regge formula}
\end{equation}
where $S_R$ is the Regge action of a tetrahedron with side length determined by the
spins of the 6j symbol, and V is its volume. Changing the sum into an integration and
using this, we have 
\begin{equation}
I = 
\int d{\mathbf j}\int d{\mathbf i}\int di_1^y\ 
\overline{\Phi({\mathbf j},{\mathbf i})}\ 
\Phi({\mathbf j},{\mathbf i^{m'_n}})\ 
 (-1)^{j_{\scr{13}}+j_{\scr{14}}+i^{\scr{x}}_1+i^{\scr{y}}_1} \ 
 \frac{e^{i (S_R+\frac{\pi}{4}) }+e^{-i (S_R+\frac{\pi}{4}) }}{\sqrt{12\pi V}}.
\end{equation}
Inserting the explicit form of the state (\ref{Cpartenza}) gives
\begin{equation}
\begin{split}
I = &
\int d{\mathbf j}\int d{\mathbf i}\ \ 
		e^{-\frac{1}{j_0}\sum \alpha_{\scr{(ij)(mr)}}\delta j_{ij}\delta j_{mr}		-\sum_{n\ne 1}\frac{3(\delta i_n)^2}{2j_0}-\frac{3(\delta i_1)^2}{4j_0}
		-i\left(\sum_a\frac{3}{4 j_0} \delta j_{an}-\frac{\pi}{2}\right)\delta i^x_1}\ 
		 \\ & \ \ \ \cdot   
\int di_1^{y}\ 				e^{-\frac{3(\delta i^{y}_1)^2}{4j_0}}
		e^{i\left(\sum_a\frac{3}{4 j_0} \delta j_{an}-\frac{\pi}{2}\right)
		\delta i_1^{y}}\ 
 \frac{e^{i (S_R+\pi\delta i_1^{y}+\frac{\pi}{4}) }+e^{-i (S_R-\pi\delta i_1^{y}+\frac{\pi}{4}) }}{\sqrt{12\pi V}}.
\end{split}
\label{integrale}
\end{equation}

In the limit, only the first terms in the expansion of the Regge action around the 
maximum of the peak of the Gaussian matter. We thus Taylor expand the Regge action 
in its six entries $j_{1n},i^{\scr{x}}_1,i^{\scr{y}}_1$ around 
the background values  $j_0$ and $i_0$. 
\begin{equation}
\begin{split}
	S_j[j_{\scr{na}}]=&
	\left.\frac{\partial{S_R}}{\partial{j_{\scr{1n}}}}\right|_{j_0,i_0}\delta j_{\scr{1n}}+
	\left.\frac{\partial{S_R}}{\partial{i^{\scr{x}}_1}}\right|_{j_0,i_0}\delta i^{\scr{x}}_1+
	\left.\frac{\partial{S_R}}{\partial{i^{\scr{y}}_1}}\right|_{j_0,i_0}\delta i^{\scr{y}}_1+
{\rm higher\ order\ terms}.
\end{split}
	\end{equation}
The key point now is that the first of these terms is a rapidly oscillating phase factor in the 
$ j_{\scr{1n}}$ variable. The Gaussian $ j_{\scr{1n}}$ integration in (\ref{integrale}) is suppressed by this phase factor.  More precisely, the integral is like a Fourier transform in the $j_{\scr{1n}}$ variable, of a gaussian centered around a large value of $j_0$ with variance proportional $\sqrt{j_0}$; this Fourier transform is then a gaussian with variance $1/\sqrt{j_0}$, which goes to zero in the $j_0\rightarrow\infty$ limit.  {\em QED.}

\subsubsection{Change of basis}

For later convenience, let us also give here the expression of the state 
\eqref{Cpartenza} under the transformation induced by the change of basis 
associated to a change of pairing. Say we change from the basis $i^y$ to the basis
$i^x$ in the node $n=1$. Then directly from (\ref{c23}) we have 
\begin{equation}
\begin{split}
		\Phi'_{\mathbf q}[{\mathbf j}, i^{\scr{x}}_1,i_2...i_5]&
		=\ \ 
		e^{-\frac{1}{2j_0}\sum \alpha_{\scr{(ij)(mr)}}\delta j_{ij}\delta j_{mr}+i\sum\Phi \delta j_{ij}}
		e^{-\sum_{n\neq1}\left(\frac{3(\delta i_n)^2}{4j_0}-i\left(\sum_a\frac{3}{4 j_0} \delta j_{an}-\frac{\pi}{2}\right) \delta i_n\right)}
	 \\
		 	&\cdot
			\sum_{i^{\scr{y}}_1} e^{-\left(\frac{3(\delta i^{\scr{y}}_1)^2}{4j_0}-i\left(\sum_a\frac{3}{4 j_0}\;\delta j_{a1}-\frac{\pi}{2}\right)\delta i^y_1\right)}
		 (-1)^{j_{\scr{13}}+j_{\scr{14}}+i^{\scr{x}}_1+i^{\scr{y}}_1} \;\sqrt{d_{ i^{\scr{x}}_1}d_{ i^{\scr{y}}_1}}
		 \left\{\begin{array}{ccc}                      j_{\scr{12}}  & j_{\scr{13}} & i^{\scr{x}}_1 \\
		                                                j_{\scr{15}} & j_{\scr{14}} & i^{\scr{y}}_1        
		 \end{array}\right\} 
	\end{split}
	\label{ca1}
\end{equation}
where, we recall, the sum over intertwiners is under the condition \eqref{esistenza dei j} that that gives $(-1)^{j_{\scr{13}}+j_{\scr{14}}+i^{\scr{x}}_1}=1$
We can evaluate the sum in the large $j_0$ limit by approximating it again with an integral. Inserting the asymptotic expansion of the $6j$ symbol, we have 
\begin{equation}
	\begin{split}
			\Phi'_{\mathbf q}({\mathbf j}, i^{\scr{x}}_1,i_2...i_5)
		&=\ \ 
			e^{-\frac{1}{2j_0}\sum \alpha_{\scr{(ij)(mr)}}\delta j_{ij}\delta j_{mr}+i\sum\Phi \delta j_{ij}}
		e^{-\sum_{n\neq1}\left(\frac{3(\delta i_n)^2}{4j_0}-i\left(\sum_a\frac{3}{4 j_0} \delta j_{an}-\frac{\pi}{2}\right) \delta i_n\right)} e^{i\pi i_0}
	 \\
	 &\hspace{-1em} \cdot  \int{d\delta i^{\scr{y}}_1}
	 e^{-\left(\frac{3(\delta i^{\scr{y}}_1)^2}{4j_0}-i\left(\sum_a\frac{3}{4 j_0}\;\delta j_{a1}-\frac{\pi}{2}\right)\delta i^y_1\right)} 
	 \sqrt{d_{ i^{\scr{x}}_1}d_{ i^{\scr{y}}_1}}\;\frac{e^{i (S_R+\pi\delta i^y_1+\frac{\pi}{4}) }+e^{-i (S_R-\pi\delta i^y_1+\frac{\pi}{4}) }}{\sqrt{12\pi V}}.
				 \end{split}
			 \label{ca2}
\end{equation}

This can be computed expanding the Regge action to second order around $j_0$ and $i_0$.
As shown in the Appendix \ref{change of basis}, the result is 
\begin{equation}
	\begin{split}
			\Phi'_{\mathbf q}({\mathbf j}, i^{\scr{x}}_1,i_2,...,i_5)			=\Phi({\mathbf j}, i^{\scr{x}}_1,i_2,...,i_5)
		 N_1\; e^{-i S[j_{\scr {1a}}]}e^{-2i\left( \sum_a \frac{3}{4j_0}\;\delta j^{\scr{a1}}\right)\delta i^{x}_1}, 
		 			 \end{split}\label{cafine}
			 \end{equation}
where $N_1$ is a normalization constant with $\left|N_1\right|^2=1$, and 
$S[j_{\scr {1a}}]$ is the expansion of the Regge Action linked to the tetrahedron 
associated with the \{6j\} symbol \eqref{regge formula} up to the second order 
\textit{only in the link variables}, that is
\begin{equation}
	S[j_{\scr{na}}]=\left.\frac{\partial{S_R}}{\partial{j_{\scr{1n}}}}\right|_{j_0,i_0}\delta j_{\scr{1n}}+
	\left.\frac{\partial^2{S_R}}{\partial{j_{\scr{1n}}}\partial{j_{\scr{1n'}}}}\right|_{j_0,i_0}\delta j_{\scr{1n}}\delta j_{\scr{1n'}}
+\frac{1}{2}\left.\frac{\partial^2{S_R}}{\partial^2{j_{\scr{1n}}}}\right|_{j_0,i_0}(\delta j_{\scr{1n}})^2
\label{S}. 
	\end{equation}
This result follows from the choice \eqref{tunings1} and \eqref{tunings2} of the parameters in  \eqref{cbis}.  In particular, the value $\chi_n=\frac{\pi}{2}$ makes the intertwiner phase 
equal, with opposite sign, to the term $\exp{-i\left(\left.\frac{\partial{S_R}}{\partial{i^{\scr{y}}_1}}\right|_{j_0,i_0}\delta i^{\scr{y}}_1-\pi\delta i^y_1\right)}$, namely the term in the expansion of the Regge action $S_R$ linear in the variable $\delta i^{\scr{y}}_1$.  This selects one of the two exponentials in the asymptotic expansion \eqref{regge formula}, while the rapidly oscillating phase factor in the variables $\delta i^y_1$ cancels the other. 

The same calculation gives the $i^z\to i^x$ change of variable
			 \begin{equation}
	\begin{split}
			\Phi''_{\mathbf q}({\mathbf j}, i^{\scr{x}}_1,i_2,...,i_5) 			=\Phi({\mathbf j}, i^{\scr{x}}_1,i_2,...,i_5)
		 N_1\; e^{-i S'[j_{\scr {1a}}]}e^{-2i\left( \sum_a \frac{3}{4j_0}\;\delta j^{\scr{a1}}\right)\delta i^{x}_1}, 
		 			 \end{split}\label{cbfine}
			 \end{equation}
with the same constant $N_1$ as above. The only differences between \eqref{cafine} and \eqref{cbfine} is that the arguments of the 6-$j$ symbol enter with a different order, so that $S'(j_{12},j_{13},j_{14},j_{15},)=S(j_{12},j_{13},j_{15},j_{14})$.  

Using these results, we can explicitly rewrite the state (\ref{c}) in our preferred basis. We obtain easily 
\begin{equation}
	\left|\Psi_{\mathbf q}\right\rangle=4^5\sum_{\mathbf{j}, \mathbf{i}}	\Phi({\mathbf j}, {\mathbf i})
	\prod_{n=1}^5
		G[\delta j_{na},\delta i_n]
			\left| \mathbf j,\mathbf i\right\rangle,  
			\label{cinunabasex5}
\end{equation}
where
\begin{equation}
	G[\delta j_{na},\delta i_n]=\left(1+
					 N_1 e^{-2i\left( \sum_a \frac{3}{4j_0}\;\delta j^{\scr{an}}\right)\delta i^{x}_n}\left(e^{-i S[j_{\scr {na}}]}+e^{-i S'[j_{\scr {na}}]}\right)\right). 
					 \label{g}
\end{equation}

\subsection{Mean values and variances}

With these preliminary completed, we can now check that mean values and relative 
fluctuations of areas and angles have the right behavior in  the large scale limit. With 
the notation 
\begin{equation}
	\langle{O}\rangle := \frac{\left\langle \Psi_{\mathbf q} \right|O\left|\Psi_{\mathbf q}\right\rangle}{\left\langle \Psi_{\mathbf q}|\Psi_{\mathbf q}\right\rangle}
\quad \textrm{and}\quad
\Delta O = \sqrt{\langle{O^2}\rangle-  \langle{O}\rangle^2}
\end{equation}
we demand
\begin{equation}
	\langle{j_{ni}}\rangle=j_0\quad \textrm{and}\quad
	\frac{\Delta j_{ni}}{\langle{j_{ni}}\rangle}
 \rightarrow 0 \quad \textrm{when $j_0\rightarrow \infty$}, 
 \label {condizionispin}
\end{equation}
as in  \cite{scattering3}, as well as  
\begin{equation}
	\langle{i^{m_n}_{n}}\rangle=i_0\quad \textrm{and}\quad
	\frac{\Delta i^{m_n}_{n}}{\langle{i^{m_n}_{n}}\rangle}
 \rightarrow 0 \quad \textrm{when $j_0\rightarrow \infty$}.
 \label{condizioniintertwiner}
\end{equation}
Notice that we demand this for all $m_n$, namely for each node {\em in each pairing}.  

It is easy to show that the state \eqref{c} satysfies \eqref{condizionispin}. Because of the vanishing of the interference terms proven above, in large $j_0$ limit the mean values reduce to the average of the mean values on each diagonal term.  
\begin{eqnarray}
	\langle j_{ni}\rangle & \approx &
	  \frac{\sum_{m_n} \sum_{\mathbf j} \sum_{\mathbf i^{m_n}_{n}}
	j_{ni}\left|\Phi[{\mathbf j\;\mathbf i^{m_n}_{n}}]\right|^2 }{\sum_{m_n} 
	\sum_{\mathbf j}\sum_{\mathbf i^{m_n}_{n}}
	\left|\Phi[{\mathbf j\;\mathbf i^{m_n}_{n}}]\right|^2}\nonumber \\  	&\approx&
	\frac{\sum_{m_n}\int d\delta \mathbf{j}\,d\delta{\mathbf i^{m_n}_{n}}j_{ni}\;	e^{-\frac{1}{j_0}\sum \alpha_{\scr{(ij)(mr)}}\delta j_{ij}\delta j_{mr}}
			e^{-\sum_{n}\frac{3(\delta i^{\scr{m_n}}_n)^2}{2j_0}}}{\sum_{m_n}\int d\delta \mathbf{j}\,d\delta{\mathbf i^{m_n}_{n}}	e^{-\frac{1}{j_0}\sum \alpha_{\scr{(ij)(mr)}}\delta j^{\scr{ij}}\delta j_{mr}}
			e^{-\sum_{n}\frac{3(\delta i^{\scr{m_n}}_n)^2}{2j_0}}}=j_0.
\end{eqnarray}
The calculation of the variance and mean value in the intertwiner variable is a bit more complicated. It is convenient to express the state in the pairing of the relevant 
variable using \eqref{cafine} and  \eqref{cbfine}. With this, we have 
\begin{eqnarray}
\langle  i^x_{1} \rangle&\approx&\frac{\sum_{m_n\neq m_1}\sum_{\mathbf j} \sum_{\mathbf i^{m_n}_{n\neq1}}
\sum_{i^x_1}
		i^x_{1}\left(\left|\Phi_{\mathbf q} \right|^2+\left|\Phi'_{\mathbf q} \right|^2+\left|\Phi''_{\mathbf q} \right|^2\right) }
		{\sum_{m_n} \sum_{\mathbf j} \sum_{\mathbf i^{m_n}_{n}}
		\left|\Phi_{\mathbf q} \right|^2}\nonumber \\
		&\approx&
		 3\  \frac{\sum_{m_n\neq m_1}\sum_{\mathbf j} \sum_{\mathbf i^{m_n}_{n\neq1}}\sum_{i^x_1}
		i^x_{1}\left|\Phi_{\mathbf q} \right|^2 }
		{\sum_{m_n} \sum_{\mathbf j}\sum_{\mathbf i^{m_n}_{n}}
		\left|\Phi_{\mathbf q} \right|^2}
		=
		\frac{\sum_{m_n\neq m_1}\sum_{\mathbf j} \sum_{\mathbf i^{m_n}_{n\neq1}}\sum_{i^x_1}
		i^x_{1}\left|\Phi_{\mathbf q} \right|^2 }
		{\sum_{m_n\neq m_1}\sum_{\mathbf j} \sum_{\mathbf i^{m_n}_{n\neq1}}\sum_{i^x_1}
		\left| \Phi_{\mathbf q} \right|^2}=i_0, 
\end{eqnarray}
where  we have used the  \eqref{cafine} and \eqref{cbfine} and the fact that the constant $N_1$ in these expression satisfies $\left|N_1\right|^2=1$. The same procedure can be used to compute the variance and check that \eqref{condizioniintertwiner} is satisfied.

\section{Calculation of the propagator}

We are now ready to compute all components of the propagator (\ref{partenzaA}).  Consider this quantity for a fixed value of $m,n,i,j,k,l$.    Because of the sum in (\ref{g}), the propagator can be written in the form:
\begin{equation}\begin{split}
{\mathbf G} _{{\mathbf q}\, n,m}^{\scriptscriptstyle ij,kl} =  
4^5
\sum_{\mathbf j} 	\sum_{\mathbf{i}_n}	&\ \ 
\Phi({\mathbf j}, {\mathbf i})
	\prod_{n=1}^5
		G[\delta j_{na},\delta i_n]\\
&\cdot  \langle W | \big(E^{\scriptscriptstyle(ni)}_n  \cdot 
E^{\scriptscriptstyle(nj)}_n-n^{\scriptscriptstyle(ni)}\cdot n^{\scriptscriptstyle(nj)}\big)
\big(E^{\scriptscriptstyle(mk)}_m  \cdot 
E^{\scriptscriptstyle(ml)}_m-n^{\scriptscriptstyle(mk)}\cdot n^{\scriptscriptstyle(ml)}\big) 
			\left| \mathbf j,\mathbf i_{n}\right\rangle,  
\end{split}
\label{partenzaB}
\end{equation}
For a given value of $m,n,i,j,k,l$, we now
can fix the reference choice of pairing so that $(ij)$ (if different) are paired 
at the node $n$ and $(kl)$ (if different) are paired at the node $m$. With this
choice of basis the action of the operators is diagonal, and we have 
\begin{eqnarray}
{\mathbf G} _{{\mathbf q}\, n,m}^{\scriptscriptstyle ij,kl} &=& 4^5
\sum_{\mathbf j} 	\sum_{\mathbf{i}_n}	
\Phi({\mathbf j}, {\mathbf i})
	\prod_{n=1}^5
		G[\delta j_{na},\delta i_n]
 \big(D_n^{ij}-n^{\scriptscriptstyle(ni)}\cdot n^{\scriptscriptstyle(nj)}\big)
\big(D_m^{kl}-n^{\scriptscriptstyle(mk)}\cdot n^{\scriptscriptstyle(ml)}\big) 
			 \langle W | \mathbf j,\mathbf i\rangle,  
\label{partenzaC}
\end{eqnarray}
We use the same form of the  the Barret-Crane vertex as in  \cite{scattering1,scattering2}. This is given by  
\begin{equation}
	 \langle W | \mathbf j,\mathbf i\rangle := W({\mathbf j},{\mathbf i})=W({\mathbf j}) \prod_n \langle i_{BC}|i_n\rangle=W({\mathbf j}) \prod_{n} (2i_n+1) ,
	\label {Wbc}
\end{equation}
where $W({\mathbf j}) $ is the Barrett-Crane vertex, which a functions of the ten spins alone.
In the large distance limit, $\prod_{n} (2i_n+1) = 2i_0^5$, hence 
\begin{equation}
	W({\mathbf j},{\mathbf i})=2i_0^5\ W({\mathbf j}).  
\end{equation}
Using this, \Ref{partenzaC} becomes 
\begin{eqnarray}
{\mathbf G} _{{\mathbf q}\, n,m}^{\scriptscriptstyle ij,kl} &=& 
\sum_{\mathbf j}  W({\mathbf j})	\sum_{\mathbf{i}^{\scr{x}}_n}	
\Phi({\mathbf j}, {\mathbf i})
	\prod_{n=1}^5
		G[\delta j_{na},\delta i_n]
 \big(D_n^{ij}-n^{\scriptscriptstyle(ni)}\cdot n^{\scriptscriptstyle(nj)}\big)
\big(D_m^{kl}-n^{\scriptscriptstyle(mk)}\cdot n^{\scriptscriptstyle(ml)}\big) , 
\label{partenzaC1}
\end{eqnarray}
where we have absorbed numerical factors and $i_0^5$ in the normalization of the state. Each factor $G[\delta j_{na},\delta i_n]$ in this expression has the form $(1+N e^{iS}+Ne^{iS'})$. The terms with the exponents contain rapidly oscillating phases in the spin variables, which again suppress the integral in the large $j_0$ limit. Therefore we can drop these factors. 

The value of the eigenvalues $D_n^{ij}$ is given in \Ref{diag}. The value of the product of normals is given in \Ref{normali di background}. Using these, we have 
\begin{equation}
D_n^{\scriptscriptstyle ij}-n^{\scriptscriptstyle(ni)}\cdot n^{\scriptscriptstyle(nj)}=\frac{\left(C(i_n)-C(i_0)\right)-\left(C(j^{\scriptscriptstyle(ni)})-C(j_0)\right)-\left(C(j^{\scriptscriptstyle(nj)})-C(j_0)\right)}{2}.
\label {gemello3}
\end{equation}
Expanding  up to second order around the background  values $j_0$ and $i_0$ 
\begin{equation}
C(j_{j})-C(j_{0})=(\delta j_{j})^2+2	\delta j_{j}j_0+\delta j_{j},
\end{equation}
we obtain, in the large $j_0$ limit 
\begin{equation}
	D_n^{\scriptscriptstyle ij}-n^{\scriptscriptstyle(ni)}\cdot n^{\scriptscriptstyle(nj)}=\delta i_n \;i_0-\delta j_{j}j_0-\delta j_{nk}j_0.
	\label{ultimo gemello}
\end{equation}
Inserting this in \eqref{partenzaC1} we have 
\begin{eqnarray}
{\mathbf G} _{{\mathbf q}\, n,m}^{\scriptscriptstyle ij,kl} &=& j_0^2
\sum_{\mathbf j}  W({\mathbf j})	\sum_{\mathbf{i}^{\scr{x}}_n}	
\left(\frac{2}{\sqrt{3}}\; \delta i_n-\delta j_{ni}-
\delta j_{nk}\right)\left(\frac{2}{\sqrt{3}}\;\delta i_m-\delta j_{mk}-\delta j_{ml}\right)
\Phi({\mathbf j} , {\mathbf i}).
\label{partenzaD}
\end{eqnarray}

In the case in which two of the indices of the propagator are parallel, say $i=j$, this reduces easily to 
\begin{eqnarray}
{\mathbf G} _{{\mathbf q}\, n,m}^{\scriptscriptstyle ii,kl} &=&2 j_0^2
\sum_{\mathbf j}  W({\mathbf j})	\sum_{\mathbf{i}^{\scr{x}}_n}	
\delta j_{ni}\left(\frac{2}{\sqrt{3}}\delta i_m-\delta j_{mk}-\delta j_{ml}\right)
\Phi({\mathbf j} , {\mathbf i}).
\label{partenzaD1}
\end{eqnarray}

While if $i=j$ and $k=l$ we recover the diagonal terms, 
\begin{eqnarray}
{\mathbf G} _{{\mathbf q}\, n,m}^{\scriptscriptstyle ii,kk} &=& 4 j_0^2
\sum_{\mathbf j}  W({\mathbf j})	\sum_{\mathbf{i}^{\scr{x}}_n}	
\delta j_{ni}\delta j_{mk}\
\Phi({\mathbf j}, {\mathbf i}).
\label{partenzaD2}
\end{eqnarray}

We can now evaluate \eqref{partenzaD}. Inserting the explicit form of the state gives
\begin{equation}
	\begin{split}
		{\mathbf G} _{{\mathbf q}\, n,m}^{\scriptscriptstyle ij,kl}&= C j^2_0 \sum_{\delta{\mathbf j},\delta{\mathbf i}}W({\mathbf j})  \left(\frac{2}{\sqrt{3}}\; \delta i_n-\delta j_{ni}-\delta j_{nk}\right)\left(\frac{2}{\sqrt{3}}\;\delta i_m-\delta j_{mk}-\delta j_{ml}\right)
	 \\ 
	&\cdot\;		e^{-\frac{1}{2j_0}\sum \alpha_{\scr{(ij)(mr)}}\delta j_{ij}\delta j_{mr}+i\sum\Phi \delta j_{ij}}
										 e^{-\sum_{n}\left(\frac{3(\delta i_n)^2}{4j_0}-i\left(\sum_a\frac{3}{4 j_0}\;\delta j_{an}+\frac{\pi}{2}\right)\delta i_n\right)}.
				\label{partenza3}
		\end{split}
\end{equation}
Using the asymptotic expression for the  BC vertex, we can proceed like in \cite{scattering1} and \cite{scattering2}. The rapidly oscillating phase term in the spins selects one of the  the factors of this expansion, giving 
\begin{equation}
\begin{split}
	{\mathbf G} _{{\mathbf q}\, n,m}^{\scriptscriptstyle ij,kl}
		&={\cal N}j^2_0\sum_{\delta j^{\scr{(ab)}},\, \delta i_{\alpha}}\prod_{a < b} {\rm dim}({j^{(ab)}}) \left(\frac{2}{\sqrt{3}}\; \delta i_n-\delta j_{ni}-\delta j_{nk}\right)\left(\frac{2}{\sqrt{3}}\;\delta i_m-\delta j_{mk}-\delta j_{ml}\right)
	  \\
	&\cdot \;e^{-\frac{1}{2j_0} {\left(\alpha+iG j_0\right)_{(ij)(mn)}}\ 
	\delta j_{ij}\delta j_{mn} }
		 e^{-\sum_{n}\left(\frac{3(\delta i_n)^2}{4j_0}-i\left(\sum_a\frac{3}{4 j_0}\;\delta j^{\scr{an}}+\frac{\pi}{2}\right)\delta i_n\right)},
		\label{partenza4}
\end{split}
\end{equation}
where the phase factor $i \Phi \sum_{pq}  j_{pq}$ in \Ref{partenza3} has been absorbed by the corresponding phase factor in the asymptotic expansion of the 10$j$ symbol $W({\mathbf j})$ (see
\cite{Freidel:2002mj,Baez:2002rx}), as in \cite{scattering2,scattering3}.   Here $G$ is the matrix of the second derivatives of the Regge
action (see \cite{scattering2,scattering3}) and should not be confused with the $G$ used in
the Appendix.   Finally, 
\begin{equation}
\begin{split}
	{\mathbf G} _{{\mathbf q}\, n,m}^{\scriptscriptstyle ij,kl}
		&={\cal N'}j_0^2\sum_{\delta j^{\scr{(ab)}},\, \delta i_{\alpha}} \left(\frac{2}{\sqrt{3}}\; \delta i_n-\delta j_{ni}-\delta j_{nk}\right)\left(\frac{2}{\sqrt{3}}\;\delta i_m-\delta j_{mk}-\delta j_{ml}\right)
	\cdot  \\
	&\cdot \;e^{-\frac{1}{2j_0} {\left(\alpha+iG j_0\right)_{(ij)(mn)}}\ 
	\delta j_{ij}\delta j_{mn} }
	e^{-\sum_{n}\left(\frac{3(\delta i_n)^2}{4j_0}-i\left(\sum_a\frac{3}{4 j_0}\;\delta j_{an}+\frac{\pi}{2}\right)\delta i_n\right)}.
						\label{partenza6}
\end{split}
\end{equation}
We can rearrange this expression introducing the 15 components vector $\delta I^\alpha=(\delta j^{ab},\delta i_n)$ and $\Theta^{\alpha}=(0,\chi_{i_n})$ and the $15 \times 15$ correlation matrix 
\begin{equation}
M=\left(\begin{array}[c]{cc}
\textrm{\large{A}}_{10\times10}&\textrm{\large{C}}_{10\times5}	\\
	\textrm{\large{C}}^T_{5\times10}
	&\textrm{\large{S}}_{5\times5}	\end{array}\right), 
\end{equation}
where $A_{{\scr{ab}}\,{\scr{cd}}}=\frac{1}{2} {\left(\alpha+iG j_0\right)}_{{\scr{ab}}\,{\scr{cd}}}$ is a $10 \times 10$ matrix and $S_{nm}=I_{nm} \frac{3}{4}$ is a diagonal $5 \times 5$ matrix and C is a $10 \times 5$ matrix and $C^T$ is its transpose,  
 and evaluate it approximating the sum with an integral 
\begin{equation}
\begin{split}
	{\mathbf G} _{{\mathbf q}\, n,m}^{\scriptscriptstyle ij,kl}
		&={\cal N'}j_0^2\int {d\delta I^{\alpha}} \left(\frac{2}{\sqrt{3}}\; \delta i_n-\delta j_{ni}-\delta j_{j}\right)\left(\frac{2}{\sqrt{3}}\;\delta i_m-\delta j_{mk}-\delta j_{ml}\right)
 \;e^{-\frac{M_{\alpha\beta}}{j_0}\delta I^{\alpha}\delta I^{\beta}}\ 
			e^{i\Theta_{\alpha} \delta I^{\alpha}}	. 
\label{partenza7}
\end{split}
\end{equation}
The matrix M is invertible and independent from $j_0$. 
Direct calculation using \eqref{trasfmomenti} gives a sum 
of terms of the kind
\begin{equation}
\frac{e^{-j_0\Theta M^{-1}\Theta}}{\sqrt{\det M}}	\left(j_0^3M^{-1}_{\alpha\beta}-j_0^4M^{-1}_{\alpha\gamma}\Theta^{\gamma}  M^{-1}_{\beta\delta}\Theta^{\delta} \right).
\label{termini sbagliati}
\end{equation}
These terms go to zero fast  in the $j_0\rightarrow\infty$ limit, and therefore do not match the
expected large distance behavior of the propagator. 
  
One could hope to circumvent the problem behaviour thanks to the normalization factor. Including this explicitly we have
\begin{eqnarray}
{\tilde{\mathbf G}} _{{\mathbf q}\, n,m}^{\scriptscriptstyle ij,kl} &=& \frac{\langle W | \big(E^{\scriptscriptstyle(ni)}_n \cdot E^{\scriptscriptstyle(nj)}_n-n^{\scriptscriptstyle(ni)}\cdot n^{\scriptscriptstyle(nj)}\big)
\big(E^{\scriptscriptstyle(mk)}_m  \cdot E^{\scriptscriptstyle(ml)}_m-n^{\scriptscriptstyle(mk)}\cdot n^{\scriptscriptstyle(ml)}\big) |\Psi_{\mathbf q} \rangle}{\langle W|\Psi_{\mathbf q} \rangle}.  \label{partenzanormalizzata}
\end{eqnarray}
The denominator gives 
\begin{equation}
\langle W|\Psi_{\mathbf q} \rangle=\frac{e^{-j_0\Theta M^{-1}\Theta}}{\sqrt{\det M}}.
	\label{normalizzazione}
\end{equation}
Terms of the kind \eqref{termini sbagliati} are still pathological, since they give 
\begin{equation}
	\left(\frac{M^{-1}_{\alpha\beta}}{j_0}-M^{-1}_{\alpha\gamma}\Theta^{\gamma}  M^{-1}_{\beta\delta}\Theta^{\delta} \right)
\label{termini sbagliati2}
\end{equation}
in the limit.  In conclusion, the calculation presented does not appear to give the correct low energy propagator. 

\section{Conclusions}

The calculation presented above is based on a number of assumptions on the form of the 
boundary state.  Could the negative result that we have obtained be 
simply the result of these assumptions being too strict, or otherwise wrong? Could,
in particular, a different boundary state give the correct low energy behavior?  
Although we do not have a real proof, we do not think that this is the case.  
The original aim of the research program motivating this article was to find such 
a state; the negative result we report here has initially come as a disappointment,
and we have fought against it at long. We have eventually got to the conclusion that 
the problem is more substantial, and is related to the BC vertex itself, at least as it is 
used in the present approach.  There are several indications pointing to this conclusion.

First, the trivial intertwiner dependence of the Barrett-Crane structure clashes with the
intertwiner dependence of the boundary state that is needed to have a good 
semiclassical behavior.   Since the variables associated to the angles between faces
do not commute with one another, the boundary state cannot be sharp on a 
classical configuration. In order for a state peaked on a given angle to be also 
peaked on the other non-commuting angles, the state must have a phase 
dependence from intertwiners and spin variables.   Following the general 
structure of quantum mechanics, one then expect the transition amplitude
matching between coherent states to include a phase factor exactly 
balancing those phases.   This is the case for instance for the free propagator
of a non-relativistic quantum particles, as well as for the phases associated
to the angles between tetrahedra in the calculation illustrated in 
 \cite{scattering1,scattering2}. However, no such phase factor appears in the
 BC vertex.   In particular, the phase factor $i\frac\pi2 \sum_p i_p$ present in the boundary state (necessary to have the complete symmetry of the state) is not matched by a corresponding factor in the vertex amplitude. This factor gives the rapidly oscillating term that suppresses the sum. 
 
Second, as already mentioned, there is in fact a structural 
difficulty, already pointed out in \cite{scattering1,scattering2}, with the definition
\Ref{Wbc} of the amplitude, and we think that this difficulty is at the roots of the
problem.  Let us illustrate this difficulty in detail.  

There are two possible interpretations of equation \Ref{Wbc}.  The first is that this is true is
one particular basis, namely for $i_n=i_n^x$.  Let us discard this possibility, which would imply that the BC the vertex itself would depend on a specific choice of pairing. The second is that it is (simultaneously) true in all possible bases, that is
\begin{equation}
	 \langle W | \mathbf j,\mathbf i^{m_n}\rangle = W({\mathbf j}) \prod_{n} (2i^{m_n}+1) 
	\label {Wbc2}
\end{equation}
for any choice of pairing, namely for any choice of $m_n$.  This is indeed the definition of the
vertex that we have implicitly used.  However, defined in this way, the vertex $ \langle W |$ is
not a {\em linear} functional on the state space. This is immediately evident by expressing, say  $\langle i_1^y |$ on the $\langle i_1^x |$ basis.  

We can say this in other words. The Barrett-Crane intertwiner is defined as a sum of simple 
$SO(4)$ intertwiners, that we can write as  
\begin{eqnarray}
i_{\rm BC}  &=& 
\sum_{i^x}\ \  (2i^x+1) |i^x,i^x\rangle
=
\sum_{i^y}\ \  (2i^y+1) |i^y,i^y\rangle
\nonumber\\ &=&
\sum_{i^x}\ \  (2i^x+1)
\begin{array}{c}\setlength{\unitlength}{1 pt}
\begin{picture}(50,40)          \put(10,10){\line(1,1){10}}\put(10,30){\line(1,-1){10}}          \put(30,20){\line(1,1){10}}\put(30,20){\line(1,-1){10}}          \put(20,20){\line(1,0){10}}\put(22,25){$i^x$}          \put(20,20){\circle*{3}}\put(30,20){\circle*{3}}\end{picture}
\begin{picture}(50,40)          \put(10,10){\line(1,1){10}}\put(10,30){\line(1,-1){10}}          \put(30,20){\line(1,1){10}}\put(30,20){\line(1,-1){10}}          \put(20,20){\line(1,0){10}}\put(22,25){$i^x$}          \put(20,20){\circle*{3}}\put(30,20){\circle*{3}}\end{picture}\end{array}
=\ \ \ 
\sum_{i^y}\ \  (2i^y+1)
\begin{array}{c}\setlength{\unitlength}{1 pt}\begin{picture}(40,40)      \put(10,10){\line(1,1){10}}\put(10,40){\line(1,-1){10}}      \put(20,30){\line(1,1){10}}\put(20,20){\line(1,-1){10}}      \put(20,20){\line(0,1){10}}\put(22,22){$i^y$}      \put(20,20){\circle*{3}}\put(20,30){\circle*{3}}\end{picture}\begin{picture}(40,40)      \put(10,10){\line(1,1){10}}\put(10,40){\line(1,-1){10}}      \put(20,30){\line(1,1){10}}\put(20,20){\line(1,-1){10}}      \put(20,20){\line(0,1){10}}\put(22,22){$i^y$}      \put(20,20){\circle*{3}}\put(20,30){\circle*{3}}\end{picture}\end{array}.
\label{ibcvero}
\end{eqnarray}
Hence
\begin{equation}
\langle i_{BC}|i^m,i^m\rangle=(2i^m+1)
\end{equation}
whatever is $m$. 
Since the simple SO(4) intertwiner $|i^x,i^x\rangle$ diagonalizes the same geometrical quantity as 
the $SO(3)$ intertwiner $|i^x\rangle$, it is tempting to physically identify the two and write 
\begin{equation}
\langle i_{BC}|i^m\rangle=(2i^m+1).
\end{equation}
But there is no state $\langle i_{BC}|$ in the the SO(3) intertwiner space that has this property. 
In other words, there is a mismatch between the linear structures of SO(4) and SO(3) in building up the theory that we have used. 

In the companion paper \cite{II}, we show that, perhaps surprisingly, a vertex with a suitable asymptotic behavior can overcame all these difficulties.  

\newpage
\appendix

\section{Recoupling theory}\label{recoupling}

We give here the definitions at the basis of recoupling theory and the graphical notation
that is used in the text. Our main reference source is \cite{BS}. 

\begin{itemize}\setlength{\itemsep}{.2in}

\item{\em Wigner 3j-symbols.} These are represented by a 3-valent node, the three lines stand for the angular momenta wich are coupled by the  3j-symbol.
We denote the anti-clockwise orientation with a + sign and the clockwise orientation with a sign -. in index notation $v^{\alpha\beta\gamma}$: 
\begin{equation}
\begin{array}{cccc}
\left(\begin{array}{ccc}     a & b & c \\          \alpha & \beta & \gamma \end{array}\right)&=
	\begin{array}{c}
	\ifx\JPicScale\undefined\def\JPicScale{1}\fi
\psset{unit=\JPicScale mm}
\psset{linewidth=0.2,dotsep=1,hatchwidth=0.3,hatchsep=1.5,shadowsize=1,dimen=middle}
\psset{dotsize=0.7 2.5,dotscale=1 1,fillcolor=black}
\psset{arrowsize=1 2,arrowlength=1,arrowinset=0.25,tbarsize=0.7 5,bracketlength=0.15,rbracketlength=0.15}
\begin{pspicture}(0,0)(25,17.5)
\psline(10,10)(17.5,10)
\psline(17.5,10)(22.5,15)
\psline(17.5,10)(22.5,5)
\rput(10,12.5){$a\, \alpha$}
\rput(25,2.5){$b\,\beta$}
\rput(25,17.5){$c\,\gamma$}
\rput(24,10){$+$}
\end{pspicture}
\end{array}&=
\begin{array}{c}
	\ifx\JPicScale\undefined\def\JPicScale{1}\fi
\psset{unit=\JPicScale mm}
\psset{linewidth=0.2,dotsep=1,hatchwidth=0.3,hatchsep=1.5,shadowsize=1,dimen=middle}
\psset{dotsize=0.7 2.5,dotscale=1 1,fillcolor=black}
\psset{arrowsize=1 2,arrowlength=1,arrowinset=0.25,tbarsize=0.7 5,bracketlength=0.15,rbracketlength=0.15}
\begin{pspicture}(0,0)(25,17.5)
\psline(10,10)(17.5,10)
\psline(17.5,10)(22.5,15)
\psline(17.5,10)(22.5,5)
\rput(10,12.5){$a\, \alpha$}
\rput(25,2.5){$c\,\gamma$}
\rput(25,17.5){$b\,\beta$}
\rput(24,10){-}
\end{pspicture}
\end{array}
\end{array}\label{3j}
\end{equation}
The symmetry relation $v^{\alpha \beta\gamma}=(-1)^{a+b+c}\;v^{\alpha\gamma\beta}$
\begin{equation}
	\left(\begin{array}{ccc}     a & b & c \\          \alpha & \beta & \gamma \end{array}\right)=
	(-1)^{a+b+c}\left(\begin{array}{ccc}     a & c & b \\          \alpha & \gamma & \beta \end{array}\right)
\end{equation}
implies

\begin{equation}
			\begin{array}{c}
\ifx\JPicScale\undefined\def\JPicScale{1}\fi
\psset{unit=\JPicScale mm}
\psset{linewidth=0.2,dotsep=1,hatchwidth=0.3,hatchsep=1.5,shadowsize=1,dimen=middle}
\psset{dotsize=0.7 2.5,dotscale=1 1,fillcolor=black}
\psset{arrowsize=1 2,arrowlength=1,arrowinset=0.25,tbarsize=0.7 5,bracketlength=0.15,rbracketlength=0.15}
\begin{pspicture}(0,0)(18,19)
\psline(0,10)(10,10)
\psline(10,10)(17,3)
\psline(10,10)(17,17)
\rput(13,10){+}
\rput(-1,12){$a$}
\rput(18,19){$b$}
\rput(18,1){$c$}
\rput(18,1){}
\end{pspicture}
\end{array}\;\;=(-1)^{a+b+c}\;\;
		\begin{array}{c}
\ifx\JPicScale\undefined\def\JPicScale{1}\fi
\psset{unit=\JPicScale mm}
\psset{linewidth=0.2,dotsep=1,hatchwidth=0.3,hatchsep=1.5,shadowsize=1,dimen=middle}
\psset{dotsize=0.7 2.5,dotscale=1 1,fillcolor=black}
\psset{arrowsize=1 2,arrowlength=1,arrowinset=0.25,tbarsize=0.7 5,bracketlength=0.15,rbracketlength=0.15}
\begin{pspicture}(0,0)(18,19)
\psline(0,10)(10,10)
\psline(10,10)(17,3)
\psline(10,10)(17,17)
\rput(13,10){-}
\rput(-1,12){$a$}
\rput(18,19){$b$}
\rput(18,1){$c$}
\rput(18,1){}
\end{pspicture}
\end{array}
\label{cambio orientazione}
\end{equation}

\item{\em The Kroneker delta.}

\begin{equation}
\delta_{ab}\;\delta^{\alpha}_{\beta}=\;\;	\begin{array}{c}	\ifx\JPicScale\undefined\def\JPicScale{1}\fi
\psset{unit=\JPicScale mm}
\psset{linewidth=0.2,dotsep=1,hatchwidth=0.3,hatchsep=1.5,shadowsize=1,dimen=middle}
\psset{dotsize=0.7 2.5,dotscale=1 1,fillcolor=black}
\psset{arrowsize=1 2,arrowlength=1,arrowinset=0.25,tbarsize=0.7 5,bracketlength=0.15,rbracketlength=0.15}
\begin{pspicture}(0,10)(20,12.5)
\psline(20,10)(0,10)
\rput(0,12.5){$a\alpha$}
\rput(20,12.5){$b\beta$}
\rput(15,10){}
\end{pspicture}
	\end{array}. 
\end{equation}

\item{\em Anti-symmetric or ``metric" tensor.} (1-j symbol). In vector notation: ${}^a\epsilon_{\alpha\beta}$
\begin{equation}
	\left(\begin{array}{c} a\\ 
	\alpha \beta
	\end{array}\right)= (-1)^{a+\alpha}\;\delta_{\alpha -\beta}
\end{equation}
in graphical notation:  
\begin{equation}
	\delta_{ab}\left(\begin{array}{c} a\\ 
	\alpha \beta
	\end{array}\right)=\;\;\begin{array}{c}
	\ifx\JPicScale\undefined\def\JPicScale{1}\fi
\psset{unit=\JPicScale mm}
\psset{linewidth=0.2,dotsep=1,hatchwidth=0.3,hatchsep=1.5,shadowsize=1,dimen=middle}
\psset{dotsize=0.7 2.5,dotscale=1 1,fillcolor=black}
\psset{arrowsize=1 2,arrowlength=1,arrowinset=0.25,tbarsize=0.7 5,bracketlength=0.15,rbracketlength=0.15}
\begin{pspicture}(0,10)(20,12.5)
\psline{<-}(10,10)(20,10)
\psline(10,10)(0,10)
\rput(0,12.5){$a\alpha$}
\rput(20,12.5){$b\beta$}
\rput(15,10){}
\end{pspicture}
	\end{array}
\end{equation}
the relations $\epsilon^{\alpha '\beta}\epsilon_{\alpha\beta}=\delta^{\alpha '}_{\;\alpha}$ and $\epsilon^{\alpha '\beta}\epsilon_{\beta\alpha}=-\delta^{\alpha '}_{\;\alpha}$, for the fundamental representation, read, for generic representations 

\begin{equation}
\sum_{\beta}	\left(\begin{array}{c} a\\ 
	\alpha' \beta
	\end{array}\right)
	\left(\begin{array}{c} a\\ 
	\alpha \beta
	\end{array}\right)=\delta^{\alpha'}_{\;\alpha}
	\end{equation}

\begin{equation}
\begin{array}{c}
	\ifx\JPicScale\undefined\def\JPicScale{1}\fi
\psset{unit=\JPicScale mm}
\psset{linewidth=0.3,dotsep=1,hatchwidth=0.3,hatchsep=1.5,shadowsize=1,dimen=middle}
\psset{dotsize=0.7 2.5,dotscale=1 1,fillcolor=black}
\psset{arrowsize=1 2,arrowlength=1,arrowinset=0.25,tbarsize=0.7 5,bracketlength=0.15,rbracketlength=0.15}
\begin{pspicture}(0,0)(20,5)
\psline{<-}(8,2)(0,2)
\psline{<-}(12,2)(20,2)
\psline(20,2)(0,2)
\rput(0,5){$a\alpha$}
\rput(20,5){$a\alpha'$}
\end{pspicture}\end{array}=
\begin{array}{c}
	\ifx\JPicScale\undefined\def\JPicScale{1}\fi
\psset{unit=\JPicScale mm}
\psset{linewidth=0.3,dotsep=1,hatchwidth=0.3,hatchsep=1.5,shadowsize=1,dimen=middle}
\psset{dotsize=0.7 2.5,dotscale=1 1,fillcolor=black}
\psset{arrowsize=1 2,arrowlength=1,arrowinset=0.25,tbarsize=0.7 5,bracketlength=0.15,rbracketlength=0.15}
\begin{pspicture}(0,0)(20,5)
\psline(20,2)(0,2)
\rput(0,5){$a\alpha$}
\rput(20,5){$a\alpha'$}
\end{pspicture}\end{array}
\label{opposte}
\end{equation}	
and
\begin{equation}
\sum_{\beta}	\left(\begin{array}{c} a\\ 
	\alpha' \beta
	\end{array}\right)
	\left(\begin{array}{c} a\\ 
	 \beta\alpha
	\end{array}\right)=(-1)^{2a}\;\delta^{\alpha'}_{\;\alpha}
\end{equation}

	\begin{equation}
\begin{array}{c}
	\ifx\JPicScale\undefined\def\JPicScale{1}\fi
\psset{unit=\JPicScale mm}
\psset{linewidth=0.3,dotsep=1,hatchwidth=0.3,hatchsep=1.5,shadowsize=1,dimen=middle}
\psset{dotsize=0.7 2.5,dotscale=1 1,fillcolor=black}
\psset{arrowsize=1 2,arrowlength=1,arrowinset=0.25,tbarsize=0.7 5,bracketlength=0.15,rbracketlength=0.15}
\begin{pspicture}(0,0)(20,5)
\psline{<-}(7,2)(0,2)
\psline{<-}(13,2)(4,2)
\rput(0,5){$a\alpha$}
\psline(20,2)(0,2)
\rput(20,5){$a\alpha'$}
\end{pspicture}\end{array}=(-1)^{2a}\;\;\;
\begin{array}{c}
	\ifx\JPicScale\undefined\def\JPicScale{1}\fi
\psset{unit=\JPicScale mm}
\psset{linewidth=0.3,dotsep=1,hatchwidth=0.3,hatchsep=1.5,shadowsize=1,dimen=middle}
\psset{dotsize=0.7 2.5,dotscale=1 1,fillcolor=black}
\psset{arrowsize=1 2,arrowlength=1,arrowinset=0.25,tbarsize=0.7 5,bracketlength=0.15,rbracketlength=0.15}
\begin{pspicture}(0,0)(20,5)
\psline(20,2)(0,2)\rput(0,5){$a\alpha$}
\rput(20,5){$a\alpha'$}
\end{pspicture}
\end{array}\label{doppie}
\end{equation}

From the properties of the 3j symbols it follows: in vector notation: $v^{\alpha\beta\gamma}=\;v_{\alpha\beta\gamma}$; in graphical notation:

\begin{equation}
		\begin{array}{c}
\ifx\JPicScale\undefined\def\JPicScale{1}\fi
\psset{unit=\JPicScale mm}
\psset{linewidth=0.2,dotsep=1,hatchwidth=0.3,hatchsep=1.5,shadowsize=1,dimen=middle}
\psset{dotsize=0.7 2.5,dotscale=1 1,fillcolor=black}
\psset{arrowsize=1 2,arrowlength=1,arrowinset=0.25,tbarsize=0.7 5,bracketlength=0.15,rbracketlength=0.15}
\begin{pspicture}(0,0)(18,19)
\psline(0,10)(10,10)
\psline(10,10)(17,3)
\psline(10,10)(17,17)
\rput(13,10){+}
\rput(-1,12){$a$}
\rput(18,19){$b$}
\rput(18,1){$c$}
\rput(18,1){}
\end{pspicture}
\end{array}\;\;=
\begin{array}{c}
\ifx\JPicScale\undefined\def\JPicScale{1}\fi
\psset{unit=\JPicScale mm}
\psset{linewidth=0.2,dotsep=1,hatchwidth=0.3,hatchsep=1.5,shadowsize=1,dimen=middle}
\psset{dotsize=0.7 2.5,dotscale=1 1,fillcolor=black}
\psset{arrowsize=1 2,arrowlength=1,arrowinset=0.25,tbarsize=0.7 5,bracketlength=0.15,rbracketlength=0.15}
\begin{pspicture}(0,0)(18,19)
\psline(0,10)(10,10)
\psline(10,10)(17,3)
\psline(10,10)(17,17)
\psline{<-}(12,12)(17,17)
\psline{<-}(12,8)(17,3)
\psline{<-}(7,10)(0,10)
\rput(13,10){+}
\rput(-1,12){$a$}
\rput(18,19){$b$}
\rput(18,1){$c$}
\rput(18,1){}
\end{pspicture}
\end{array}=
\begin{array}{c}
\ifx\JPicScale\undefined\def\JPicScale{1}\fi
\psset{unit=\JPicScale mm}
\psset{linewidth=0.2,dotsep=1,hatchwidth=0.3,hatchsep=1.5,shadowsize=1,dimen=middle}
\psset{dotsize=0.7 2.5,dotscale=1 1,fillcolor=black}
\psset{arrowsize=1 2,arrowlength=1,arrowinset=0.25,tbarsize=0.7 5,bracketlength=0.15,rbracketlength=0.15}
\begin{pspicture}(0,0)(18,19)
\psline(0,10)(10,10)
\psline(10,10)(17,3)
\psline(10,10)(17,17)
\rput(13,10){+}
\rput(-1,12){$a$}
\rput(18,19){$b$}
\rput(18,1){$c$}
\psline{<-}(5,10)(10,10)
\psline{<-}(14,6)(10,10)
\psline{<-}(14,14)(10,10)
\end{pspicture}	
\end{array}.
\label{arrowsuscenti}
\end{equation}

Trace of the identity 
\begin{equation}
{}^a\delta^{\alpha}_{\;\;\alpha}=\begin{array}{c}
	\ifx\JPicScale\undefined\def\JPicScale{0.8}\fi
	\psset{unit=\JPicScale mm}
	\psset{linewidth=0.3,dotsep=1,hatchwidth=0.3,hatchsep=1.5,shadowsize=1,dimen=middle}
	\psset{dotsize=0.7 2.5,dotscale=1 1,fillcolor=black}
	\psset{arrowsize=1 2,arrowlength=1,arrowinset=0.25,tbarsize=0.7 5,bracketlength=0.15,rbracketlength=0.15}
	\begin{pspicture}(0,0)(17.5,25)
	\rput{0}(10,15){\psellipse[](0,0)(7.5,-7.5)}
	\rput(10,25){$a$}
	\end{pspicture}
	\end{array}=2a+1
\end{equation}

\item{\em First orthogonality relation for 3j-symbols.} 
\begin{equation}
	\sum_{\alpha,\beta}\left(\begin{array}{ccc}     a & b & c \\          \alpha & \beta & \gamma \end{array}\right)
	\left(\begin{array}{ccc}     a & b & c' \\          \alpha & \beta  & \gamma' \end{array}\right)=\frac{1}{2c+1}\;\delta_{cc'}\; \delta^{\gamma}_{\;\gamma'}
\end{equation}
\begin{equation}
\begin{array}{c}
		\ifx\JPicScale\undefined\def\JPicScale{0.8}\fi
	\psset{unit=\JPicScale mm}
	\psset{linewidth=0.2,dotsep=1,hatchwidth=0.3,hatchsep=1.5,shadowsize=1,dimen=middle}
	\psset{dotsize=0.7 2.5,dotscale=1 1,fillcolor=black}
	\psset{arrowsize=1 2,arrowlength=1,arrowinset=0.25,tbarsize=0.7 5,bracketlength=0.15,rbracketlength=0.15}
	\begin{pspicture}(0,0)(50,16)
	\rput{0}(24.5,7){\psellipse[](0,0)(10.5,-7)}
	\rput(25,2){$a$}
	\rput(25,16){$b	$}
	\rput(37,4){-}
	\rput(12,4){+}
	\psline(14,7)(0,7)
	\psline(35,7)(49,7)
	\rput(3,9){$c$}
	\rput(46,10){$c'$}
		\end{pspicture}
\end{array}
=\frac{1}{2c+1}
\begin{array}{c}	\ifx\JPicScale\undefined\def\JPicScale{1}\fi
\psset{unit=\JPicScale mm}
\psset{linewidth=0.2,dotsep=1,hatchwidth=0.3,hatchsep=1.5,shadowsize=1,dimen=middle}
\psset{dotsize=0.7 2.5,dotscale=1 1,fillcolor=black}
\psset{arrowsize=1 2,arrowlength=1,arrowinset=0.25,tbarsize=0.7 5,bracketlength=0.15,rbracketlength=0.15}
\begin{pspicture}(0,10)(20,12.5)
\psline(20,10)(0,10)
\rput(4,12.5){c$\gamma$}
\rput(20,12.5){c$'$$\gamma'$}
\rput(15,10){}
\end{pspicture}
	\end{array}
	\label{second ortogonality}
\end{equation}
This implies
\begin{equation}
\begin{array}{c}
		\ifx\JPicScale\undefined\def\JPicScale{0.8}\fi
	\psset{unit=\JPicScale mm}
	\psset{linewidth=0.3,dotsep=1,hatchwidth=0.3,hatchsep=1.5,shadowsize=1,dimen=middle}
	\psset{dotsize=0.7 2.5,dotscale=1 1,fillcolor=black}
	\psset{arrowsize=1 2,arrowlength=1,arrowinset=0.25,tbarsize=0.7 5,bracketlength=0.15,rbracketlength=0.15}
	\begin{pspicture}(0,0)(35,16)
	\rput{0}(16.5,7){\psellipse[](0,0)(10.5,-7)}
	\psline(27,7)(6,7)
	\rput(3,7){-}
	\rput(30,7){+}
	\rput(16,2){$a$}
	\rput(16,9){$c$}
	\rput(16,16){$b$}
	\rput(35,7){}
	\rput(40,7){}
	\rput[l](35,7){}
	\end{pspicture}
\end{array}
=1 \label{theta=1}
\end{equation}

\item{\em Second orthogonality relation.}
\begin{equation}
		\sum_{c\gamma}(2c+1)\left(\begin{array}{ccc}     a & b & c \\          \alpha & \beta & \gamma \end{array}\right)
	\left(\begin{array}{ccc}     a & b & c \\          \alpha' & \beta'  & \gamma \end{array}\right)=\;\delta^{\alpha}_{\;\alpha'}\; \delta^{\beta}_{\;\beta'}
\end{equation}
Graphically
\begin{equation}
\sum_{c}(2c+1)
\begin{array}{c}
	\ifx\JPicScale\undefined\def\JPicScale{0.9}\fi
\psset{unit=\JPicScale mm}
\psset{linewidth=0.2,dotsep=1,hatchwidth=0.3,hatchsep=1.5,shadowsize=1,dimen=middle}
\psset{dotsize=0.7 2.5,dotscale=1 1,fillcolor=black}
\psset{arrowsize=1 2,arrowlength=1,arrowinset=0.25,tbarsize=0.7 5,bracketlength=0.15,rbracketlength=0.15}
\begin{pspicture}(0,0)(27,18)
\psline(3,3)(10,10)
\psline(10,10)(3,17)
\psline(10,10)(20,10)
\psline(20,10)(27,17)
\psline(20,10)(27,3)
\rput(11,8){+}
\rput(19,8){-}
\rput(15,12){c}
\rput(5,18){$a\alpha$}
\rput(5,2){$b\beta$}
\rput(25,18){$a\alpha'$}
\rput(25,2){$b\beta'$}
\end{pspicture}\end{array}
=
\begin{array}{c}
	\ifx\JPicScale\undefined\def\JPicScale{0.9}\fi
\psset{unit=\JPicScale mm}
\psset{linewidth=0.2,dotsep=1,hatchwidth=0.3,hatchsep=1.5,shadowsize=1,dimen=middle}
\psset{dotsize=0.7 2.5,dotscale=1 1,fillcolor=black}
\psset{arrowsize=1 2,arrowlength=1,arrowinset=0.25,tbarsize=0.7 5,bracketlength=0.15,rbracketlength=0.15}
\begin{pspicture}(0,0)(27,18)
\psline(3,5)(27,5)
\psline(3,15)(27,15)
\rput(5,18){$a\alpha$}
\rput(5,2){$b\beta$}
\rput(25,18){$a\alpha'$}
\rput(25,2){$b\beta'$}
\end{pspicture}\end{array}
\end{equation}

\item{\em 6j symbol.} 
\begin{align}\hspace{-2em}
	\left\{\begin{array}{ccc}     a  & b & e \\      d & c & f                 \end{array}\right\}\!=\!
	\sum_{\alpha\epsilon\gamma}(-1)^{a+e+c-\alpha-\epsilon-\gamma} 
 &\left(\begin{array}{ccc}    a  & f & c \\          \alpha & \phi & -\gamma \end{array}\right)\!\!
\left(\begin{array}{ccc}     c  & d & e \\          \gamma & \delta & -\epsilon \end{array}\right)\!\! 
\left(\begin{array}{ccc}     e  & b & a \\          \epsilon & \beta & -\alpha \end{array}\right) \!\!
\left(\begin{array}{ccc}     b  & d & f \\          \beta & \delta & \phi \end{array}\right) 
\nonumber
\\\nonumber
\\
&\begin{array}{c}
	\ifx\JPicScale\undefined\def\JPicScale{1.4}\fi
\psset{unit=\JPicScale mm}
\psset{linewidth=0.1,dotsep=1,hatchwidth=0.3,hatchsep=1.5,shadowsize=1,dimen=middle}
\psset{dotsize=0.7 2.5,dotscale=1 1,fillcolor=black}
\psset{arrowsize=1 2,arrowlength=1,arrowinset=0.25,tbarsize=0.7 5,bracketlength=0.15,rbracketlength=0.15}
\begin{pspicture}(0,0)(26,25)
\psline(4,23)(14,5)
\psline(14,5)(24,23)
\psline(4,23)(24,23)
\psline(14,16)(14,5)
\psline(14,5)(14,17)
\psline(4,23)(14,17)
\psline(14,17)(24,23)
\rput(14,3){+}
\rput(14,3){}
\rput(14,2){}
\rput(14,3){+}
\rput(26,25){+}
\rput(2,25){+}
\rput(14,19){+}
\rput(21,13){$a$}
\rput(7,13){$c$}
\rput(14,25){$e$}
\rput(17,21){$b$}
\rput(11,21){$d$}
\rput(15,13){$f$}
\psline{<-}(14,23)(24,23)
\psline{<-}(9,14)(4,23)
\psline{<-}(19.37,14.63)(14,5)
\end{pspicture}\end{array}\label{6j}
\end{align}

\item{\em The 4j coefficient, or 4-valent node.}
\begin{eqnarray}
&\left(\begin{array}{cccc}            a  & c & b & d\\               \alpha & \gamma & \beta & \delta       \end{array}\right)=\sum_{\epsilon}(-1)^{e-\epsilon} \left(\begin{array}{ccc}     e  & a & c \\          \epsilon & \alpha & \gamma \end{array}\right)
\left(\begin{array}{ccc}     e  & b & d \\          -\epsilon & \beta & \delta \end{array}\right) 
\\ \nonumber
\\ \nonumber
&\begin{array}{c}
		\ifx\JPicScale\undefined\def\JPicScale{1}\fi
\psset{unit=\JPicScale mm}
\psset{linewidth=0.3,dotsep=1,hatchwidth=0.3,hatchsep=1.5,shadowsize=1,dimen=middle}
\psset{dotsize=0.7 2.5,dotscale=1 1,fillcolor=black}
\psset{arrowsize=1 2,arrowlength=1,arrowinset=0.25,tbarsize=0.7 5,bracketlength=0.15,rbracketlength=0.15}
\begin{pspicture}(0,0)(28,19)
\psline(10,10)(20,10)
\psline(20,10)(27,3)
\psline(20,10)(27,17)
\rput(23,10){+}
\rput(28,1){}
\psline{<-}(16,10)(10,10)
\psline(10,10)(3,3)
\psline(10,10)(3,17)
\rput(2,19){$a$}
\rput(2,1){$c$}
\rput(28,19){$d$}
\rput(28,1){$b$}
\rput(7,10){+}
\rput(15,13){$e$}
\end{pspicture}
\end{array}
\end{eqnarray}

\item{\em Recoupling theorem.} 

\begin{eqnarray}
&\left(\begin{array}{cccc}            a  & c & b & d\\               \alpha & \gamma & \beta & \delta       \end{array}\right)=	\sum_f \dim{f} \,(-1)^{b+c+e+f}  
\left\{\begin{array}{ccc}                      a  & b & f \\                      d & c & e                  \end{array}\right\} 
\left(\begin{array}{cccc}            a  & b & c & d\\               \alpha & \beta & \gamma & \delta       \end{array}\right)
\\ \nonumber
\\ \nonumber
&\begin{array}{c}
		\ifx\JPicScale\undefined\def\JPicScale{1}\fi
\psset{unit=\JPicScale mm}
\psset{linewidth=0.3,dotsep=1,hatchwidth=0.3,hatchsep=1.5,shadowsize=1,dimen=middle}
\psset{dotsize=0.7 2.5,dotscale=1 1,fillcolor=black}
\psset{arrowsize=1 2,arrowlength=1,arrowinset=0.25,tbarsize=0.7 5,bracketlength=0.15,rbracketlength=0.15}
\begin{pspicture}(0,0)(28,19)
\psline(10,10)(20,10)
\psline(20,10)(27,3)
\psline(20,10)(27,17)
\rput(23,10){+}
\rput(28,1){}
\psline{<-}(16,10)(10,10)
\psline(10,10)(3,3)
\psline(10,10)(3,17)
\rput(2,19){$a$}
\rput(2,1){$c$}
\rput(28,19){$d$}
\rput(28,1){$b$}
\rput(7,10){+}
\rput(15,13){$e$}
\end{pspicture}
\end{array}
=
\sum_f \dim{f} \quad
\begin{array}{c}
\ifx\JPicScale\undefined\def\JPicScale{1}\fi
\psset{unit=\JPicScale mm}
\psset{linewidth=0.3,dotsep=1,hatchwidth=0.3,hatchsep=1.5,shadowsize=1,dimen=middle}
\psset{dotsize=0.7 2.5,dotscale=1 1,fillcolor=black}
\psset{arrowsize=1 2,arrowlength=1,arrowinset=0.25,tbarsize=0.7 5,bracketlength=0.15,rbracketlength=0.15}
\begin{pspicture}(0,0)(19,19)
\psline(3,3)(3,17)
\psline(3,17)(17,17)
\psline(3,3)(17,3)
\psline(17,17)(17,3)
\psline(3,17)(17,3)
\psline(3,3)(17,17)
\psline{<-}(3,11)(3,3)
\psline{<-}(17,11)(17,3)
\rput(2,1){+}
\rput(2,19){+}
\rput(18,19){-}
\rput(18,1){-}
\rput(1,10){$e$}
\rput(19,10){$f$}
\rput(10,1){$a$}
\rput(10,19){$d$}
\rput(9,14){$b$}
\rput(8,6){$c$}
\end{pspicture}
\end{array}\quad
\begin{array}{c}
\ifx\JPicScale\undefined\def\JPicScale{1}\fi
\psset{unit=\JPicScale mm}
\psset{linewidth=0.3,dotsep=1,hatchwidth=0.3,hatchsep=1.5,shadowsize=1,dimen=middle}
\psset{dotsize=0.7 2.5,dotscale=1 1,fillcolor=black}
\psset{arrowsize=1 2,arrowlength=1,arrowinset=0.25,tbarsize=0.7 5,bracketlength=0.15,rbracketlength=0.15}
\begin{pspicture}(0,0)(18,26)
\psline(3,3)(10,10)
\psline(10,10)(17,3)
\psline(10,10)(10,17)
\psline(3,24)(10,17)
\psline(17,24)(10,17)
\psline{<-}(10,15)(10,10)
\rput(13,14){$f$}
\rput(10,7){+}
\rput(10,20){+}
\rput(2,1){$a$}
\rput(18,1){$b$}
\rput(2,26){$d$}
\rput(18,26){$c$}
\end{pspicture}
\end{array}
\label{recoupling theorem}
\end{eqnarray}

\item{\em Inverse transformation.} 

\begin{equation}
\begin{array}{c}
\ifx\JPicScale\undefined\def\JPicScale{1}\fi
\psset{unit=\JPicScale mm}
\psset{linewidth=0.3,dotsep=1,hatchwidth=0.3,hatchsep=1.5,shadowsize=1,dimen=middle}
\psset{dotsize=0.7 2.5,dotscale=1 1,fillcolor=black}
\psset{arrowsize=1 2,arrowlength=1,arrowinset=0.25,tbarsize=0.7 5,bracketlength=0.15,rbracketlength=0.15}
\begin{pspicture}(0,0)(18,26)
\psline(3,3)(10,10)
\psline(10,10)(17,3)
\psline(10,10)(10,17)
\psline(3,24)(10,17)
\psline(17,24)(10,17)
\psline{<-}(10,15)(10,10)
\rput(13,14){$f$}
\rput(10,7){+}
\rput(10,20){+}
\rput(2,1){$a$}
\rput(18,1){$b$}
\rput(2,26){$d$}
\rput(18,26){$c$}
\end{pspicture}
\end{array}
=\sum_m
\dim{m} \,(-1)^{b+c+f+m}  
\left\{\begin{array}{ccc}                      a  & c & m \\                      d & b & f                  \end{array}\right\} 
\begin{array}{c}
		\ifx\JPicScale\undefined\def\JPicScale{1}\fi
\psset{unit=\JPicScale mm}
\psset{linewidth=0.3,dotsep=1,hatchwidth=0.3,hatchsep=1.5,shadowsize=1,dimen=middle}
\psset{dotsize=0.7 2.5,dotscale=1 1,fillcolor=black}
\psset{arrowsize=1 2,arrowlength=1,arrowinset=0.25,tbarsize=0.7 5,bracketlength=0.15,rbracketlength=0.15}
\begin{pspicture}(0,0)(28,19)
\psline(10,10)(20,10)
\psline(20,10)(27,3)
\psline(20,10)(27,17)
\rput(23,10){+}
\rput(28,1){}
\psline{<-}(16,10)(10,10)
\psline(10,10)(3,3)
\psline(10,10)(3,17)
\rput(2,19){$a$}
\rput(2,1){$c$}
\rput(28,19){$d$}
\rput(28,1){$b$}
\rput(7,10){+}
\rput(15,13){$m$}
\end{pspicture}
\end{array}
\label{inverse}
\end{equation}

\item{\em Orthogonality relation for the 6j symbols.}
\begin{equation}
\sum_{f}
\dim{m} 
\dim{f}
\left\{\begin{array}{ccc}                      a  & b & f \\                      d & c & e                  \end{array}\right\}
\left\{\begin{array}{ccc}                      a  & c & m \\                      d & b & f                  \end{array}\right\} 
=\delta_{em}
	\label{orthogonality}
\end{equation}

\item{\em Biedenharn-Elliot identity.} 

\begin{equation}\begin{split}
	\sum_x \dim{x} \,(-1)^{a+b+c+d+e+f+g+h+i+x}  
\left\{\begin{array}{ccc}                      e  & f & x \\                      b & a & i                  \end{array}\right\}  \left\{\begin{array}{ccc}                      a  & b & x \\                      c & d & h                  \end{array}\right\}
\left\{\begin{array}{ccc}                      d  & c & x \\                      f & e & g                  \end{array}\right\}&\\
=\left\{\begin{array}{ccc}                      g  & h & i \\                      a & e & d                  \end{array}\right\}
\left\{\begin{array}{ccc}                      g  & h & i \\                      b & f & c                  \end{array}\right\}
\end{split}
\label{BEidentity}
\end{equation}

\item{\em The ``basic rule".}

\begin{equation}
\begin{split}
	&\sum_{\delta\epsilon\phi}(-1)^{d+e+f-\delta-\epsilon-\phi}
	 \left(\begin{array}{ccc}     d  & e & c \\          -\delta & \epsilon & \gamma \end{array}\right)
	\left(\begin{array}{ccc}     e  & f & a \\          -\epsilon & \phi & \alpha \end{array}\right) 
	\left(\begin{array}{ccc}     f  & d & b \\          -\phi & \delta & \beta \end{array}\right) \\
	\\ &\hspace{5em}=
\left\{\begin{array}{ccc}                      a  & b & c \\                      d & e & f                  \end{array}\right\} 	
	\left(\begin{array}{ccc}     a  & b & c \\          \alpha & \beta & \gamma \end{array}\right) 
\end{split}
\end{equation}

\begin{equation}\label{basic rule}
\begin{array}{c}
	\ifx\JPicScale\undefined\def\JPicScale{1}\fi
\psset{unit=\JPicScale mm}
\psset{linewidth=0.3,dotsep=1,hatchwidth=0.3,hatchsep=1.5,shadowsize=1,dimen=middle}
\psset{dotsize=0.7 2.5,dotscale=1 1,fillcolor=black}
\psset{arrowsize=1 2,arrowlength=1,arrowinset=0.25,tbarsize=0.7 5,bracketlength=0.15,rbracketlength=0.15}
\begin{pspicture}(0,0)(32,29)
\psline(3,3)(10,10)
\psline(10,22)(3,29)
\psline(10,22)(16,16)
\psline(10,10)(16,16)
\psline(9,9)(9,23)
\psline(16,16)(31,16)
\rput(10,8){-}
\rput(10,24){}
\rput(10,24){-}
\rput(17,14){-}
\rput(2,1){$a$}
\rput(2,31){$c$}
\rput(32,14){$b$}
\rput(6,16){$e$}
\rput(15,11){$f$}
\rput(15,21){$d$}
\rput(9,16){}
\psline{<-}(12,20)(16,16)
\psline{<-}(13,13)(9,9)
\psline{<-}(9,15)(9,23)
\end{pspicture}
\end{array}=\quad
\begin{array}{c}
	\ifx\JPicScale\undefined\def\JPicScale{1}\fi
\psset{unit=\JPicScale mm}
\psset{linewidth=0.3,dotsep=1,hatchwidth=0.3,hatchsep=1.5,shadowsize=1,dimen=middle}
\psset{dotsize=0.7 2.5,dotscale=1 1,fillcolor=black}
\psset{arrowsize=1 2,arrowlength=1,arrowinset=0.25,tbarsize=0.7 5,bracketlength=0.15,rbracketlength=0.15}
\begin{pspicture}(0,0)(29,25)
\psline(3,3)(3,23)
\psline(11,13)(3,23)
\psline(11,13)(3,3)
\psline(11,13)(27,13)
\psline(3,23)(27,13)
\psline{<-}(15,18)(27,13)
\psline(27,13)(3,3)
\psline{<-}(15,8)(3,3)
\psline{<-}(3,12)(3,23)
\rput(8,13){+}
\rput(29,13){+}
\rput(13,21){$d$}
\rput(13,5){$f$}
\rput(1,11){$e$}
\rput(14,15){$b$}
\rput(6,16){$c$}
\rput(6,10){$a$}
\rput(2,25){+}
\rput(2,1){+}
\end{pspicture}
\end{array}\begin{array}{c}
	\ifx\JPicScale\undefined\def\JPicScale{1}\fi
\psset{unit=\JPicScale mm}
\psset{linewidth=0.3,dotsep=1,hatchwidth=0.3,hatchsep=1.5,shadowsize=1,dimen=middle}
\psset{dotsize=0.7 2.5,dotscale=1 1,fillcolor=black}
\psset{arrowsize=1 2,arrowlength=1,arrowinset=0.25,tbarsize=0.7 5,bracketlength=0.15,rbracketlength=0.15}
\begin{pspicture}(0,0)(23,19)
\psline(3,3)(10,10)
\psline(3,17)(10,10)
\psline(10,10)(21,10)
\rput(7,10){+}
\rput(2,1){$a$}
\rput(2,19){$c$}
\rput(23,10){$b$}
\end{pspicture}
\end{array}
\end{equation}

\end{itemize}

\section{Analytic expressions for 6j symbols}\label{6j symbols}
From \cite{BS}.
	\begin{eqnarray}
	\left\{\begin{array}{ccc}     a  & b & e \\      d & c & f                 \end{array}\right\}=(-1)^{a+b+c+d}\Delta(a,b,e)\Delta(a,c,f)\Delta(b,d,f)\Delta(c,d,e)\sum_z (-1)^z \frac{f(z)}{z!}
	\label{6jracahformula}
	\end{eqnarray}
where 
\begin{equation}
	\Delta(a,b,c)=\sqrt{\frac{(a+b-c)!(a+c-b)!(b+c-a)!}{(a+b+c+1)!}}
	\label{Delta}
\end{equation}
	and
\begin{equation}
	f(z)=
	\frac{(a + b + c + d + 1 - z)!}{(e\!+\!f\!-\!a\!-\!d\!+\!z)!(e\!+\!f\!-\!b\!-\!c\!+\!z)!(a\!+\!b\!-\!e\!-\!z)!(c\!+\!d\!-\!e\!-\!z)!(a\!+\!c\!-\!f\!-\!z)!(b\!+\!d\!-\!e\!-\!f)!}
\end{equation}
The sum is extended to all the positive integer z, such that no factorial has negative argument.

The definition \eqref{6jracahformula} implies some restrictions on the arguments of the 6j:

In particular the $\Delta(a,b,c)$ restricts the arguments to satisfy the triangle inequalities
\begin{equation}
	(a + b - c) \geq 0 \quad (a - b + c) \geq 0 \quad (-a + b + c) \geq 0
\end{equation}
and $a+b+c$ has to be an integer number.

The expression \eqref{6jracahformula} reduces to the following simple expressions used in the calculation
   \begin{equation}
	\left\{\begin{array}{ccc}     a  & a & 1 \\      b & b & e                 \end{array}\right\}=
	\frac{(-1)^{a+b+e+1}}{2}\frac{C^2(a)+C^2(b)-C^2(e)}{\sqrt{C^2(a)\dim(a)C^2(b)\dim(b)}} 
	\label{6j1}
	 \end{equation}
 \begin{equation}
	\left\{\begin{array}{ccc}     e  & e-1 & 1 \\      a & a & b                 \end{array}\right\}=
	 \frac{(-1)^{a+b+e}}{2}\sqrt{\frac{(a+b+e+1)(a-b+e)(-a+b+e)(a+b-e+1)}{C^2(a)\; \di (a)\; e\; \di(e)\; \di(e-1)}}
 \label{6j-1}
 \end{equation}
 \begin{equation}
	\left\{\begin{array}{ccc}     e  & e+1 & 1 \\      a & a & b                 \end{array}\right\}=
	 \frac{(-1)^{a+b+e+1}}{2}\sqrt{\frac{(a+b+e+2)(a-b+e+1)(-a+b+e+1)(a+b-e)}{C^2(a)\; \di (a)\; (e+1)\; \di(e)\; \di(e+1)}}
 \label{6j+1}
 \end{equation}

The 6j symbol is invariant for interchange of any two columns, and also for interchange of the upper and lower arguments in each of any two columns:		
\begin{equation}
	\left\{\begin{array}{ccc}     a  & b & e \\      d & c & f                 \end{array}\right\}
	=	\left\{\begin{array}{ccc}     a  & e & b \\      d & f & c                 \end{array}\right\}
	=	\left\{\begin{array}{ccc}     e  & a & b \\      f & d & c                 \end{array}\right\}
		=	\left\{\begin{array}{ccc}     a  & c & f \\      d & b & e                 \end{array}\right\}
			=	\left\{\begin{array}{ccc}     d  & c & e \\      a & b & f                 \end{array}\right\}, etc
\end{equation}

We have also used the trivial facts  
	\[(-1)^a=(-1)^{-a} \quad\forall a\in \mathbb{Z},\qquad (-1)^{2a}=1 \quad\forall a\in \mathbb{Z},\qquad (-1)^{3s}=(-1)^{-s} \quad\forall s\in \frac{\mathbb{Z}}{2}
\]
		in the calculations involving the 6j symbols

\section{Grasping operators}\label{operators}
		
		The operator $E^{a}(\vec{x})n_a^{\scr{(ni)}}$ is the ``grasping operator" that acts on the spinnetwork's link dual to the triangle with normal $n_a^{\scr{(ni)}}$. Let say that this link is in the $j$ representation ; $E^{a}(\vec{x})n_a^{\scr{ni}}$ will acts inserting an $SU(2)$ generator in the same representation \cite{book} or equivalently, by inserting an intertwiner between the $(j)$ rep and the rep 1, namely a 3j symbol not normalized:
\begin{equation}
E^{\scr{(ni)}}(\vec{x})^{i\alpha}_{\;\;\;\beta}\; = i\; ^{(j)}J^{i\alpha}_{\;\;\;\beta}=i N^j v^{i\alpha}_{\;\;\;\beta}
\label{Etensor}
\end{equation}
where $^{(j)}J^{i\alpha}_{\;\;\;\beta}$ is the $SU(2)$ generator in the $j$ representation $(i=-1,0,1), (\alpha,\beta=-j,..,j)$, $N^j$ is a normalization factor and $v^{i\alpha\beta}$ is the normalized 3j symbol. The action of the operator $E^{\scr{(ni)}}$ is then determined by the representation of the links on which it acts; in the following we will call $E^{\scr{(j)}}$ an operator acting on the link with rep $j$.

Graphically, with our conventions
\begin{equation}
	E^{\scriptscriptstyle(j)}_n=i\,N^{(j)}
\begin{array}{c}
	\ifx\JPicScale\undefined\def\JPicScale{1}\fi
	\psset{unit=\JPicScale mm}
	\psset{linewidth=0.3,dotsep=1,hatchwidth=0.3,hatchsep=1.5,shadowsize=1,dimen=middle}
	\psset{dotsize=0.7 2.5,dotscale=1 1,fillcolor=black}
	\psset{arrowsize=1 2,arrowlength=1,arrowinset=0.25,tbarsize=0.7 5,bracketlength=0.15,rbracketlength=0.15}
	\begin{pspicture}(0,0)(19,9)
	\psline{->}(2,2)(18,2)
	\psline(10,2)(10,7)
	\rput(10,0){-}
	\rput(1,4){$j$}
	\rput(19,4){$j$}
	\rput(11,9){1}
	\end{pspicture}
\end{array}
\label{Egraph}
\end{equation}
(Note the arrow that reflect the lowered magnetic index).

To fix the normalization factor $N^{j}$ is enough to square the expression \eqref{Etensor}, use \eqref{second ortogonality}
\begin{equation}
	^{j}J^{2\;\alpha}_{\;\;\;\;\beta}= C^2(j)\,^{j}I^{\alpha}_{\;\;\;\beta}=\frac{(N^j)^2}{\dim j}\, ^{j}I^{\alpha}_{\;\;\;\beta}
\end{equation}
and take the trace of the previous equation (where $^{j}I^{\alpha}_{\;\;\;\beta}$ is the identity in the rep $j$ ), obtaining
\begin{equation}
	N^j=\sqrt{j(j+1) \dim j}
\end{equation}

Our triangulated manifold consist of a 4-simplex made of 5 tetrahedron $t_n$, bounded by triangles $t_{nm}$. In the dual picture the 4 symplex is represented by the pentagonal net where the tetrahedra are the 4-valent nodes $n$, labeled by the intertwiners $i_n$ in a given pairing, and the triangles are the links $nm$ labeled by the spin numbers $j^{\scr{nm}}$.   

In our calculation we act with the operator $E^a(\vec{x})n^{\scriptscriptstyle(nl)}_a$ on the tetrahedron $t_n$ in the direction $n^{\scriptscriptstyle(nl)}_a$ orthogonal to the triangle $t_{nl}$; in the dual picture we are then acting on the 4-valent nodes $n$ and precisely on the link $j^{\scr{ni}}$.
To enlighten the notation, fixed a node $n$, we will call the four possible colorings corresponding to the 4 directions $ni$ with $a, b , c, d$ where the letter indicate the representation of the links.
Graphically the action of a single grasping operator operating on the link $a$ for example is 
\begin{equation}
	E^{\scriptscriptstyle(a)}_n \begin{array}{c}
		\ifx\JPicScale\undefined\def\JPicScale{1}\fi
\psset{unit=\JPicScale mm}
\psset{linewidth=0.3,dotsep=1,hatchwidth=0.3,hatchsep=1.5,shadowsize=1,dimen=middle}
\psset{dotsize=0.7 2.5,dotscale=1 1,fillcolor=black}
\psset{arrowsize=1 2,arrowlength=1,arrowinset=0.25,tbarsize=0.7 5,bracketlength=0.15,rbracketlength=0.15}
\begin{pspicture}(0,0)(28,19)
\psline(10,10)(20,10)
\psline(20,10)(27,3)
\psline(20,10)(27,17)
\rput(23,10){+}
\rput(28,1){}
\psline{<-}(16,10)(10,10)
\psline(10,10)(3,3)
\psline(10,10)(3,17)
\rput(2,19){$b$}
\rput(2,1){$a$}
\rput(28,19){$c$}
\rput(28,1){$d$}
\rput(7,10){$+$}
\rput(15,13){$e$}
\end{pspicture}
\end{array}=\;
i N^{\scr{a}}
\begin{array}{c}
\ifx\JPicScale\undefined\def\JPicScale{1}\fi
\psset{unit=\JPicScale mm}
\psset{linewidth=0.3,dotsep=1,hatchwidth=0.3,hatchsep=1.5,shadowsize=1,dimen=middle}
\psset{dotsize=0.7 2.5,dotscale=1 1,fillcolor=black}
\psset{arrowsize=1 2,arrowlength=1,arrowinset=0.25,tbarsize=0.7 5,bracketlength=0.15,rbracketlength=0.15}
\begin{pspicture}(0,0)(29,18)
\psline(11,9)(21,9)
\psline(21,9)(28,2)
\psline(21,9)(28,16)
\rput(24,9){+}
\rput(29,0){}
\psline{<-}(17,9)(11,9)
\psline(11,9)(4,2)
\psline(11,9)(4,16)
\rput(3,18){$b$}
\rput(3,0){$a$}
\rput(29,18){$c$}
\rput(29,0){$d$}
\rput(8,9){+}
\rput(16,11){$e$}
\psline(6,4)(3,7)
\psline{<-}(8,6)(6,4)
\rput(1,8){1}
\rput(10,6){$a$}
\rput(7,3){-}
\end{pspicture}\end{array}
\end{equation}
The action of our operators $E^{\scr(ni)}_n \cdot E^{\scr(nj)}_n\;$ on a node in a fixed pairing can then produce four different result depending on the two directions $n^{\scr{ni}}$,$n^{\scr{nj}}$
\begin{equation}
\begin{split}
	E^{\scr(a)}_n \cdot E^{\scr(a)}_n\;\begin{array}{c}
		\ifx\JPicScale\undefined\def\JPicScale{1}\fi
\psset{unit=\JPicScale mm}
\psset{linewidth=0.3,dotsep=1,hatchwidth=0.3,hatchsep=1.5,shadowsize=1,dimen=middle}
\psset{dotsize=0.7 2.5,dotscale=1 1,fillcolor=black}
\psset{arrowsize=1 2,arrowlength=1,arrowinset=0.25,tbarsize=0.7 5,bracketlength=0.15,rbracketlength=0.15}
\begin{pspicture}(3,0)(26,19)
\psline(10,10)(20,10)
\psline(20,10)(27,3)
\psline(20,10)(27,17)
\rput(23,10){+}
\rput(28,1){}
\psline{<-}(16,10)(10,10)
\psline(10,10)(3,3)
\psline(10,10)(3,17)
\rput(2,19){$b$}
\rput(2,1){$a$}
\rput(28,19){$c$}
\rput(28,1){$d$}
\rput(7,10){+}
\rput(15,13){$e$}
\end{pspicture}
\end{array}
&=- (N^{\scr{a}})^2
\begin{array}{c}
\ifx\JPicScale\undefined\def\JPicScale{1}\fi
\psset{unit=\JPicScale mm}
\psset{linewidth=0.3,dotsep=1,hatchwidth=0.3,hatchsep=1.5,shadowsize=1,dimen=middle}
\psset{dotsize=0.7 2.5,dotscale=1 1,fillcolor=black}
\psset{arrowsize=1 2,arrowlength=1,arrowinset=0.25,tbarsize=0.7 5,bracketlength=0.15,rbracketlength=0.15}
\begin{pspicture}(7,0)(39,26)
\psline(23,17)(33,17)
\psline(33,17)(40,10)
\psline(33,17)(40,24)
\rput(36,17){+}
\psline{<-}(29,17)(23,17)
\psline(23,17)(16,24)
\rput(15,26){$b$}
\rput(41,26){$c$}
\rput(41,8){$d$}
\rput(20,17){+}
\rput(28,19){$e$}
\rput(28,19){}
\psline(23,17)(12,6)
\psline(23,17)(6,0)
\rput{0}(14,8){\psarc[](0,0){5.66}{45}{225}}
\psline{<-}(11,13)(10,12)
\psline{<-}(15,9)(14,8)
\psline{<-}(20,14)(19,13)
\rput(11,3){-}
\rput(19,11){-}
\rput(7,11){1}
\rput(8,-1){$a$}
\rput(14,5){$a$}
\rput(22,13){$a$}
\end{pspicture}\end{array}
= - (N^{\scr{a}})^2
\begin{array}{c}
\ifx\JPicScale\undefined\def\JPicScale{1}\fi
\psset{unit=\JPicScale mm}
\psset{linewidth=0.3,dotsep=1,hatchwidth=0.3,hatchsep=1.5,shadowsize=1,dimen=middle}
\psset{dotsize=0.7 2.5,dotscale=1 1,fillcolor=black}
\psset{arrowsize=1 2,arrowlength=1,arrowinset=0.25,tbarsize=0.7 5,bracketlength=0.15,rbracketlength=0.15}
\begin{pspicture}(7,0)(39,26)
\psline(23,17)(33,17)
\psline(33,17)(40,10)
\psline(33,17)(40,24)
\rput(36,17){+}
\psline{<-}(29,17)(23,17)
\psline(23,17)(16,24)
\rput(15,26){$b$}
\rput(41,26){$c$}
\rput(41,8){$d$}
\rput(20,17){+}
\rput(28,19){$e$}
\psline(23,17)(12,6)
\psline(23,17)(6,0)
\rput{0}(14,8){\psarc[](0,0){5.66}{45}{225}}
\psline{<-}(11,13)(10,12)
\psline{<-}(15,9)(14,8)
\psline{<-}(20,14)(19,13)
\rput(11,3){-}
\rput(19,11){-}
\rput(7,11){1}
\rput(8,-1){$a$}
\rput(14,5){$a$}
\rput(22,13){$a$}
\psline{<-}(16,10)(18,12)
\psline{<-}(22,16)(20,14)
\psline{<-}(13.18,13.71)(14.63,13.58)
\end{pspicture}
\end{array}=\\
&=(-1)^{2a+1}(N^{\scr{a}})^2
\begin{array}{c}
	\ifx\JPicScale\undefined\def\JPicScale{1}\fi
\psset{unit=\JPicScale mm}
\psset{linewidth=0.3,dotsep=1,hatchwidth=0.3,hatchsep=1.5,shadowsize=1,dimen=middle}
\psset{dotsize=0.7 2.5,dotscale=1 1,fillcolor=black}
\psset{arrowsize=1 2,arrowlength=1,arrowinset=0.25,tbarsize=0.7 5,bracketlength=0.15,rbracketlength=0.15}
\begin{pspicture}(7,0)(39,26)
\psline(23,17)(33,17)
\psline(33,17)(40,10)
\psline(33,17)(40,24)
\rput(36,17){+}
\rput(41,8){}
\psline{<-}(29,17)(23,17)
\psline(23,17)(16,24)
\rput(15,26){$b$}
\rput(41,26){$c$}
\rput(41,8){$d$}
\rput(20,17){+}
\rput(28,19){$e$}
\rput(28,19){}
\psline(23,17)(12,6)
\psline(23,17)(6,0)
\rput{0}(14,8){\psarc[](0,0){5.66}{45}{225}}
\rput(11,3){-}
\rput(19,11){-}
\rput(7,11){1}
\rput(8,-1){$a$}
\rput(14,5){$a$}
\rput(22,13){$a$}
\end{pspicture}
\end{array}
=C^2(a)
\begin{array}{c}
		\ifx\JPicScale\undefined\def\JPicScale{1}\fi
\psset{unit=\JPicScale mm}
\psset{linewidth=0.3,dotsep=1,hatchwidth=0.3,hatchsep=1.5,shadowsize=1,dimen=middle}
\psset{dotsize=0.7 2.5,dotscale=1 1,fillcolor=black}
\psset{arrowsize=1 2,arrowlength=1,arrowinset=0.25,tbarsize=0.7 5,bracketlength=0.15,rbracketlength=0.15}
\begin{pspicture}(0,0)(28,19)
\psline(10,10)(20,10)
\psline(20,10)(27,3)
\psline(20,10)(27,17)
\rput(23,10){+}
\rput(28,1){}
\psline{<-}(16,10)(10,10)
\psline(10,10)(3,3)
\psline(10,10)(3,17)
\rput(2,19){$b$}
\rput(2,1){$a$}
\rput(28,19){$c$}
\rput(28,1){$d$}
\rput(7,10){+}
\rput(15,13){$e$}
\end{pspicture}
\end{array}
\end {split}
\label{aa}
\end{equation}
where in the last equalities we have used the relation \eqref{arrowsuscenti},\eqref{opposte},\eqref{doppie} to eliminate the arrows and the \eqref{cambio orientazione} to solve the loop using \eqref{second ortogonality}.
The other possible case is
\begin{equation}
\begin{split}
	&E^{\scr(a)}_n \cdot E^{\scr(b)}_n\;\begin{array}{c}
		\ifx\JPicScale\undefined\def\JPicScale{1}\fi
\psset{unit=\JPicScale mm}
\psset{linewidth=0.3,dotsep=1,hatchwidth=0.3,hatchsep=1.5,shadowsize=1,dimen=middle}
\psset{dotsize=0.7 2.5,dotscale=1 1,fillcolor=black}
\psset{arrowsize=1 2,arrowlength=1,arrowinset=0.25,tbarsize=0.7 5,bracketlength=0.15,rbracketlength=0.15}
\begin{pspicture}(4,0)(25,19)
\psline(10,10)(20,10)
\psline(20,10)(27,3)
\psline(20,10)(27,17)
\rput(23,10){+}
\rput(28,1){}
\psline{<-}(16,10)(10,10)
\psline(10,10)(3,3)
\psline(10,10)(3,17)
\rput(2,19){$b$}
\rput(2,1){$a$}
\rput(28,19){$c$}
\rput(28,1){$d$}
\rput(7,10){+}
\rput(15,13){$e$}
\end{pspicture}
\end{array}
=
-N^{\scr(a)} N^{\scr(b)}
\begin{array}{c}
	\ifx\JPicScale\undefined\def\JPicScale{1}\fi
\psset{unit=\JPicScale mm}
\psset{linewidth=0.3,dotsep=1,hatchwidth=0.3,hatchsep=1.5,shadowsize=1,dimen=middle}
\psset{dotsize=0.7 2.5,dotscale=1 1,fillcolor=black}
\psset{arrowsize=1 2,arrowlength=1,arrowinset=0.25,tbarsize=0.7 5,bracketlength=0.15,rbracketlength=0.15}
\begin{pspicture}(2,0)(27,23)
\psline(12,11)(22,11)
\psline(22,11)(29,4)
\psline(22,11)(29,18)
\rput(25,11){+}
\psline{<-}(18,11)(12,11)
\rput(1,23){$b$}
\rput(1,-1){$a$}
\rput(30,20){$c$}
\rput(30,2){$d$}
\rput(13,13){+}
\rput(17,13){$e$}
\rput(3,11){1}
\rput(6,19){+}
\rput(6,3){-}
\psline(2,1)(12,11)
\psline(5,18)(5,4)
\psline{<-}(5,10)(5,18)
\psline(12,11)(2,21)
\psline{<-}(9,8)(5,4)
\psline{<-}(9,14)(5,18)
\end{pspicture}
\end{array}=(-1)^{a+b+e}
N^{\scr(a)} N^{\scr(b)}
\begin{array}{c}
	\ifx\JPicScale\undefined\def\JPicScale{1}\fi
\psset{unit=\JPicScale mm}
\psset{linewidth=0.3,dotsep=1,hatchwidth=0.3,hatchsep=1.5,shadowsize=1,dimen=middle}
\psset{dotsize=0.7 2.5,dotscale=1 1,fillcolor=black}
\psset{arrowsize=1 2,arrowlength=1,arrowinset=0.25,tbarsize=0.7 5,bracketlength=0.15,rbracketlength=0.15}
\begin{pspicture}(2,0)(27,23)
\psline(12,11)(22,11)
\psline(22,11)(29,4)
\psline(22,11)(29,18)
\rput(25,11){+}
\rput(30,2){}
\psline{<-}(18,11)(12,11)
\rput(1,23){$b$}
\rput(1,-1){$a$}
\rput(30,20){$c$}
\rput(30,2){$d$}
\rput(17,13){$e$}
\rput(3,11){1}
\rput(6,19){-}
\psline(2,1)(12,11)
\psline(5,18)(5,4)
\psline{<-}(5,10)(5,18)
\psline(12,11)(2,21)
\psline{<-}(9,8)(5,4)
\rput(6,3){-}
\psline{<-}(8,15)(12,11)
\rput(13,13){-}
\end{pspicture}
\end{array}=\\
&=(-1)^{a+b+e}
N^{\scr(a)} N^{\scr(b)}
\left\{\begin{array}{ccc}     b  & e & a \\      a & 1 & b                 \end{array}\right\}
\begin{array}{c}
		\ifx\JPicScale\undefined\def\JPicScale{1}\fi
\psset{unit=\JPicScale mm}
\psset{linewidth=0.3,dotsep=1,hatchwidth=0.3,hatchsep=1.5,shadowsize=1,dimen=middle}
\psset{dotsize=0.7 2.5,dotscale=1 1,fillcolor=black}
\psset{arrowsize=1 2,arrowlength=1,arrowinset=0.25,tbarsize=0.7 5,bracketlength=0.15,rbracketlength=0.15}
\begin{pspicture}(4,0)(26,19)
\psline(10,10)(20,10)
\psline(20,10)(27,3)
\psline(20,10)(27,17)
\rput(23,10){+}
\rput(28,1){}
\psline{<-}(16,10)(10,10)
\psline(10,10)(3,3)
\psline(10,10)(3,17)
\rput(2,19){$b$}
\rput(2,1){$a$}
\rput(28,19){$c$}
\rput(28,1){$d$}
\rput(7,10){+}
\rput(15,13){$e$}
\end{pspicture}
\end{array}=\frac{C^2(e)-C^2(a)-C^2(b)}{2}\begin{array}{c}
		\ifx\JPicScale\undefined\def\JPicScale{1}\fi
\psset{unit=\JPicScale mm}
\psset{linewidth=0.3,dotsep=1,hatchwidth=0.3,hatchsep=1.5,shadowsize=1,dimen=middle}
\psset{dotsize=0.7 2.5,dotscale=1 1,fillcolor=black}
\psset{arrowsize=1 2,arrowlength=1,arrowinset=0.25,tbarsize=0.7 5,bracketlength=0.15,rbracketlength=0.15}
\begin{pspicture}(-2,0)(25,19)
\psline(10,10)(20,10)
\psline(20,10)(27,3)
\psline(20,10)(27,17)
\rput(23,10){+}
\rput(28,1){}
\psline{<-}(16,10)(10,10)
\psline(10,10)(3,3)
\psline(10,10)(3,17)
\rput(2,19){$b$}
\rput(2,1){$a$}
\rput(28,19){$c$}
\rput(28,1){$d$}
\rput(7,10){+}
\rput(15,13){$e$}
\end{pspicture}
\end{array}
\end{split}
\label{ab}
\end{equation}
where we have changed the orientations of the 3-valent nodes to simplify the loop, using the basic identity \eqref{basic rule},and used the symmetry properties of 6$j$ symbols and its explicit expression \eqref{6j1}

The other possible action is
\begin{equation}
\begin{split}
	&E^{\scr(a)}_n \cdot E^{\scr(c)}_n\;\begin{array}{c}
		\ifx\JPicScale\undefined\def\JPicScale{1}\fi
\psset{unit=\JPicScale mm}
\psset{linewidth=0.3,dotsep=1,hatchwidth=0.3,hatchsep=1.5,shadowsize=1,dimen=middle}
\psset{dotsize=0.7 2.5,dotscale=1 1,fillcolor=black}
\psset{arrowsize=1 2,arrowlength=1,arrowinset=0.25,tbarsize=0.7 5,bracketlength=0.15,rbracketlength=0.15}
\begin{pspicture}(0,0)(28,19)
\psline(10,10)(20,10)
\psline(20,10)(27,3)
\psline(20,10)(27,17)
\rput(23,10){+}
\rput(28,1){}
\psline{<-}(16,10)(10,10)
\psline(10,10)(3,3)
\psline(10,10)(3,17)
\rput(2,19){$b$}
\rput(2,1){$a$}
\rput(28,19){$c$}
\rput(28,1){$d$}
\rput(7,10){+}
\rput(15,13){$e$}
\end{pspicture}
\end{array}
=
-N^{\scr(a)} N^{\scr(c)}
\begin{array}{c}
\ifx\JPicScale\undefined\def\JPicScale{1}\fi
\psset{unit=\JPicScale mm}
\psset{linewidth=0.3,dotsep=1,hatchwidth=0.3,hatchsep=1.5,shadowsize=1,dimen=middle}
\psset{dotsize=0.7 2.5,dotscale=1 1,fillcolor=black}
\psset{arrowsize=1 2,arrowlength=1,arrowinset=0.25,tbarsize=0.7 5,bracketlength=0.15,rbracketlength=0.15}
\begin{pspicture}(0,0)(33,25)
\psline(12,13)(22,13)
\psline(22,13)(29,6)
\psline(22,13)(32,23)
\rput(25,13){+}
\rput(30,4){}
\psline{<-}(18,13)(12,13)
\psline(12,13)(2,3)
\psline(12,13)(5,20)
\rput(9,13){+}
\rput(17,15){$e$}
\rput(17,15){}
\rput(1,1){$a$}
\rput(4,22){$b$}
\rput(33,25){$c$}
\rput(30,4){$d$}
\psline(30,21)(4,5)
\rput(5,4){+}
\rput(29,22){+}
\psline{<-}(9,10)(4,5)
\psline{<-}(25,16)(30,21)
\psline{<-}(14,11)(4,5)
\rput(12,7){$1$}
\end{pspicture}
\end{array}
=\\
&=N^{\scr(a)} N^{\scr(c)}
\sum_x 
(-1)^{a+d+e+x}\dim{x}
\left\{\begin{array}{ccc}     b  & d & x \\      c & a & e                 \end{array}\right\}
\begin{array}{c}
	\ifx\JPicScale\undefined\def\JPicScale{1}\fi
\psset{unit=\JPicScale mm}
\psset{linewidth=0.3,dotsep=1,hatchwidth=0.3,hatchsep=1.5,shadowsize=1,dimen=middle}
\psset{dotsize=0.7 2.5,dotscale=1 1,fillcolor=black}
\psset{arrowsize=1 2,arrowlength=1,arrowinset=0.25,tbarsize=0.7 5,bracketlength=0.15,rbracketlength=0.15}
\begin{pspicture}(0,0)(20,25)
\psline(3,0)(10,7)
\psline(10,7)(17,0)
\psline(10,7)(10,14)
\psline{<-}(10,12)(10,7)
\rput(13,11){$x$}
\rput(10,4){+}
\rput(8,13){+}
\rput(2,-2){$b$}
\rput(18,-2){$d$}
\rput(0,25){$c$}
\rput(20,25){$a$}
\psline(10,14)(19,23)
\psline(10,14)(1,23)
\psline(16,20)(4,20)
\psline{<-}(9,20)(16,20)
\psline{<-}(8,16)(4,20)
\psline{<-}(14,18)(10,14)
\rput(2,20){-}
\rput(18,20){-}
\rput(10,23){1}
\end{pspicture}
\end{array}=\\
&=N^{\scr(a)} N^{\scr(c)}
\sum_x 
(-1)^{a+c+x}(-1)^{a+d+e+x}\dim{x}
\left\{\begin{array}{ccc}     b  & d & x \\      c & a & e                 \end{array}\right\}
\left\{\begin{array}{ccc}     c  & a & x \\      a & c & 1                 \end{array}\right\}
\begin{array}{c}
\ifx\JPicScale\undefined\def\JPicScale{1}\fi
\psset{unit=\JPicScale mm}
\psset{linewidth=0.3,dotsep=1,hatchwidth=0.3,hatchsep=1.5,shadowsize=1,dimen=middle}
\psset{dotsize=0.7 2.5,dotscale=1 1,fillcolor=black}
\psset{arrowsize=1 2,arrowlength=1,arrowinset=0.25,tbarsize=0.7 5,bracketlength=0.15,rbracketlength=0.15}
\begin{pspicture}(0,0)(18,26)
\psline(3,3)(10,10)
\psline(10,10)(17,3)
\psline(10,10)(10,17)
\psline(3,24)(10,17)
\psline(17,24)(10,17)
\psline{<-}(10,15)(10,10)
\rput(13,14){$x$}
\rput(10,7){+}
\rput(10,20){+}
\rput(2,1){$b$}
\rput(18,1){$d$}
\rput(2,26){$c$}
\rput(18,26){$a$}
\end{pspicture}
\end{array}
=\\
&=N^{\scr(a)} N^{\scr(c)}
\sum_x 
(-1)^{a+c+x}(-1)^{a+d+e+x}\dim{x}\sum_m\dim{m}(-1)^{a+d+m+x} \cdot \\
&\quad
\cdot
\left\{\begin{array}{ccc}     b  & d & x \\      c & a & e                 \end{array}\right\}
\left\{\begin{array}{ccc}     c  & a & x \\      a & c & 1                 \end{array}\right\}
\left\{\begin{array}{ccc}     b  & a & m \\      c & d & x                 \end{array}\right\}
\begin{array}{c}
	\ifx\JPicScale\undefined\def\JPicScale{1}\fi
\psset{unit=\JPicScale mm}
\psset{linewidth=0.3,dotsep=1,hatchwidth=0.3,hatchsep=1.5,shadowsize=1,dimen=middle}
\psset{dotsize=0.7 2.5,dotscale=1 1,fillcolor=black}
\psset{arrowsize=1 2,arrowlength=1,arrowinset=0.25,tbarsize=0.7 5,bracketlength=0.15,rbracketlength=0.15}
\begin{pspicture}(0,0)(28,19)
\psline(10,10)(20,10)
\psline(20,10)(27,3)
\psline(20,10)(27,17)
\rput(23,10){+}
\rput(28,1){}
\psline{<-}(16,10)(10,10)
\psline(10,10)(3,3)
\psline(10,10)(3,17)
\rput(2,19){$b$}
\rput(2,1){$a$}
\rput(28,19){$c$}
\rput(28,1){$d$}
\rput(7,10){+}
\rput(15,13){$m$}
\end{pspicture}
\end{array}=\\
&=-N^{\scr(a)} N^{\scr(c)}(-1)^{3d+a+b-c}
\sum_m \dim{m}
\left\{\begin{array}{ccc}     e  & m & 1 \\      a & a & b                 \end{array}\right\}
\left\{\begin{array}{ccc}     e  & m & 1 \\      c & c & d                 \end{array}\right\}
\begin{array}{c}
	\ifx\JPicScale\undefined\def\JPicScale{1}\fi
\psset{unit=\JPicScale mm}
\psset{linewidth=0.3,dotsep=1,hatchwidth=0.3,hatchsep=1.5,shadowsize=1,dimen=middle}
\psset{dotsize=0.7 2.5,dotscale=1 1,fillcolor=black}
\psset{arrowsize=1 2,arrowlength=1,arrowinset=0.25,tbarsize=0.7 5,bracketlength=0.15,rbracketlength=0.15}
\begin{pspicture}(0,0)(28,19)
\psline(10,10)(20,10)
\psline(20,10)(27,3)
\psline(20,10)(27,17)
\rput(23,10){+}
\rput(28,1){}
\psline{<-}(16,10)(10,10)
\psline(10,10)(3,3)
\psline(10,10)(3,17)
\rput(2,19){$b$}
\rput(2,1){$a$}
\rput(28,19){$c$}
\rput(28,1){$d$}
\rput(7,10){+}
\rput(15,13){$m$}
\end{pspicture}
\end{array}
\end{split}
\label{ac}
\end{equation}
In the derivation of the result we have used, in order, the recoupling theorem \eqref{recoupling theorem} to change the pairing of the node, the basic rule \eqref{basic rule} to solve the loop,  the inverse transformation \eqref{inverse} to put the graph on the starting pairing and the Biedenharn-Elliot identity \eqref{BEidentity}, having adjusted the sign factors, using the triangles inequalities of the 3j symbols defining the 6j.
To analyze the result we have to look at the existence conditions of the \{6j\}(Appendix \ref{6j symbols}) concluding that $m$ can only take the values $e-1$, $e$, $e+1$, the final resul is then
\begin{equation}
	\begin{split}
	&	E^{\scr(a)}_n \cdot E^{\scr(c)}_n\;\begin{array}{c}
			\ifx\JPicScale\undefined\def\JPicScale{1}\fi
	\psset{unit=\JPicScale mm}
	\psset{linewidth=0.3,dotsep=1,hatchwidth=0.3,hatchsep=1.5,shadowsize=1,dimen=middle}
	\psset{dotsize=0.7 2.5,dotscale=1 1,fillcolor=black}
	\psset{arrowsize=1 2,arrowlength=1,arrowinset=0.25,tbarsize=0.7 5,bracketlength=0.15,rbracketlength=0.15}
	\begin{pspicture}(0,0)(28,19)
	\psline(10,10)(20,10)
	\psline(20,10)(27,3)
	\psline(20,10)(27,17)
	\rput(23,10){+}
	\rput(28,1){}
	\psline{<-}(16,10)(10,10)
	\psline(10,10)(3,3)
	\psline(10,10)(3,17)
	\rput(2,19){$b$}
	\rput(2,1){$a$}
	\rput(28,19){$c$}
	\rput(28,1){$d$}
	\rput(7,10){+}
	\rput(15,13){$e$}
	\end{pspicture}
	\end{array}
	=\\
	&=X^{\scr{ac}}_e
	\begin{array}{c}
			\ifx\JPicScale\undefined\def\JPicScale{1}\fi
	\psset{unit=\JPicScale mm}
	\psset{linewidth=0.3,dotsep=1,hatchwidth=0.3,hatchsep=1.5,shadowsize=1,dimen=middle}
	\psset{dotsize=0.7 2.5,dotscale=1 1,fillcolor=black}
	\psset{arrowsize=1 2,arrowlength=1,arrowinset=0.25,tbarsize=0.7 5,bracketlength=0.15,rbracketlength=0.15}
	\begin{pspicture}(0,0)(28,19)
	\psline(10,10)(20,10)
	\psline(20,10)(27,3)
	\psline(20,10)(27,17)
	\rput(23,10){+}
	\rput(28,1){}
	\psline{<-}(16,10)(10,10)
	\psline(10,10)(3,3)
	\psline(10,10)(3,17)
	\rput(2,19){$b$}
	\rput(2,1){$a$}
	\rput(28,19){$c$}
	\rput(28,1){$d$}
	\rput(7,10){+}
	\rput(15,13){$e$}
	\end{pspicture}
	\end{array}+
	Y^{\scr{ac}}_{e}
	\begin{array}{c}
			\ifx\JPicScale\undefined\def\JPicScale{1}\fi
	\psset{unit=\JPicScale mm}
	\psset{linewidth=0.3,dotsep=1,hatchwidth=0.3,hatchsep=1.5,shadowsize=1,dimen=middle}
	\psset{dotsize=0.7 2.5,dotscale=1 1,fillcolor=black}
	\psset{arrowsize=1 2,arrowlength=1,arrowinset=0.25,tbarsize=0.7 5,bracketlength=0.15,rbracketlength=0.15}
	\begin{pspicture}(0,0)(28,19)
	\psline(10,10)(20,10)
	\psline(20,10)(27,3)
	\psline(20,10)(27,17)
	\rput(23,10){+}
	\rput(28,1){}
	\psline{<-}(16,10)(10,10)
	\psline(10,10)(3,3)
	\psline(10,10)(3,17)
	\rput(2,19){$b$}
	\rput(2,1){$a$}
	\rput(28,19){$c$}
	\rput(28,1){$d$}
	\rput(7,10){+}
	\rput(15,13){$e-1$}
	\end{pspicture}
	\end{array}
	+
	Z^{\scr{ac}}_{e}
	\begin{array}{c}
			\ifx\JPicScale\undefined\def\JPicScale{1}\fi
	\psset{unit=\JPicScale mm}
	\psset{linewidth=0.3,dotsep=1,hatchwidth=0.3,hatchsep=1.5,shadowsize=1,dimen=middle}
	\psset{dotsize=0.7 2.5,dotscale=1 1,fillcolor=black}
	\psset{arrowsize=1 2,arrowlength=1,arrowinset=0.25,tbarsize=0.7 5,bracketlength=0.15,rbracketlength=0.15}
	\begin{pspicture}(0,0)(28,19)
	\psline(10,10)(20,10)
	\psline(20,10)(27,3)
	\psline(20,10)(27,17)
	\rput(23,10){+}
	\rput(28,1){}
	\psline{<-}(16,10)(10,10)
	\psline(10,10)(3,3)
	\psline(10,10)(3,17)
	\rput(2,19){$b$}
	\rput(2,1){$a$}
	\rput(28,19){$c$}
	\rput(28,1){$d$}
	\rput(7,10){+}
	\rput(15,13){$e+1$}
	\end{pspicture}
	\end{array}
	\end{split}
	\label{ac1}
\end{equation}
The form of the coefficient form is easily calculated inserting the explicit expression of the \{6j\} symbols given in Appendix \ref{6j symbols}
\begin{equation}
	\begin{split}
	X^{\scr{ac}}_e&=-N^{\scr(a)} N^{\scr(c)}(-1)^{3d+a+b-c}
	\dim(e)
		\left\{\begin{array}{ccc}     e  & e & 1 \\      a & a & b                 \end{array}\right\}
	\left\{\begin{array}{ccc}     e  & e & 1 \\      c & c & d                 \end{array}\right\}=\\
		&=-\frac{(-1)^{2(a+b+e)}}{4}\frac{\left(C^2(b)-C^2(a)-C^2(e)\right)\left(C^2(d)-C^2(c)-C^2(e)\right)}{C^2(e)}
\end{split}
\end{equation}
\begin{equation}
	\begin{split}
	Y^{\scr{ac}}_{e}&=-N^{\scr(a)} N^{\scr(c)}(-1)^{3d+a+b-c}
	\dim(e-1)
		\left\{\begin{array}{ccc}     e  & e- 1 & 1 \\      a & a & b                 \end{array}\right\}
	\left\{\begin{array}{ccc}     e  & e- 1 & 1 \\      c & c & d                 \end{array}\right\}=\\
		&=-\frac{(-1)^{2(a+b+e)}}{4e \dim(e)}\sqrt{(a+b+e+1)(a-b+e)(-a+b+e)(a+b-e+1)}\;\cdot\\
		&\cdot\sqrt{(c+d+e+1)(-c+d+e)(c-d+e)(c+d-e+1)}
\end{split}
\end{equation}
\begin{equation}
	\begin{split}
	Z^{\scr{ac}}_{e}&=-N^{\scr(a)} N^{\scr(c)}(-1)^{3d+a+b-c}
	\dim(e+1)
		\left\{\begin{array}{ccc}     e  & e+ 1 & 1 \\      a & a & b                 \end{array}\right\}
	\left\{\begin{array}{ccc}     e  & e+ 1 & 1 \\      c & c & d                 \end{array}\right\}=\\
		&=-\frac{(-1)^{2(a+b+e+1)}}{4(e+1) \dim(e)}\;\sqrt{(a+b+e+2)(a-b+e+1)(-a+b+e+1)(a+b-e)} \\
		&\cdot\sqrt{(c+d+e+2)(-c+d+e+1)(c-d+e+1)(c+d-e)}
\end{split}
\end{equation}

Note that by definition $(a+b+e)$ is an integer, so there aren't sign factors appearing in these expressions.

The last term is
\begin{equation}
\begin{split}
	&E^{\scr(a)}_n \cdot E^{\scr(d)}_n\;\begin{array}{c}
		\ifx\JPicScale\undefined\def\JPicScale{1}\fi
\psset{unit=\JPicScale mm}
\psset{linewidth=0.3,dotsep=1,hatchwidth=0.3,hatchsep=1.5,shadowsize=1,dimen=middle}
\psset{dotsize=0.7 2.5,dotscale=1 1,fillcolor=black}
\psset{arrowsize=1 2,arrowlength=1,arrowinset=0.25,tbarsize=0.7 5,bracketlength=0.15,rbracketlength=0.15}
\begin{pspicture}(0,0)(28,19)
\psline(10,10)(20,10)
\psline(20,10)(27,3)
\psline(20,10)(27,17)
\rput(23,10){+}
\rput(28,1){}
\psline{<-}(16,10)(10,10)
\psline(10,10)(3,3)
\psline(10,10)(3,17)
\rput(2,19){$b$}
\rput(2,1){$a$}
\rput(28,19){$c$}
\rput(28,1){$d$}
\rput(7,10){+}
\rput(15,13){$e$}
\end{pspicture}
\end{array}
=-N^{\scr(a)} N^{\scr(d)}
\begin{array}{c}
\ifx\JPicScale\undefined\def\JPicScale{1}\fi
\psset{unit=\JPicScale mm}
\psset{linewidth=0.3,dotsep=1,hatchwidth=0.3,hatchsep=1.5,shadowsize=1,dimen=middle}
\psset{dotsize=0.7 2.5,dotscale=1 1,fillcolor=black}
\psset{arrowsize=1 2,arrowlength=1,arrowinset=0.25,tbarsize=0.7 5,bracketlength=0.15,rbracketlength=0.15}
\begin{pspicture}(0,0)(33,22)
\psline(12,13)(22,13)
\psline(22,13)(32,3)
\rput(25,13){+}
\rput(30,4){}
\psline{<-}(18,13)(12,13)
\psline(12,13)(2,3)
\psline(12,13)(5,20)
\rput(9,13){+}
\rput(17,15){$e$}
\rput(17,15){}
\rput(1,1){$a$}
\rput(4,22){$b$}
\rput(30,22){$c$}
\rput(33,1){$d$}
\psline(29,6)(5,6)
\rput(5,4){+}
\rput(29,4){-}
\psline{<-}(9,10)(4,5)
\psline{<-}(25,10)(30,5)
\psline{<-}(18,6)(5,6)
\psline(22,13)(29,20)
\rput(17,3){1}
\end{pspicture}
\end{array}
=\\
&=-N^{\scr(a)} N^{\scr(d)}(-1)^{c+d+e}
\begin{array}{c}
\ifx\JPicScale\undefined\def\JPicScale{1}\fi
\psset{unit=\JPicScale mm}
\psset{linewidth=0.3,dotsep=1,hatchwidth=0.3,hatchsep=1.5,shadowsize=1,dimen=middle}
\psset{dotsize=0.7 2.5,dotscale=1 1,fillcolor=black}
\psset{arrowsize=1 2,arrowlength=1,arrowinset=0.25,tbarsize=0.7 5,bracketlength=0.15,rbracketlength=0.15}
\begin{pspicture}(0,0)(33,25)
\psline(12,13)(22,13)
\psline(22,13)(29,6)
\psline(22,13)(32,23)
\rput(25,13){+}
\rput(30,4){}
\psline{<-}(18,13)(12,13)
\psline(12,13)(2,3)
\psline(12,13)(5,20)
\rput(9,13){+}
\rput(17,15){$e$}
\rput(17,15){}
\rput(1,1){$a$}
\rput(4,22){$b$}
\rput(33,25){$d$}
\rput(30,4){$c$}
\psline(30,21)(4,5)
\rput(5,4){+}
\rput(29,22){+}
\psline{<-}(9,10)(4,5)
\psline{<-}(25,16)(30,21)
\psline{<-}(14,11)(4,5)
\rput(12,7){$1$}
\end{pspicture}
\end{array}	=\\
&=-N^{\scr(a)} N^{\scr(d)}(-1)^{3c+a+b-d}(-1)^{c+d+e}
\sum_m \dim{m}
\left\{\begin{array}{ccc}     e  & m & 1 \\      a & a & b                 \end{array}\right\}
\left\{\begin{array}{ccc}     e  & m & 1 \\      d & d & c                 \end{array}\right\}
\begin{array}{c}
	\ifx\JPicScale\undefined\def\JPicScale{1}\fi
\psset{unit=\JPicScale mm}
\psset{linewidth=0.3,dotsep=1,hatchwidth=0.3,hatchsep=1.5,shadowsize=1,dimen=middle}
\psset{dotsize=0.7 2.5,dotscale=1 1,fillcolor=black}
\psset{arrowsize=1 2,arrowlength=1,arrowinset=0.25,tbarsize=0.7 5,bracketlength=0.15,rbracketlength=0.15}
\begin{pspicture}(3,0)(28,19)
\psline(10,10)(20,10)
\psline(20,10)(27,3)
\psline(20,10)(27,17)
\rput(23,10){+}
\rput(28,1){}
\psline{<-}(16,10)(10,10)
\psline(10,10)(3,3)
\psline(10,10)(3,17)
\rput(2,19){$b$}
\rput(2,1){$a$}
\rput(28,19){$d$}
\rput(28,1){$c$}
\rput(7,10){+}
\rput(15,13){$m$}
\end{pspicture}
\end{array}=\\
&=-N^{\scr(a)} N^{\scr(d)}(-1)^{a+b+e}
\sum_m (-1)^{c+d+m}\dim{m}
\left\{\begin{array}{ccc}     e  & m & 1 \\      a & a & b                 \end{array}\right\}
\left\{\begin{array}{ccc}     e  & m & 1 \\      d & d & c                 \end{array}\right\}
\begin{array}{c}
	\ifx\JPicScale\undefined\def\JPicScale{1}\fi
\psset{unit=\JPicScale mm}
\psset{linewidth=0.3,dotsep=1,hatchwidth=0.3,hatchsep=1.5,shadowsize=1,dimen=middle}
\psset{dotsize=0.7 2.5,dotscale=1 1,fillcolor=black}
\psset{arrowsize=1 2,arrowlength=1,arrowinset=0.25,tbarsize=0.7 5,bracketlength=0.15,rbracketlength=0.15}
\begin{pspicture}(3,0)(26,19)
\psline(10,10)(20,10)
\psline(20,10)(27,3)
\psline(20,10)(27,17)
\rput(23,10){+}
\rput(28,1){}
\psline{<-}(16,10)(10,10)
\psline(10,10)(3,3)
\psline(10,10)(3,17)
\rput(2,19){$b$}
\rput(2,1){$a$}
\rput(28,19){$c$}
\rput(28,1){$d$}
\rput(7,10){+}
\rput(15,13){$m$}
\end{pspicture}
\end{array}
\end{split}
\label{ad},
\end{equation}
The result is obtained flipping the two link's $c$ and $d$ to recast the graph in the form \eqref{ac}, using the previous result and flipping back the graph in the summation.
Keeping in mind that the product of \{6j\}appearing in the non diagonal terms is left unchanged by the change $c\rightarrow d$, the final result is then the same as \eqref{ac1} apart from the sign of the non-diagonal terms and the change $c\rightarrow d$ in the diagonal one 
\begin{equation}
	\begin{split}
	&	E^{\scr(a)}_n \cdot E^{\scr(d)}_n\;\begin{array}{c}
			\ifx\JPicScale\undefined\def\JPicScale{1}\fi
	\psset{unit=\JPicScale mm}
	\psset{linewidth=0.3,dotsep=1,hatchwidth=0.3,hatchsep=1.5,shadowsize=1,dimen=middle}
	\psset{dotsize=0.7 2.5,dotscale=1 1,fillcolor=black}
	\psset{arrowsize=1 2,arrowlength=1,arrowinset=0.25,tbarsize=0.7 5,bracketlength=0.15,rbracketlength=0.15}
	\begin{pspicture}(0,0)(28,19)
	\psline(10,10)(20,10)
	\psline(20,10)(27,3)
	\psline(20,10)(27,17)
	\rput(23,10){+}
	\rput(28,1){}
	\psline{<-}(16,10)(10,10)
	\psline(10,10)(3,3)
	\psline(10,10)(3,17)
	\rput(2,19){$b$}
	\rput(2,1){$a$}
	\rput(28,19){$c$}
	\rput(28,1){$d$}
	\rput(7,10){+}
	\rput(15,13){$e$}
	\end{pspicture}
	\end{array}
	=\\
	&=X^{\scr{ad}}_e
	\begin{array}{c}
			\ifx\JPicScale\undefined\def\JPicScale{1}\fi
	\psset{unit=\JPicScale mm}
	\psset{linewidth=0.3,dotsep=1,hatchwidth=0.3,hatchsep=1.5,shadowsize=1,dimen=middle}
	\psset{dotsize=0.7 2.5,dotscale=1 1,fillcolor=black}
	\psset{arrowsize=1 2,arrowlength=1,arrowinset=0.25,tbarsize=0.7 5,bracketlength=0.15,rbracketlength=0.15}
	\begin{pspicture}(0,0)(28,19)
	\psline(10,10)(20,10)
	\psline(20,10)(27,3)
	\psline(20,10)(27,17)
	\rput(23,10){+}
	\rput(28,1){}
	\psline{<-}(16,10)(10,10)
	\psline(10,10)(3,3)
	\psline(10,10)(3,17)
	\rput(2,19){$b$}
	\rput(2,1){$a$}
	\rput(28,19){$c$}
	\rput(28,1){$d$}
	\rput(7,10){+}
	\rput(15,13){$e$}
	\end{pspicture}
	\end{array}-
	Y^{\scr{ad}}_{e}
	\begin{array}{c}
			\ifx\JPicScale\undefined\def\JPicScale{1}\fi
	\psset{unit=\JPicScale mm}
	\psset{linewidth=0.3,dotsep=1,hatchwidth=0.3,hatchsep=1.5,shadowsize=1,dimen=middle}
	\psset{dotsize=0.7 2.5,dotscale=1 1,fillcolor=black}
	\psset{arrowsize=1 2,arrowlength=1,arrowinset=0.25,tbarsize=0.7 5,bracketlength=0.15,rbracketlength=0.15}
	\begin{pspicture}(0,0)(28,19)
	\psline(10,10)(20,10)
	\psline(20,10)(27,3)
	\psline(20,10)(27,17)
	\rput(23,10){+}
	\rput(28,1){}
	\psline{<-}(16,10)(10,10)
	\psline(10,10)(3,3)
	\psline(10,10)(3,17)
	\rput(2,19){$b$}
	\rput(2,1){$a$}
	\rput(28,19){$c$}
	\rput(28,1){$d$}
	\rput(7,10){+}
	\rput(15,13){$e-1$}
	\end{pspicture}
	\end{array}
	-
	Z^{\scr{ad}}_{e}
	\begin{array}{c}
			\ifx\JPicScale\undefined\def\JPicScale{1}\fi
	\psset{unit=\JPicScale mm}
	\psset{linewidth=0.3,dotsep=1,hatchwidth=0.3,hatchsep=1.5,shadowsize=1,dimen=middle}
	\psset{dotsize=0.7 2.5,dotscale=1 1,fillcolor=black}
	\psset{arrowsize=1 2,arrowlength=1,arrowinset=0.25,tbarsize=0.7 5,bracketlength=0.15,rbracketlength=0.15}
	\begin{pspicture}(0,0)(28,19)
	\psline(10,10)(20,10)
	\psline(20,10)(27,3)
	\psline(20,10)(27,17)
	\rput(23,10){+}
	\rput(28,1){}
	\psline{<-}(16,10)(10,10)
	\psline(10,10)(3,3)
	\psline(10,10)(3,17)
	\rput(2,19){$b$}
	\rput(2,1){$a$}
	\rput(28,19){$c$}
	\rput(28,1){$d$}
	\rput(7,10){+}
	\rput(15,13){$e+1$}
	\end{pspicture}
	\end{array}
	\end{split}
	\label{ad1}
\end{equation}
where
\begin{equation}
	\begin{split}
	X^{\scr{ad}}_e&=-N^{\scr(a)} N^{\scr(d)}(-1)^{a+b+c+d+2e}
	\dim(e)
		\left\{\begin{array}{ccc}     e  & e & 1 \\      a & a & b                 \end{array}\right\}
	\left\{\begin{array}{ccc}     e  & e & 1 \\      d & d & c                 \end{array}\right\}=\\
		&=-\frac{1}{4}\frac{\left(C^2(b)-C^2(a)-C^2(e)\right)\left(C^2(c)-C^2(d)-C^2(e)\right)}{C^2(e)}
\end{split}
\end{equation}

Note that by definition
\begin{equation}
	Y_e^{\scr{ac}}=Y_e^{\scr{ad}}\qquad Z_e^{\scr{ac}}=Z_e^{\scr{ad}}
\end{equation}

The operators that we have calculated have to satisfy  
\begin{equation}
	E^{\scr(a)}_n \cdot E^{\scr(a)}_n+E^{\scr(a)}_n \cdot E^{\scr(b)}_n+E^{\scr(a)}_n \cdot E^{\scr(c)}_n+E^{\scr(a)}_n \cdot E^{\scr(d)}_n=0
\end{equation}
as a direct consequence of \eqref{dipendenzalin} which, at quantum level, implies that a 
four-valent node (by definition an intertwiner) is invariant under under the action of the group.
A direct calculation on our four-valent node shows that this is indeed the case
\begin{equation}
	\begin{split}
	&\left(E^{\scr(a)}_n \cdot E^{\scr(a)}_n+E^{\scr(a)}_n \cdot E^{\scr(b)}_n+E^{\scr(a)}_n \cdot E^{\scr(c)}_n+E^{\scr(a)}_n \cdot E^{\scr(d)}_n \right)
		\begin{array}{c}
				\ifx\JPicScale\undefined\def\JPicScale{1}\fi
		\psset{unit=\JPicScale mm}
		\psset{linewidth=0.3,dotsep=1,hatchwidth=0.3,hatchsep=1.5,shadowsize=1,dimen=middle}
		\psset{dotsize=0.7 2.5,dotscale=1 1,fillcolor=black}
		\psset{arrowsize=1 2,arrowlength=1,arrowinset=0.25,tbarsize=0.7 5,bracketlength=0.15,rbracketlength=0.15}
		\begin{pspicture}(0,0)(28,19)
		\psline(10,10)(20,10)
		\psline(20,10)(27,3)
		\psline(20,10)(27,17)
		\rput(23,10){+}
		\rput(28,1){}
		\psline{<-}(16,10)(10,10)
		\psline(10,10)(3,3)
		\psline(10,10)(3,17)
		\rput(2,19){$b$}
		\rput(2,1){$a$}
		\rput(28,19){$c$}
		\rput(28,1){$d$}
		\rput(7,10){+}
		\rput(15,13){$e$}
		\end{pspicture}
		\end{array}=\\
		&=\left(C^2(a)+\frac{C^2(e)-C^2(a)-C^2(b)}{2}+X^{\scr{ac}}_e+X^{\scr{ad}}_e\right)\begin{array}{c}
			\ifx\JPicScale\undefined\def\JPicScale{1}\fi
	\psset{unit=\JPicScale mm}
	\psset{linewidth=0.3,dotsep=1,hatchwidth=0.3,hatchsep=1.5,shadowsize=1,dimen=middle}
	\psset{dotsize=0.7 2.5,dotscale=1 1,fillcolor=black}
	\psset{arrowsize=1 2,arrowlength=1,arrowinset=0.25,tbarsize=0.7 5,bracketlength=0.15,rbracketlength=0.15}
	\begin{pspicture}(0,0)(28,19)
	\psline(10,10)(20,10)
	\psline(20,10)(27,3)
	\psline(20,10)(27,17)
	\rput(23,10){+}
	\rput(28,1){}
	\psline{<-}(16,10)(10,10)
	\psline(10,10)(3,3)
	\psline(10,10)(3,17)
	\rput(2,19){$b$}
	\rput(2,1){$a$}
	\rput(28,19){$c$}
	\rput(28,1){$d$}
	\rput(7,10){+}
	\rput(15,13){$e$}
	\end{pspicture}
	\end{array}+\\
	&+\left(Y^{\scr{ac}}_{e}-Y^{\scr{ad}}_{e}\right)\begin{array}{c}
			\ifx\JPicScale\undefined\def\JPicScale{1}\fi
	\psset{unit=\JPicScale mm}
	\psset{linewidth=0.3,dotsep=1,hatchwidth=0.3,hatchsep=1.5,shadowsize=1,dimen=middle}
	\psset{dotsize=0.7 2.5,dotscale=1 1,fillcolor=black}
	\psset{arrowsize=1 2,arrowlength=1,arrowinset=0.25,tbarsize=0.7 5,bracketlength=0.15,rbracketlength=0.15}
	\begin{pspicture}(0,0)(28,19)
	\psline(10,10)(20,10)
	\psline(20,10)(27,3)
	\psline(20,10)(27,17)
	\rput(23,10){+}
	\rput(28,1){}
	\psline{<-}(16,10)(10,10)
	\psline(10,10)(3,3)
	\psline(10,10)(3,17)
	\rput(2,19){$b$}
	\rput(2,1){$a$}
	\rput(28,19){$c$}
	\rput(28,1){$d$}
	\rput(7,10){+}
	\rput(15,13){$e-1$}
	\end{pspicture}
	\end{array}
	+\left(Z^{\scr{ac}}_{e}-Z^{\scr{ad}}_{e}\right)\begin{array}{c}
			\ifx\JPicScale\undefined\def\JPicScale{1}\fi
	\psset{unit=\JPicScale mm}
	\psset{linewidth=0.3,dotsep=1,hatchwidth=0.3,hatchsep=1.5,shadowsize=1,dimen=middle}
	\psset{dotsize=0.7 2.5,dotscale=1 1,fillcolor=black}
	\psset{arrowsize=1 2,arrowlength=1,arrowinset=0.25,tbarsize=0.7 5,bracketlength=0.15,rbracketlength=0.15}
	\begin{pspicture}(0,0)(28,19)
	\psline(10,10)(20,10)
	\psline(20,10)(27,3)
	\psline(20,10)(27,17)
	\rput(23,10){+}
	\rput(28,1){}
	\psline{<-}(16,10)(10,10)
	\psline(10,10)(3,3)
	\psline(10,10)(3,17)
	\rput(2,19){$b$}
	\rput(2,1){$a$}
	\rput(28,19){$c$}
	\rput(28,1){$d$}
	\rput(7,10){+}
	\rput(15,13){$e+1$}
	\end{pspicture}
	\end{array}=0
\end{split}
\label{somma0}
	\end{equation}
being 0 the coefficient of all the states. 

\section{Normalization of the spinnetwork states}\label{snormalization}

Following \cite{book}, we define a spinnetwork $S=(\Gamma,j_l,i_n)$  as given by a graph $\Gamma$ with a given orientation (or ordering of the links) with $L$ links and $N$ nodes, and by a representation $j_l$ associated to each to each link and an intertwiner $i_n$ to each node.
As a functional of the connection, a spin network state is given by 
\begin{equation}
	\Psi_S[A]=\left\langle A|S\right\rangle\equiv \left(\otimes_l R^{j_l}(H[A,\gamma_l])\right)\cdot\left(\otimes_n i_n\right)
\end{equation}
where the notation $\cdot$ indicates the contraction between dual spaces and $R^{j_l}(H[A,\gamma_l])$ is the $j_l$ representation of the holonomy group element $H[A,\gamma_l]$ along the curve $\gamma_l$ of the gravitation field connection $A$.
In the paper we have used states normalized in such a way that
\begin{equation}
	\left\langle S|S'\right\rangle=\delta_{S,S'} \label{ortonormali}.
\end{equation}
Following \cite {depietri1, depietri2} we can see that the scalar product reduces to the evaluation of the spinnetwork and that the definition of the spinnet state has to be properly normalized in order for \eqref{ortonormali} to be satisfied.
Here we have used three-valent intertwiners (3j-Wigner symbols \eqref{3j}) normalized to 1, so that the evaluation of the theta-graph gives 1: see \eqref{theta=1}. This means that the formula (8.7) of \cite {depietri1} defining a normalized spinnetwork state in our case reads
\begin{equation}
	|S\rangle_N=\sqrt{\prod_{e\in \mathcal{E}}\dim{j_e}}|S\rangle, 
\end{equation}
where $\mathcal{E}$ is the set of real and virtual edges (intertwiner links of the decomposition of multivalent nodes).
We can then see that the recoupling theorem \eqref{recoupling theorem} when applied to spinnetwork normalized state becomes
\begin{equation}
	\left|\begin{array}{c}
		\ifx\JPicScale\undefined\def\JPicScale{1}\fi
\psset{unit=\JPicScale mm}
\psset{linewidth=0.3,dotsep=1,hatchwidth=0.3,hatchsep=1.5,shadowsize=1,dimen=middle}
\psset{dotsize=0.7 2.5,dotscale=1 1,fillcolor=black}
\psset{arrowsize=1 2,arrowlength=1,arrowinset=0.25,tbarsize=0.7 5,bracketlength=0.15,rbracketlength=0.15}
\begin{pspicture}(0,0)(28,19)
\psline(10,10)(20,10)
\psline(20,10)(27,3)
\psline(20,10)(27,17)
\rput(23,10){+}
\rput(28,1){}
\psline{<-}(16,10)(10,10)
\psline(10,10)(3,3)
\psline(10,10)(3,17)
\rput(2,19){$a$}
\rput(2,1){$c$}
\rput(28,19){$d$}
\rput(28,1){$b$}
\rput(7,10){+}
\rput(15,13){$e$}
\end{pspicture}
\end{array}\right\rangle_N
=
\sum_f \sqrt{\dim{e}}\sqrt{\dim{f}} (-1)^{b+c+e+f}
\quad
\left\{\begin{array}{ccc}                      a  & b & f \\                      d & c & e                  \end{array}\right\} 
\quad
\left|
	\begin{array}{c}
	\ifx\JPicScale\undefined\def\JPicScale{1}\fi
	\psset{unit=\JPicScale mm}
	\psset{linewidth=0.3,dotsep=1,hatchwidth=0.3,hatchsep=1.5,shadowsize=1,dimen=middle}
	\psset{dotsize=0.7 2.5,dotscale=1 1,fillcolor=black}
	\psset{arrowsize=1 2,arrowlength=1,arrowinset=0.25,tbarsize=0.7 5,bracketlength=0.15,rbracketlength=0.15}
	\begin{pspicture}(0,0)(18,26)
	\psline(3,3)(10,10)
	\psline(10,10)(17,3)
	\psline(10,10)(10,17)
	\psline(3,24)(10,17)
	\psline(17,24)(10,17)
	\psline{<-}(10,15)(10,10)
	\rput(13,14){$f$}
	\rput(10,7){+}
	\rput(10,20){+}
	\rput(2,1){$a$}
	\rput(18,1){$b$}
	\rput(2,26){$d$}
	\rput(18,26){$c$}
	\end{pspicture}
	\end{array}
\right\rangle_N
\end{equation}

\section{Regge Action and its derivatives} \label{regge}

Following \cite{Regge}, we can write the asymptotic formula of a 6j symbol as 
\begin{equation}
	\left\{\begin{array}{ccc}     a  & b & c \\      d & e & f                 \end{array}\right\} \approx \frac{1}{\sqrt{12\pi V}}\; \cos{\left(S_R+\frac{\pi}{4}\right)}
\end{equation}
where 
\begin{equation}
	S_R=\sum_{i,j=1}^4 l_{\scr{ij}} \phi_{\scr{ij}}
\end{equation}
where $S_R$ is the Regge action of the tetrahedron 
\begin{equation}
	\begin{array}{c}
	\ifx\JPicScale\undefined\def\JPicScale{1}\fi
\psset{unit=\JPicScale mm}
\psset{linewidth=0.2,dotsep=1,hatchwidth=0.3,hatchsep=1.5,shadowsize=1,dimen=middle}
\psset{dotsize=0.7 2.5,dotscale=1 1,fillcolor=black}
\psset{arrowsize=1 2,arrowlength=1,arrowinset=0.25,tbarsize=0.7 5,bracketlength=0.15,rbracketlength=0.15}
\begin{pspicture}(0,0)(50,25)
\psline(2,2)(38,2)
\psline(2,2)(21,23)
\psline(21,23)(38,2)
\psline(21,23)(48,15)
\psline(38,2)(48,15)
\psline[linestyle=dotted](2,2)(48,15)
\rput(0,0){$1$}
\rput(40,0){$2$}
\rput(50,17){$3$}
\rput(19,25){$4$}
\rput(15,-1){$l_{\scr{34}}$}
\rput(46,7){$l_{\scr{14}}$}
\rput(36,21){$l_{\scr{12}}$}
\rput(8,13){$l_{\scr{23}}$}
\rput(19,9){$l_{\scr{24}}$}
\rput(31,15){$l_{\scr{13}}$}
\end{pspicture}
\end{array}
\end{equation}
 associated to the 6j symbol, and $\phi_{ij}=\phi_{ji}$ $(i\neq j)$ are the dihedral angle at the edge $l_{\scr{ij}}$.
 The edge lengths in terms of the 6j entries are: $l_{\scr{12}}=a+\frac{1}{2}$, $l_{\scr{13}}=b+\frac{1}{2}$, $l_{\scr{14}}=c+\frac{1}{2}$, $l_{\scr{34}}=d+\frac{1}{2}$, $l_{\scr{23}}=b+\frac{1}{2}$ and $l_{\scr{hh}}=0$, $l_{\scr{hk}}=l_{\scr{kh}}$.

The dihedral angles can be expressed in terms of the volume and the areas of the tetrahedron
\begin{equation}
	A_{\scr{i}} A_{\scr{j}} \sin{\phi_{\scr{ij}}}=\frac{3}{2}\,  l_{\scr{ij}} V
	\label{angolo diedro in termini di edge}
\end{equation}
 where $A_{\scr{i}}$ is the area of the triangle opposite to the vertex $i$ ($A_{\scr{i}}, A_{\scr{j}}$ are the areas of the triangles that share the edge $l_{\scr{ij}}$).
We are interested in the expansion of the Regge action in the variables $l_{ij}$; we can express everything in term of the edge length expressing the volume and the areas using the formula
\begin{equation}
	V^2_d=\frac{(-1)^{d+1}}{2^d (d!)^2} \det C_d
\end{equation}
where $V_d$ is the volume of a simplex of dimension $d$ and $C_d$ is the Cayley matrix of dimension $d$; in particular given 6 edges for the tetrahedron or 3 for the triangle, with the following Cayley matrix we can calculate all the quantities appearing in \eqref{angolo diedro in termini di edge}
\begin{equation}
	C_3=\begin{pmatrix}
	0&1&    1&1 &1\\
	1&0&l_1^2&l_2^2 &l_3^2\\
	1&l_1^2&0&l^2_4&l^2_5      \\
	1& l^2_2&l_4^2   &0&   l^2_6\\	
	1& l^2_3&   l_5^2       &l_6^2     &0
	      \end{pmatrix}
	      \qquad\qquad
	      	C_2=\begin{pmatrix}
	0&1&    1&1 \\
	1&0&l_1^2&l_2^2 \\
	1&l_1^2&0&l^2_3      \\
	1& l^2_2&l_3^2   &0&   \\
		      \end{pmatrix}
\end{equation}

We are interested in the asymptotic expansion of the 6j symbol that realizes the change of pairing at a given node; in the node 1 for example
 \begin{equation}
	\left\{\begin{array}{ccc}                      j_{\scr{12}}  & j_{\scr{13}} & i^{\scr{x}}_1 \\
			                                                j_{\scr{15}} & j_{\scr{14}} & i^{\scr{y}}_1        
			 \end{array}\right\} 
\end{equation}
with link variables $j_{\scr{1n}}$ centered around $j^0$ and intertwiners variables $i^{\scr{m_n}}_1$ centered around $i^0=\frac{2}{\sqrt{3}}j^0$.
Using the previous formula we can calculate the coefficients of the Regge action expansion linked to this symbol.
The relevant derivatives for our calculation are (see also the Appendix of \cite{corrections})
\begin{eqnarray}
\left.\frac{\partial{S^A_R}}{\partial{i^{\scr{x}}_1}}\right|_{j^0,i^0}=\left.\frac{\partial{S^A_R}}{\partial{i^{\scr{y}}_1}}\right|_{j^0,i^0}=\frac{\pi}{2},\\
\left.\frac{\partial^2{S^A_R}}{\partial{j_{\scr{1n}}}\partial{i^{\scr{x}}_1}}\right|_{j^0,i^0}=
\left.\frac{\partial^2{S^A_R}}{\partial{j_{\scr{1n}}}\partial{i^{\scr{y}}_1}}\right|_{j^0,i^0}=\frac{3}{4j^0},
\\
\left.\frac{\partial^2{S^A_R}}{\partial{i^{\scr{x}}_1}\partial{i^{\scr{y}}_1}}\right|_{j^0,i^0}=-\frac{\sqrt{3}}{j^0},\\
	\left.\frac{\partial^2{S^A_R}}{\partial^2{i^{\scr{x}}_1}}\right|_{j^0,i^0}
	=\left.\frac{\partial^2{S^A_R}}{\partial^2{i^{\scr{y}}_1}}\right|_{j^0,i^0}=-\frac{\sqrt{3}}{2j^0}.
\end{eqnarray}

\section{Change of pairing on the boundary state} \label{change of basis}

Here we show how one of the coefficients defined by \eqref{c2} transforms under
the change of basis determined by a different pairing. In particular, we show that with the choice of parameters in \eqref{cbis}, equation \eqref{ca1} becomes \eqref{cafine}.  Under the change of basis,
\begin{equation}
\begin{split}
		\Phi'_{\mathbf q}[{\mathbf j}, i^{\scr{x}}_1,i_2...i_5]
		=			\sum_{i^{\scr{y}}_1} 
			\Phi_{\mathbf q}[{\mathbf j}, i^{\scr{y}}_1,i_2...i_5]
		 (-1)^{j_{\scr{13}}+j_{\scr{14}}+i^{\scr{x}}_1+i^{\scr{y}}_1} \;\sqrt{d_{ i^{\scr{x}}_1}d_{ i^{\scr{y}}_1}}
		 \left\{\begin{array}{ccc}                      j_{\scr{12}}  & j_{\scr{13}} & i^{\scr{x}}_1 \\
		                                                j_{\scr{15}} & j_{\scr{14}} & i^{\scr{y}}_1        .
		 \end{array}\right\} 
	\end{split}
	\label{cainappendix}
\end{equation}
With the choice of the boundary state defined by \eqref{cbis}, this reads 
\begin{equation}
\begin{split}
		\Phi'_{\mathbf q}[{\mathbf j}, i^{\scr{x}}_1,i_2...i_5]&
		=\ \ 
		e^{-\frac{1}{2j_0}\sum \alpha_{\scr{(ij)(mr)}}\delta j_{ij}\delta j_{mr}+i\sum\Phi \delta j_{ij}}
		e^{-\sum_{n\neq1}\left(\frac{(\delta i^{\scr{m_n}}_n)^2}{4\sigma_{i^{m_n}}}+\sum_a\phi_{j_{na}\;i^{m_n}_n} \delta j^{\scr{an}}\delta i^{\scr{m_n}}_n +i \chi_{i^{m_n}_n} \delta i^{\scr{m_n}}_n\right)}	 \\
		 	&\cdot
			\sum_{i^{\scr{y}}_1} 
			e^{-\left(\frac{(\delta i^{\scr{y}}_1)^2}{4\sigma_{i^{y}_1}}+\sum_a\phi_{j_{a1}\;i^{y}_1} \delta j^{\scr{a1}}\delta i^{\scr{y}}_1 +i \chi_{i^{y}_1}\delta i^y_1\right)}
		 (-1)^{j_{\scr{13}}+j_{\scr{14}}+i^{\scr{x}}_1+i^{\scr{y}}_1} \;\sqrt{d_{ i^{\scr{x}}_1}d_{ i^{\scr{y}}_1}}
		 \left\{\begin{array}{ccc}                      j_{\scr{12}}  & j_{\scr{13}} & i^{\scr{x}}_1 \\
		                                                j_{\scr{15}} & j_{\scr{14}} & i^{\scr{y}}_1        
		 \end{array}\right\} .
	\end{split}
	\label{ca1inappendix}
\end{equation}
Expanding the 6j symbol in the large-$j$ limit, and applying the relation \eqref{esistenza dei j} we get
\begin{equation}
	\begin{split}
		\Phi'_{\mathbf q}&({\mathbf j}, i^{\scr{x}}_1,i_2,...,i_5)
			=
				e^{-\frac{1}{2j^0}\sum \alpha_{\scr{(ij)(mr)}}\delta j^{\scr{ij}}\delta j^{\scr{mr}}+i\sum\Phi \delta j^{\scr{ij}}}
		e^{-\sum_{n\neq1}\left(\frac{(\delta i^{\scr{m_n}}_n)^2}{4\sigma_{i^{m_n}}}+\sum_a\phi_{j_{na}\;i^{m_n}_n} \delta j^{\scr{an}}\delta i^{\scr{m_n}}_n +i \chi_{i^{m_n}_n} \delta i^{\scr{m_n}}_n\right)}
	 \\
	 &\ \ \ \cdot \frac{e^{i\pi i_0}}{2} \int{d\delta i^{\scr{y}}_1}
	 e^{-\left(\frac{(\delta i^{\scr{y}}_1)^2}{4\sigma_{i^{y}_1}}+\sum_a\phi_{j_{a1}\;i^{y}_1} \delta j^{\scr{a1}}\delta i^{\scr{y}}_1 +i \chi_{i^{y}_1}\delta i^y_1\right)}
	 \sqrt{d_{ i^{\scr{x}}_1}d_{ i^{\scr{y}}_1}}\;\frac{e^{i (S_R+\pi\delta i^y_1+\frac{\pi}{4}) }+e^{-i (S_R-\pi\delta i^y_1+\frac{\pi}{4}) }}{\sqrt{12\pi V}}.
				 \end{split}
			 \label{ca2inappendix}
\end{equation}
We expand the Regge action up to second order in all its 6 entries; the external link around $j^0$ and the intertwiners around $i^0$
\begin{equation}
\begin{split}
	S_R[j_{\scr{1n}},i^{\scr{y}}_1,i^{\scr{x}}_1]
		=&S_R[j^0,i^0]+\left.\frac{\partial{S_R}}{\partial{j_{\scr{1n}}}}\right|_{j^0,i^0}\delta j_{\scr{1n}}
			+\left.\frac{\partial{S_R}}{\partial{i^{\scr{x}}_1}}\right|_{j^0,i^0}\delta i^{\scr{x}}_1
			+\left.\frac{\partial{S_R}}{\partial{i^{\scr{y}}_1}}\right|_{j^0,i^0}\delta i^{\scr{y}}_1
			+\left.\frac{\partial^2{S_R}}{\partial{j_{\scr{1n}}}\partial{j_{\scr{1n'}}}}\right|_{j^0,i^0}\delta j_{\scr{1n}}\delta j_{\scr{1n'}}+\\
			&+\left.\frac{\partial^2{S_R}}{\partial{j_{\scr{1n}}}\partial{i^{\scr{x}}_1}}\right|_{j^0,i^0}\delta j_{\scr{1n}}\delta i^{\scr{x}}_1
			+\left.\frac{\partial^2{S_R}}{\partial{j_{\scr{1n}}}\partial{i^{\scr{y}}_1}}\right|_{j^0,i^0}\delta j_{\scr{1n}}\delta i^{\scr{y}}_1 
			+\left.\frac{\partial^2{S_R}}{\partial{i^{\scr{x}}_1}\partial{i^{\scr{y}}_1}}\right|_{j^0,i^0}\delta i^{\scr{x}}_1\delta i^{\scr{y}}_1
			\\
		&	+\frac{1}{2}\left.\frac{\partial^2{S_R}}{\partial^2{j_{\scr{1n}}}}\right|_{j^0,i^0}(\delta j_{\scr{1n}})^2
					+\frac{1}{2}\left.\frac{\partial^2{S_R}}{\partial^2{i^{\scr{x}}_1}}\right|_{j^0,i^0}(\delta i^{\scr{x}}_1)^2
+\frac{1}{2}\left.\frac{\partial^2{S_R}}{\partial^2{i^{\scr{y}}_1}}\right|_{j^0,i^0}(\delta i^{\scr{y}}_1)^2+...
\end{split}
\end{equation}
In the background in which we are interested, $i^0=\frac{2}{\sqrt3}j^0$ and $\left.\frac{\partial{S_R}}{\partial{i^{\scr{y}}_1}}\right|_{j^0,i^0}=\left.\frac{\partial{S_R}}{\partial{i^{\scr{x}}_1}}\right|_{j^0,i^0}=\frac{\pi}{2}$.
The value $\chi=\left.\frac{\partial{S_R}}{\partial{i^{\scr{y}}_1}}\right|_{j^0,i^0}=\frac{\pi}{2}$, yields a phase in the intertwiner variable $e^{-i\frac{\pi}{2}\delta i^{\scr{y}}_1}$ that cancels one of the two rapidly-oscillating phase factor due to the linear term of the expansion of the Regge action. In particular the linear part in the intertwiner variable of the first exponential  $e^{i(\left.\frac{\partial{S_R}}{\partial{i^{\scr{y}}_1}}\right|_{j^0,i^0}+\pi)\delta  i^{\scr{y}}_1}=e^{i\frac{3\pi}{2}\delta i^{\scr{y}}_1}$ combines with the boundary phase factor but the linear part of the second one $e^{-i(\left.\frac{\partial{S_R}}{\partial{i^{\scr{y}}_1}}\right|_{j^0,i^0}-\pi)\delta i^{\scr{y}}_1}=e^{i\frac{\pi}{2}\delta i^{\scr{y}}_1}$ is canceled: for the same mechanism described in \cite{scattering3} only the second term in the summation \eqref{ca2inappendix} survives. Denoting  $\tilde{S_R}=S_R-i\frac{\pi}{2}\delta i^{\scr{y}}_1$,
we have that \eqref{ca2inappendix} reduces to
\begin{equation}
	\begin{split}
		\Phi'_{\mathbf q}({\mathbf j}, i^{\scr{x}}_1,i_2,...,i_5)
			=&
			e^{-\frac{1}{2j^0}\sum \alpha_{\scr{(ij)(mr)}}\delta j^{\scr{ij}}\delta j^{\scr{mr}}+i\sum\Phi \delta j^{\scr{ij}}}
			e^{-\sum_{n\neq1}\left(\frac{(\delta i^{\scr{m_n}}_n)^2}{4\sigma_{i^{m_n}}}+\sum_a\phi_{j_{na}\;i^{m_n}_n} \delta j^{\scr{an}}\delta i^{\scr{m_n}}_n +i \frac{\pi}{2} \delta i^{\scr{m_n}}_n\right)}
	\;	\cdot \\
	 &\cdot \frac{e^{i\pi i_0}}{2} \int{d\delta i^{\scr{y}}_1}\;e^{-\left(\frac{(\delta i^{\scr{y}}_1)^2}{4\sigma_{i^{y}_1}}+\sum_a \phi_{j_{1a}\;i^{y}_1}\;\delta j^{\scr{a1}}\delta i^y_1+i \frac{\pi}{2}\delta i^y_1\right)}
	 \sqrt{d_{ i^{\scr{x}}_1}d_{ i^{\scr{y}}_1}}\;\frac{e^{-i (\tilde{S_R}+\frac{\pi}{4}) }}{\sqrt{12\pi V}}.
				 \end{split}
			 \end{equation}
From  \cite{corrections}, we have that denoting $\mu= \sqrt{\frac{d_{ i^{\scr{x}}_1}d_{ i^{\scr{y}}_1}}{12\pi V}}$, the dominant term is $\mu[j^0]$.  We take $\mu[j^0]$ out of the integration and 
evaluate the integral following \cite{sr}.  To simplify the notation, rename the second derivative of the Regge action $G_{j^{\scr{na}},i^{m_n}_{n}}=\left.\frac{\partial^2{S_R}}{\partial{j_{\scr{na}}}\partial{i^{\scr{m_n}}_n}}\right|_{j^0,i^0}$, $G_{i^{m_{n'}}_n,i^{m_n}_n}=\left.\frac{\partial^2{S_R}}{\partial{i^{\scr{m_{n'}}}_n}\partial{i^{\scr{m_n}}_n}}\right|_{j^0,i^0}$ and indicate with $S[j_{\scr{na}}]$ \eqref{S} the part of the Regge action that depends only on the boundary links involved in the $6j$ symbol considered and with no dependence from the intertwiners. 
 Substituting we get
\begin{equation}
	\begin{split}
	\Phi'_{\mathbf q}({\mathbf j}, i^{\scr{x}}_1,i_2,...,i_5)
			=&
			e^{-\frac{1}{2j^0}\sum \alpha_{\scr{(ij)(mr)}}\delta j^{\scr{ij}}\delta j^{\scr{mr}}+i\sum\Phi \delta j^{\scr{ij}}}
			e^{-\sum_{n\neq1}\left(\frac{(\delta i^{\scr{m_n}}_n)^2}{4\sigma_{i^{m_n}}}+\sum_a\phi_{j_{na}\;i^{m_n}_n} \delta j^{\scr{an}}\delta i^{\scr{m_n}}_n +i \frac{\pi}{2} \delta i^{\scr{m_n}}_n\right)}
	 \\
	 &\cdot \frac{e^{i\pi i_0}}{2}e^{-i\frac{\pi}{4}}\mu[j^0] e^{-i S_R[j^0,i^0]}\\
&\cdot  e^{-i S_j[j_{\scr {1a}}]}
e^{-i\frac{\pi}{2}\delta i^{x}_1}
e^{-i\left( \sum_a G_{j_{1a}\;i^{x}_1}\;\delta j^{\scr{a1}}\right)\delta i^{x}_1}
	e^{-\frac{i}{2}G_{i^{x}_1 i^{x}_1}(\delta i^{x}_1)^2}\\
	 &\cdot \int{d\delta i^{\scr{y}}_1}\;e^{-\frac{1}{2}(\frac{1}{2\sigma_{i^{y}_1}}+iG_{i^{y}_1 i^{y}_1})(\delta i^{\scr{y}}_1)^2}
	 e^{-i G_{i^{x}_1\;i^{y}_1}\;\delta i^{x}_1\delta i^y_1}\;
	 	e^{-\left(\sum_a \left(\phi_{j_{1a}\;i^{\scr{y}}_1}+iG_{j_{1a}\;i^{y}_1}\right)\delta \;j^{\scr{a1}}\right) \delta i^{\scr{y}}_1}
			 \end{split}
			 \label{caintegrale}
			 \end{equation}
The choice $\phi=-i G_{j_{1a}\;i^{y}_1}=-i\frac{3}{4j^0}$ eliminates the argument of last exponential. So that we fall into the same as calculation \cite{sr}, and we can transform the gaussian in another gaussian with the same variance. Evaluating the integral we get
		 \begin{equation}
	\begin{split}
	\Phi'_{\mathbf q}({\mathbf j}, i^{\scr{x}}_1,i_2,...,i_5)
			=&
			e^{-\frac{1}{2j^0}\sum \alpha_{\scr{(ij)(mr)}}\delta j^{\scr{ij}}\delta j^{\scr{mr}}+i\sum\Phi \delta j^{\scr{ij}}}
			e^{-\sum_{n\neq1}\left(\frac{(\delta i^{\scr{m_n}}_n)^2}{4\sigma_{i^{m_n}}}-i\left(\sum_a \frac{3}{4j^0}\delta j^{\scr{an}}-\frac{\pi}{2}\right) \delta i^{\scr{m_n}}_n\right)}   
		\;	\cdot \\
	 &\sqrt{\frac{\pi}{2(\frac{1}{2\sigma_{i^{y}_1}}+iG^A_{i^{y}_1 i^{y}_1})}}\cdot e^{i\pi i_0}e^{-i\frac{\pi}{4}}\mu[j^0]e^{-i S^A[j^0,i^0]} \\
&e^{-i S^A_j[j_{\scr {1a}}]}
e^{-i\frac{\pi}{2}\delta i^{x}_1}
e^{-i\left( \sum_a G_{j_{1a}\;i^{x}_1}\;\delta j^{\scr{a1}}\right)\delta i^{x}_1}
	\\
	 &e^{-\frac{1}{2}\left(\frac{G^2_{i^{x}_1\;i^{y}_1}}{(\frac{1}{2\sigma_{i^{y}_1}}+iG^A_{i^{y}_1 i^{y}_1})}+iG^A_{i^{x}_1 i^{x}_1}\right)(\delta i^{\scr{x}}_1)^2
	     }
	 			 \end{split}
			 \end{equation}
The Gaussian in the last equation has variance 
\begin{equation}
	\sigma_{i^{x}_1}=\frac{1}{2}\left(\frac{G^2_{i^{x}_1\;i^{y}_1}}{(\frac{1}{2\sigma_{i^{y}_1}}+iG_{i^{y}_1 i^{y}_1})}+iG_{i^{x}_1 i^{x}_1}\right)^{-1}
\end{equation}
as in \cite{sr}. Proceeding in the same way, we fix $\sigma$ so that both  $\sigma_{i^{y}_1}$ and $\sigma_{i^{x}_1}$ are real quantities. Remarkably the auxiliary tetrahedron described by $S_R$  is isosceles and in this case $\sigma_{i^{x}_1}=\sigma_{i^{y}_1}=j^0/3$

The final form of the coefficient is then
		 \begin{equation}
	\begin{split}
		\Phi'_{\mathbf q}({\mathbf j}, i^{\scr{x}}_1,i_2,...,i_5)
			=&
			e^{-\frac{1}{2j^0}\sum \alpha_{\scr{(ij)(mr)}}\delta j^{\scr{ij}}\delta j^{\scr{mr}}+i\sum\Phi \delta j^{\scr{ij}}}
			e^{-\sum_{n\neq1}\left(\frac{(\delta i^{\scr{m_n}}_n)^2}{4\sigma_{i^{m_n}}}-i\left(\sum_a \frac{3}{4j^0}\delta j^{\scr{an}}-\frac{\pi}{2}\right) \delta i^{\scr{m_n}}_n\right)}   
	\;	\cdot \\
	 &\cdot N_1 e^{-i S_j[j_{\scr {1a}}]}e^{-\frac{1}{4}\frac{1}{\sigma_{i^{x}_1}}(\delta i^{\scr{x}}_1)^2}
e^{-i\left( \sum_a G_{j_{1a}\;i^{x}_1}\;\delta j^{\scr{a1}}+\frac{\pi}{2}\right)\delta i^{x}_1}
		 			 \end{split}
			 \end{equation}
			 where 
\begin{equation}
	N_1=\sqrt{\frac{\pi}{2(\frac{1}{2\sigma_{i^{y}_1}}+iG_{i^{y}_1 i^{y}_1})}} e^{i\pi i_0}e^{-i\frac{\pi}{4}}\mu[j^0]e^{-i S_R[j^0,i^0]} 
\end{equation}
and we have the result \eqref{cafine}. 

Summarizing, the parameters \eqref{tunings1} and \eqref{tunings2} are determined by the requirement that the gaussian has the same shape in all bases. 

\section{Simple gaussian integrals used in the calculation}
\begin{align}
	&\int_{-\infty}^{+\infty}d x^D \exp{-\frac{1}{2}x^a A_{ab} x^b}=\frac{(2\pi)^{\frac{D}{2}}}{\sqrt{det A}},\\
	&\int_{-\infty}^{+\infty}d x^D x^i x^j \exp{-\frac{1}{2}x^a A_{ab} x^b}=\frac{(2\pi)^{\frac{D}{2}}}{\sqrt{det A}}\;A^{-1}_{ij},\\
	&\int_{-\infty}^{+\infty}d x^D \exp{-\frac{1}{2}x^a A_{ab} x^b}+i\theta_a x^a=\frac{(2\pi)^{\frac{D}{2}}}{\sqrt{det A}}\exp{-\frac{1}{2}\theta^a A^{-1}_{ab} \theta^b},\\
	&\int_{-\infty}^{+\infty}d x^D x^i \exp{-\frac{1}{2}x^a A_{ab} x^b}+i\theta_a x^a=\frac{(2\pi)^{\frac{D}{2}}}{\sqrt{det A}}\;i\;A^{-1}_{ia}\theta^a\; \exp{-\frac{1}{2}\theta^a A^{-1}_{ab} \theta^b},\\
	&\int_{-\infty}^{+\infty}d x^D x^i x^j \exp{-\frac{1}{2}x^a A_{ab} x^b}+i\theta_a x^a=\frac{(2\pi)^{\frac{D}{2}}}{\sqrt{det A}}\;\;(A^{-1}_{ia}\theta^a\,A^{-1}_{jb}\theta^b-A^{-1}_{ij})\; \exp{-\frac{1}{2}\theta^a A^{-1}_{ab} \theta^b}\label{trasfmomenti},\\
	&\int_{-\infty}^{+\infty}d x\; x^m \exp{-\frac{1}{2}a x^2}+i\theta x=\sqrt{\frac{2\pi}{a}}(-i)^m\;\frac{\partial^m}{\partial \theta^m}\;\exp{-\frac{1}{2a} \theta^2}.
			\end{align}



\begin{thebibliography}{99}


\bibitem{lqg1}
  A. Ashtekar,
  ``An introduction to loop quantum gravity through cosmology",
  gr-qc/0702030. 
A. Ashtekar, J. Lewandowski, 
 ``Background independent quantum gravity: A status report",
 {\em Class Quant Grav} {\bf 21} (2004) R53-R152. 
T. Thiemann, \emph{Introduction to Modern Canonical Quantum General Relativity}
(CUP,  to appear).

\bibitem{book}
C. Rovelli, {\em Quantum Gravity}, (Cambridge University Press,
Cambridge, 2004). 

\bibitem{lqg2}  
C. Rovelli, L Smolin,
 ``Knot theory and quantum gravity'' 
\textit{Phys  Rev  Lett } \textbf{61} (1988) 1155-1158. 
C. Rovelli, L. Smolin,
``Loop space representation for quantum general relativity", 
\textit{Nuclear Physics} \textbf{B331}  (1990) 80-152. 
 A. Ashtekar, C. Rovelli, L. Smolin,
 ``Weaving a classical geometry with quantum threads",
\textit{Phys Rev Lett}  {\bf 69}, 237 (1992).
C. Rovelli, L. Smolin,
 ``Discreteness of Area and Volume in Quantum Gravity", 
\textit{Nucl Phys} \textbf{B442} (1995)  593-619;
   \textit{Nucl Phys} \textbf{B456}, 734 (1995). 
A. Ashtekar, J. Lewandowski, 
``Quantum Theory of Geometry I: Area Operators" 
{\em Class  Quantum Grav} {\bf 14}
(1997) A55-A82; 
``II : Volume Operators", 
{\em Adv Theo Math Phys} {\bf 1} (1997) 388-429


\bibitem{scattering1} L. Modesto, C. Rovelli, ``Particle scattering in loop quantum
gravity", {\em Phys. Rev. Lett.}  {\bf 95} (2005), 191301, [arXiv:
gr-qc/0502036].

\bibitem{scattering2}  C. Rovelli, ``Graviton propagator from background--independent
quantum gravity" {\em Phys. Rev. Lett.} {\bf 97} (2006), 151301, [arXiv:
gr-qc/0508124].

\bibitem{scattering3}  E. Bianchi, L. Modesto, C. Rovelli, S. Speziale, ``Graviton propagator
in loop quantum gravity", {\em Class. Quant. Grav.}  23 (2006), 6989-7028;
[arXiv: gr-qc/0604044].

\bibitem{3d} S.Speziale ``Towards the graviton from spinfoams: the 3d toy model'', JHEP 0605 (2006) 039, [arXiv:gr-qc/0512102].
 
\bibitem{boundarystate} E. R. Livine, S. Speziale, ``Group Integral Techniques for the Spinfoam Graviton Propagator'', {\em JHEP} {\bf 0611} (2006) 092, [arXiv:gr-qc/0608131].

\bibitem{corrections} E. Livine, S. Speziale, J. Willis, 
``Towards the graviton from spinfoams: higher order corrections in the 3d toy model", 
{\em Phys \ Rev  D}   {\bf 75}, 024038 (2007).

\bibitem{Dittrich:2007wm}
  B.~Dittrich, L.~Freidel and S.~Speziale,
  ``Linearized dynamics from the 4-simplex Regge action'', 
  arXiv:0707.4513 [gr-qc].

\bibitem{BC} J.W. Barrett and L. Crane, ``Relativistic spin networks and quantum gravity", 
{\em J. Math. Phys.} {\bf 39}, 3296 (1998) [gr-qc/9709028].

\bibitem{baez} 
JC Baez, JD Christensen, TR Halford, DC Tsang, 
``Spin Foam Models of Riemannian Quantum Gravity", 
\textit{Class  Quant  Grav } \textbf{19} (2002) 4627-4648
[arXiv:gr-qc/0202017v4].

\bibitem{II} E. Alesci, C. Rovelli, ``The complete LQG propagator: II. Asymptotics of the vertex'' to appear. 

\bibitem{variants}
 R.~De Pietri, L.~Freidel, K.~Krasnov and C.~Rovelli,
  ``Barrett-Crane model from a Boulatov-Ooguri field theory over a  homogeneous
  space,''
  Nucl.\ Phys.\  B {\bf 574}, 785 (2000)
  [arXiv:hep-th/9907154].
D Oriti, RM Williams,
 ``Gluing 4-simplices: a derivation of the Barrett-Crane spinfoam model for Euclidean quantum gravity",
\textit{Phys  Rev } D\textbf{63}(2001) 024022. 
A.~Perez and C.~Rovelli,
  ``Finite SO(4)-state sum model of Euclidean GR,''
{\it Prepared for 9th Marcel Grossmann Meeting on Recent Developments in Theoretical and Experimental General Relativity, Gravitation and
Relativistic Field Theories (MG 9), Rome, Italy, 2-9 Jul 2000}
  A.~Perez and C.~Rovelli,
  ``A spin foam model without bubble divergences,''
  {\em Nucl.\ Phys.\  B} {\bf 599}, 255 (2001)
  [arXiv:gr-qc/0006107].


\bibitem{Reisenberger:1997sk}
  M.~P.~Reisenberger,
  ``A lattice worldsheet sum for 4-d Euclidean general relativity,''
  [arXiv:gr-qc/9711052].

\bibitem{vertice nuovo1}
J. Engle, R. Pereira, C. Rovelli ``The loop-quantum-gravity vertex-amplitude'', [arXiv:0705.2388].

\bibitem{Livine:2007vk}
  E.~R.~Livine and S.~Speziale,
  ``A new spinfoam vertex for quantum gravity,''
  [arXiv:0705.0674,  gr-qc].

\bibitem{sergei}   
     S. Alexandrov,  ``Spin foam model from canonical quantization'', arXiv:0705.3892 [gr-qc].

\bibitem{spinnetworks} C. Rovelli, L. Smolin  
``Spin Networks and Quantum Gravity'', {\em Phys.Rev. D} {\bf 52} (1995) 5743-5759, [arXiv:gr-qc/9505006].


\bibitem{sr}  S. Speziale, C. Rovelli, ``A semiclassical tetrahedron", {\em Class.Quant.Grav.} {\bf 23} (2006) 5861-5870 [arXiv: gr-qc/0606074].

\bibitem{Regge}
G.Ponzano and T.Regge, 
``Semiclassical limit of Racah coefficients'' in \textit{Spectroscopic and group
theoretical methods in physics} (Bloch ed.), North-Holland, 1968.

\bibitem{Freidel:2002mj}
  L.~Freidel and D.~Louapre,
  ``Asymptotics of 6j and 10j symbols,''
  {\em Class.\ Quant.\ Grav.}  {\bf 20}, 1267 (2003)
  [arXiv:hep-th/0209134]. 
   
\bibitem{Baez:2002rx}
  J.~W.~Barrett and R.~M.~Williams,
  ``The asymptotics of an amplitude for the 4-simplex,''
  {\em Adv.\ Theor.\ Math.\ Phys.}  {\bf 3} (1999) 209
  [arXiv:gr-qc/9809032].
    J.~C.~Baez, J.~D.~Christensen and G.~Egan,
  ``Asymptotics of 10j symbols,''
  {\em Class.\ Quant.\ Grav.}  {\bf 19} (2002) 6489
  [arXiv:gr-qc/0208010].
J.~W.~Barrett and C.~M.~Steele,
  ``Asymptotics of relativistic spin networks,''
  {\em Class.\ Quant.\ Grav.}  {\bf 20} (2003) 1341
  [arXiv:gr-qc/0209023].
    J.~D.~Christensen and G.~Egan,
  ``An efficient algorithm for the Riemannian 10j symbols,''
  {\em Class.\ Quant.\ Grav.}  {\bf 19} (2002) 1185
  [arXiv:gr-qc/0110045].
  
\bibitem{GFT}
 D.~Oriti,
  ``The group field theory approach to quantum gravity,''
  [arXiv:gr-qc/0607032].
 L Freidel, 
 ``Group Field Theory: an overview'',  
\textit{Int. J. Theor. Phys.} {\bf 44} (2005) 1769-1783.

\bibitem{BS}
D. M. Brink, G. R. Satchler, \textit{Angular Momentum} (Claredon Press, Oxford, 1968).

\bibitem{depietri1}
R. De Pietri, C. Rovelli  ``Geometry Eigenvalues and Scalar Product from recoupling Theory
in Loop Quantum Gravity'', {\em Phys.\ Rev.\ D} {\bf 54} (1996) 2664-2690 , [arXiv:gr-qc/9602023].

\bibitem{depietri2}
R. De Pietri ``On the relation between the connection and
the loop representation of quantum gravity'', {\em  Class.\  Quant.\ Grav.} {\bf 14} (1997) 53-70,   [arXiv:gr-qc/9605064].


\end{thebibliography}
\end{document}